\documentclass[twocolumn]{aastex63}

\usepackage{color,graphicx,natbib}
\usepackage{longtable}
\usepackage{graphicx}
\usepackage{amsmath}
\usepackage{amssymb}
\usepackage{bm}
\usepackage[toc,page]{appendix}
\usepackage{url}

\def\ra#1#2#3{#1$^{\rm h}$ #2$^{\rm m}$ #3$^{\rm s}$}
\def\dec#1#2#3{$#1^\circ #2' #3''$}

\shortauthors{Schroeder et al.}
\shorttitle{Radio-selected Dark GRBs}

\begin{document}

\author[0000-0001-9915-8147]{Genevieve~Schroeder}
\affiliation{Center for Interdisciplinary Exploration and Research in Astrophysics and Department of Physics and Astronomy, Northwestern University, 2145 Sheridan Road, Evanston, IL 60208-3112, USA}

\author[0000-0003-1792-2338]{Tanmoy~Laskar}
\affiliation{Department of Astrophysics/IMAPP, Radboud University, PO Box 9010,
6500 GL, The Netherlands}

\author[0000-0002-7374-935X]{Wen-fai~Fong}
\affiliation{Center for Interdisciplinary Exploration and Research in Astrophysics and Department of Physics and Astronomy, Northwestern University, 2145 Sheridan Road, Evanston, IL 60208-3112, USA}

\author[0000-0002-2028-9329]{Anya~E.~Nugent}
\affiliation{Center for Interdisciplinary Exploration and Research in Astrophysics and Department of Physics and Astronomy, Northwestern University, 2145 Sheridan Road, Evanston, IL 60208-3112, USA}

\author[0000-0002-9392-9681]{Edo~Berger}
\affiliation{Center for Astrophysics \textbar{} Harvard \& Smithsonian, 60 Garden Street, Cambridge, MA 02138-1516, USA}

\author[0000-0002-7706-5668]{Ryan~Chornock}
\affiliation{Department of Astronomy, University of California, Berkeley, CA 94720-3411, USA}

\author[0000-0002-8297-2473]{Kate~D.~Alexander}
\thanks{NASA Einstein Fellow}
\affiliation{Center for Interdisciplinary Exploration and Research in Astrophysics and Department of Physics and Astronomy, Northwestern University, 2145 Sheridan Road, Evanston, IL 60208-3112, USA}

\author[0000-0003-0123-0062]{Jennifer~Andrews}\affiliation{Steward Observatory, University of Arizona, 933 North Cherry Avenue, Tucson, AZ 85721-0065, USA}

\author{R.~Shane~Bussmann}\affiliation{Department of Astronomy, Space Science Building, Cornell University, Ithaca, NY 14853-6801, USA}

\author[0000-0003-2999-3563]{Alberto~J.~Castro-Tirado}\affiliation{Instituto de Astrof\'{\i}sica de Andaluc\'{\i}a (IAA-CSIC), Glorieta de la Astronom\'{\i}a s/n, E-18008,
Granada, Spain}

\author[0000-0001-9652-8384]{Armaan~V.~Goyal}\affiliation{Department of Astronomy, Indiana University, Bloomington, IN 47405, USA}

\author[0000-0002-5740-7747]{Charles D.~Kilpatrick}\affiliation{Center for Interdisciplinary Exploration and Research in Astrophysics and Department of Physics and Astronomy, Northwestern University, 2145 Sheridan Road, Evanston, IL 60208-3112, USA}

\author[0000-0002-4443-6725]{Maura~Lally}\affiliation{Department of Astronomy, Cornell University, Ithaca, NY 14853, USA}

\author[0000-0001-9515-478X
]{Adam~Miller}\affiliation{Center for Interdisciplinary Exploration and Research in Astrophysics and Department of Physics and Astronomy, Northwestern University, 2145 Sheridan Road, Evanston, IL 60208-3112, USA}

\author[0000-0002-0370-157X]{Peter~Milne}\affiliation{University of Arizona, Steward Observatory, 933 N. Cherry Avenue, Tucson, AZ 85721, USA}

\author[0000-0001-8340-3486]{Kerry~Paterson}\affiliation{Max-Planck-Institut f\"ur Astronomie (MPIA), Königstuhl 17, 69117 Heidelberg, Germany}

\author[0000-0003-3937-0618]{Alicia~Rouco~Escorial}\affiliation{Center for Interdisciplinary Exploration and Research in Astrophysics and Department of Physics and Astronomy, Northwestern University, 2145 Sheridan Road, Evanston, IL 60208-3112, USA}

\author[0000-0002-3019-4577 ]{Michael~C.~Stroh}\affiliation{Center for Interdisciplinary Exploration and Research in Astrophysics and Department of Physics and Astronomy, Northwestern University, 2145 Sheridan Road, Evanston, IL 60208-3112, USA}

\author[0000-0003-0794-5982]{Giacomo~Terreran}\affiliation{Las Cumbres Observatory, 6740 Cortona Drive, Suite 102, Goleta, CA 93117-5575, USA}

\author[0000-0003-1152-518X]{Bevin Ashley~Zauderer}\affiliation{National Science Foundation, 2415 Eisenhower Avenue, Alexandria, Virginia 22314, USA }

\title{A Radio-selected Population of Dark, Long Gamma-ray Bursts: Comparison to the Long Gamma-ray Burst Population and Implications for Host Dust Distributions}

\begin{abstract}
    We present cm-band and mm-band afterglow observations of five long-duration $\gamma$-ray bursts (GRBs; GRB\,130131A, 130420B, 130609A, 131229A, 140713A) with dust-obscured optical afterglow emission, known as
    ``dark'' GRBs. We detect the radio afterglow of two of the dark GRBs (GRB\,130131A and 140713A), along with a tentative detection of a third (GRB\,131229A) with the Karl G. Jansky Very Large Array (VLA). Supplemented by three additional VLA-detected dark GRBs from the literature, we present uniform modeling of their broadband afterglows. We derive high line-of-sight dust extinctions of $A_{V, \rm GRB} \gtrsim 2.2 - 10.6~{\rm mag}$. Additionally, we model the host galaxies of the six bursts in our sample, and derive host galaxy dust extinctions of $A_{V, \rm Host} \approx 0.3-4.7~{\rm mag}$. Across all tested $\gamma$-ray (fluence and duration) and afterglow properties (energy scales, geometries and circumburst densities), we find dark GRBs to be representative of more typical unobscured long GRBs, except in fluence, for which observational biases and inconsistent classification may influence the dark GRB distribution.
    Additionally, we find that $A_{V, \rm GRB}$ is not related to a uniform distribution of dust throughout the host, nor to the extremely local environment of the burst, indicating that a larger scale patchy dust distribution is the cause of the high line-of-sight extinction. 
    Since radio observations are invaluable to revealing heavily dust-obscured GRBs, we make predictions for the detection of radio emission from host star formation with the next generation VLA. 
\end{abstract}

\section{Introduction}
\label{sec:Introduction}
% Dark GRBs are best GRBs

Long duration $\gamma$-ray bursts (GRBs), the most luminous, energetic transients in the universe (e.g. \citealt{Racusin2008}), are associated with the death of massive stars \citep{MacFadyen1999} and Type Ic supernovae (e.g. \citealt{Hjorth2003, Woosley2006}). The relativistic shocks produced by these bursts interact with the surrounding medium, leading to the production of broad-band synchrotron emission from radio to X-rays, known as the ``afterglow''.

A subset of long GRBs have suppressed optical emission, earning them the moniker of optically ``dark'' GRBs, with the first documented dark GRB being GRB\,970828 \citep{Groot1998}. It is estimated that $\sim 10\%$-$50\%$ of the long GRB population are among the dark GRB class \citep{Jakobsson2004, Cenko2009_darkGRBs, Fynbo2009, Greiner2011, Melandri2012, Perley2013_DarkGRB_AV}. One of the prominent causes of the darkness is attributed to dust extinction along the line-of-sight to the GRB, though other causes, such as intrinsically faint and rapidly fading bursts (e.g. \citealt{Berger2002}), or ${\rm Ly}\alpha$ absorption from a high redshift ($z \gtrsim 6$) origin \citep{Haislip2006, Salvaterra2009, Tanvir2009, Cucchiara2011} can also produce optically faint or undectectable afterglows. For the dust-extinguished population of dark GRBs, the amount of extinction ($A_{V, \rm GRB}$) along the line-of-sight is of interest as it can provide insight on the amount of dust and distribution within their host galaxies \citep{Greiner2011, Zafar2011, Kruhler2011, Zauderer2013, Perley2013_DarkGRB_AV}. Typically, $A_{V, \rm GRB}$ is estimated using simple power law arguments extending from the X-rays to the optical bands (e.g. \citealt{Perley2013_DarkGRB_AV}). However, the addition of a detected radio afterglow can allow for proper afterglow modeling and provide a robust measurement of $A_{V, \rm GRB}$. 

Since the first radio afterglow of a GRB was discovered in 1997 (GRB\,970508, \citealt{Frail1997_FirstRadio}), radio follow-up has been vital to our general understanding of GRB afterglow behavior, helping to constrain the energetics and environment of GRBs. A fairly comprehensive census of radio follow-up of 304 GRB afterglows found that only $\sim 30\%$ of GRB afterglows have radio detections \citep{ChandraFrail2012}. However, it was postulated that the low fraction of radio detections is likely attributed to the limited sensitivity of prior generations of radio telescopes  \citep{ChandraFrail2012, Osborne2021}, and therefore more sensitive radio facilities could increase the fraction of detected radio afterglows of GRBs. The upgrade to the Karl G. Jansky Very Large Array (VLA), which concluded in 2012, increased the sensitivity of the radio array by a factor of $\sim 10$ \citep{EVLA2011}, and has allowed for well-sampled, multi-band, follow-up of long GRB radio afterglows (including, but not limited to: \citealt{ChandraFrail2012, Laskar2013_lbz13, Zauderer2013, Laskar2014_tl14, Laskar2015_tl15, Laskar2016_160509A_lab16, Alexander2017_alb17, LaskarBC2018_GRBsI_lbc18, LaskarBM2018_GRBsII_lbm18, LaskarAB2018_lab18}). Radio observations of dust obscured dark GRBs are critical for properly modeling the afterglow of these bursts, especially when the optical afterglow is not detected (e.g. \citealt{Jakobsson2005, CastroTirado2007, Rol2007, Zauderer2013, VanderHorst2015, Higgins2019, KangasFruchter2021}), as the low-frequency afterglow can constrain burst energetics and environment density through determination of the synchrotron break frequencies (e.g. \citealt{Sari1998, GS2002}). 

In addition to the properties of the afterglow of dark GRBs, there has been much interest in the host galaxies of dark GRBs and how they compare to the typical long GRB population (e.g. \citealt{Kruhler2011, Perley2013_DarkGRB_AV}). The {\it Swift} X-ray Telescope ({\it Swift}/XRT) typically provides afterglow positions of $\sim 2\arcsec$ \citep{Evans2009}, while the detection of a radio afterglow with the VLA can often provide unambiguous association to a host galaxy via sub-arcsecond precision, especially in the case of optically-faint or non-detected dark GRBs (e.g. \citealt{Zauderer2013}). Previous studies based on small numbers have found that the hosts of dust-obscured bursts are overall more massive, more luminous, more star forming, and dustier than other long GRB hosts (i.e. \citealt{Kruhler2011, Perley2013_DarkGRB_AV}). Long GRBs in general are already associated with star forming galaxies (i.e. \citealt{Djorgovski1998, Christensen2004, Japelj2016_LGRB_Bat6_II, Palmerio2019_LGRB_Bat6_III}), and the obscuration of dark GRBs may point to obscured star formation in their hosts \citep{Blain2000, Ramirez-Ruiz2002}.

Here, we present the multi-wavelength observations for five dark GRBs, including VLA observations for all five bursts. We present the discovery of the VLA radio afterglow for one burst (GRB\,130131A), present a nominal VLA radio detection for another burst (GRB\,131229A), present new radio and millimeter detections for an additional event (GRB\,140713A), and new upper limits for two events. For our three VLA detected/nomiminally detected dark GRBs, we uniformly model the afterglows and host galaxies, and include uniform models of the afterglow and host galaxies of three VLA detected dark bursts from the literature. We proceed to compare the afterglow and host properties of dark GRBs to the larger population of typical long GRBs that have not been classified as dark. 
In Section~\ref{sec:Sample} we describe our sample and our criteria for classifying GRBs as dark. In Section~\ref{sec:afterglowmodeling} we describe our methods for self-consistent modeling of all available broadband afterglow data and we apply our afterglow modeling to the relevant bursts in our sample to extract their local environment, burst energetics and microphysical parameters. In Section~\ref{sec:Prospector} we present host galaxy modeling for the bursts with robust host galaxy associations and spectroscopic or photometric data. In Section~\ref{sec:Discussion} we compare our dark GRB sample to the broader population of long GRBs in terms of their $\gamma$-ray, afterglow, and host properties, and consider the detectability of obscured star formation in long GRB host galaxies. We conclude in Section \ref{sec:Conclusions}. We present the details of the multi-wavelenth observations and data reduction in Appendix~\ref{appendix:Observations}. In this paper, we employ the $\Lambda$CDM cosmological parameters of $H_{0} = 68~{\rm km \, s}^{-1} \, {\rm Mpc}^{-1}$, $\Omega_{M} = 0.31$, $\Omega_{\rm \Lambda} = 0.69$.

\section{Sample Selection and Classification Method}
\label{sec:Sample}

Our primary goal is to uniformly model the afterglows and host galaxies of dark GRBs. We focus on dark GRBs with VLA observations taken after the upgrade \citep{EVLA2011}, specifically those taken with our Programs 13A-046, 13A-541, and 14A-344 (PI: Berger)\footnote{We exclude GRB\,130606A as the darkness of this burst is likely attributed to its high redshift of $z \approx 5.91$ \citep{2013GCN.14796_130606A, Chornock2013_130606A, LittleJohns2015}}. This sample includes five dark GRBs, two of which have unambiguous radio afterglow detections, and one of which has a tentative radio afterglow detection. We supplement this sample with three other dark GRBs with VLA detections, GRB\,110709B \citep{Zauderer2013}, 111215A \citep{Zauderer2013, VanderHorst2015}\footnote{While the upgrade to the VLA completed at the end of 2012, the C-band (4-8~GHz) and K-band (18-26.5~GHz) receivers were upgraded by 2011 \citep{EVLA2011}. As the majority of the VLA observations of GRB\,110709B and GRB\,111215A were taken with these upgraded receivers, we include these bursts in our sample.}, and GRB\,160509A (\citealt{Laskar2016_160509A_lab16}, classifying this latter burst as dark for the first time in Section \ref{sec:160509A_literature}). As 14 dark GRBs have been observed by the VLA since its upgrade (\citealt{2012GCN.12895_GRB120119A, Zauderer2013, Laskar2013_130606A, Veres2015_130907A, Horesh2015_130925A,  Laskar2016_160509A_lab16}, the VLA Data Archive, This Work), our sample of 8 comprises over half of the known dark GRBs with upgraded VLA observations. The eight 8 dark GRBs featured in this paper are listed in Table~\ref{tab:Obs_summary}. 

Standard measures of classifying long GRBs as ``dark'' in the standard synchrotron model (\citealt{GS2002} and Section~\ref{sec:afterglowmodeling}), involve the optical-to-X-ray spectral index $\beta_{\rm OX}$ and the X-ray spectral index $\beta_{\rm X}$. One such method of classification uses the expected spectral index between the optical and X-ray afterglow fluxes of $-p/2 < \beta_{\rm OX} < (1-p)/2 $, where $p$ is the power-law index of the electron energy distribution, $F_{\nu} \propto \nu^{\beta}$, and the spectral index is dependent on the location of the cooling frequency ($\nu_{\rm c}$) in relation the optical and X-ray bands. As $p > 2$ is typically expected (e.g. \citealt{Sari1998}), the shallowest expected $\beta_{\rm OX} = -0.5$, where $p = 2$. Therefore, if $\beta_{\rm OX} > -0.5$, then the shallow spectral slope indicates the optical afterglow has been suppressed and the burst can be considered dark \citep{Jakobsson2004}. An alternate definition of afterglow darkness is $\beta_{\rm OX} > \beta_{\rm X} + 0.5$, corresponding to an optical flux that is even lower than the shallowest possible extrapolation from the X-rays, $\nu_{\rm c} \approx \nu_{\rm X}$, in the synchrotron framework \citep{vanderHorst2009}. For the purposes of this paper, we will consider any burst ``dark'' if they meet the \citet{Jakobsson2004} criterion of $\beta_{\rm OX} > -0.5$, though we will note whether the bursts in our sample meet the \citet{vanderHorst2009} classification criterion as well. 

To determine $\beta_{\rm OX}$, we interpolate the X-ray light curve to the times of the optical observations, using least squares fits to the X-ray light curves to calculate the X-ray temporal index $\alpha_{\rm X}$ (where $F_{\nu} \propto t^{\alpha}$), and calculate $\beta_{\rm OX}$, correcting the optical observations for Galactic extinction in the direction of the burst \citep{sf11}. To determine $\beta_{\rm X}$, we create time-sliced spectra from the {\it Swift} tool
\footnote{\url{https://www.swift.ac.uk/user_objects/docs.php\#specform}}, which calculates the photon index, $\Gamma_{\rm X}$ \citep{Evans2009},
from which we derive $\beta_{\rm X} \equiv 1 - \Gamma_{\rm X}$. On a per-burst basis, we exclude any times over which the X-ray light curve exhibits flaring activity super-imposed on the power-law afterglow, as this emission is not likely to originate from the external shock \citep{Burrows2007, Margutti2010}. 

\begin{figure*}
\centering
\includegraphics[width=0.3\textwidth]{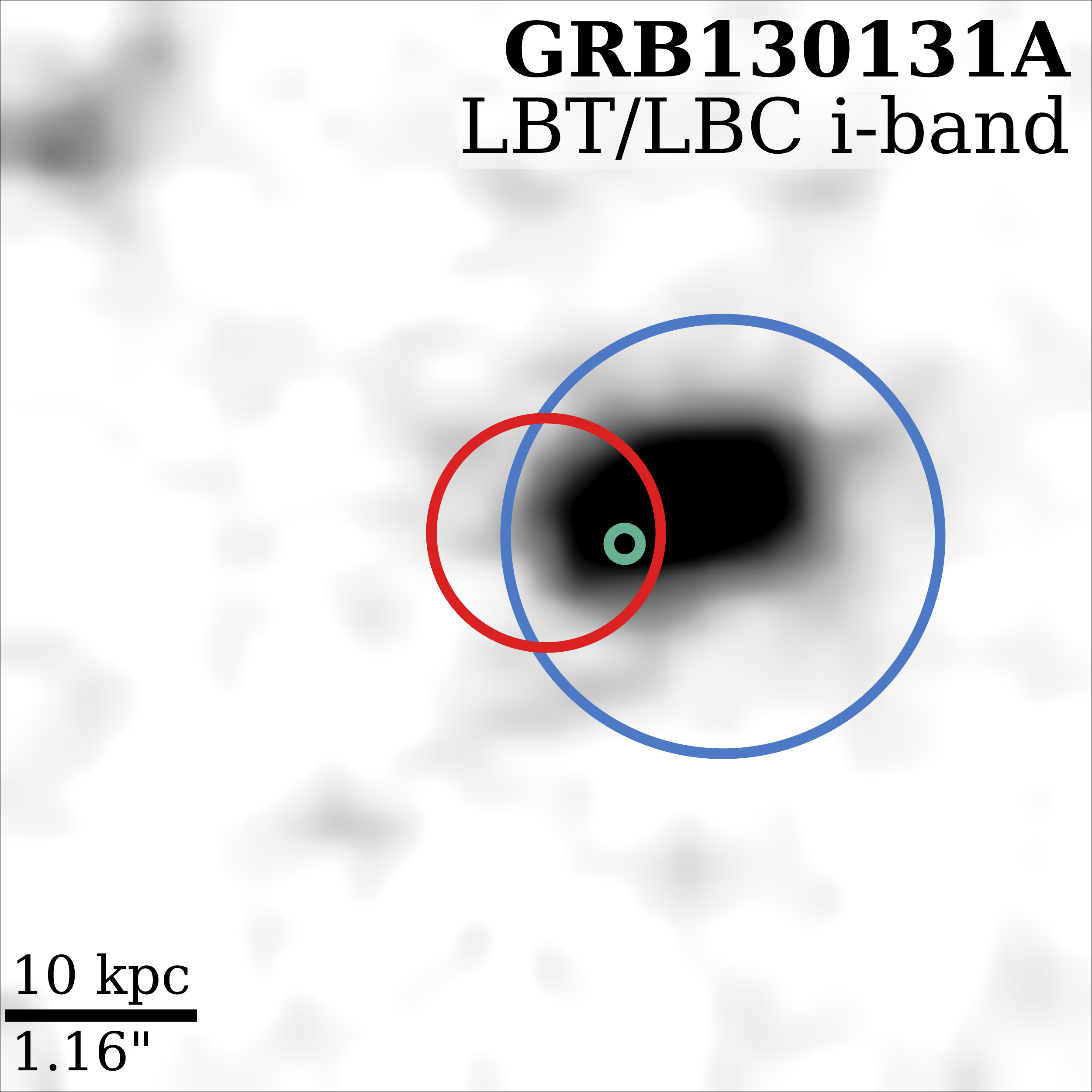}
\includegraphics[width=0.3\textwidth]{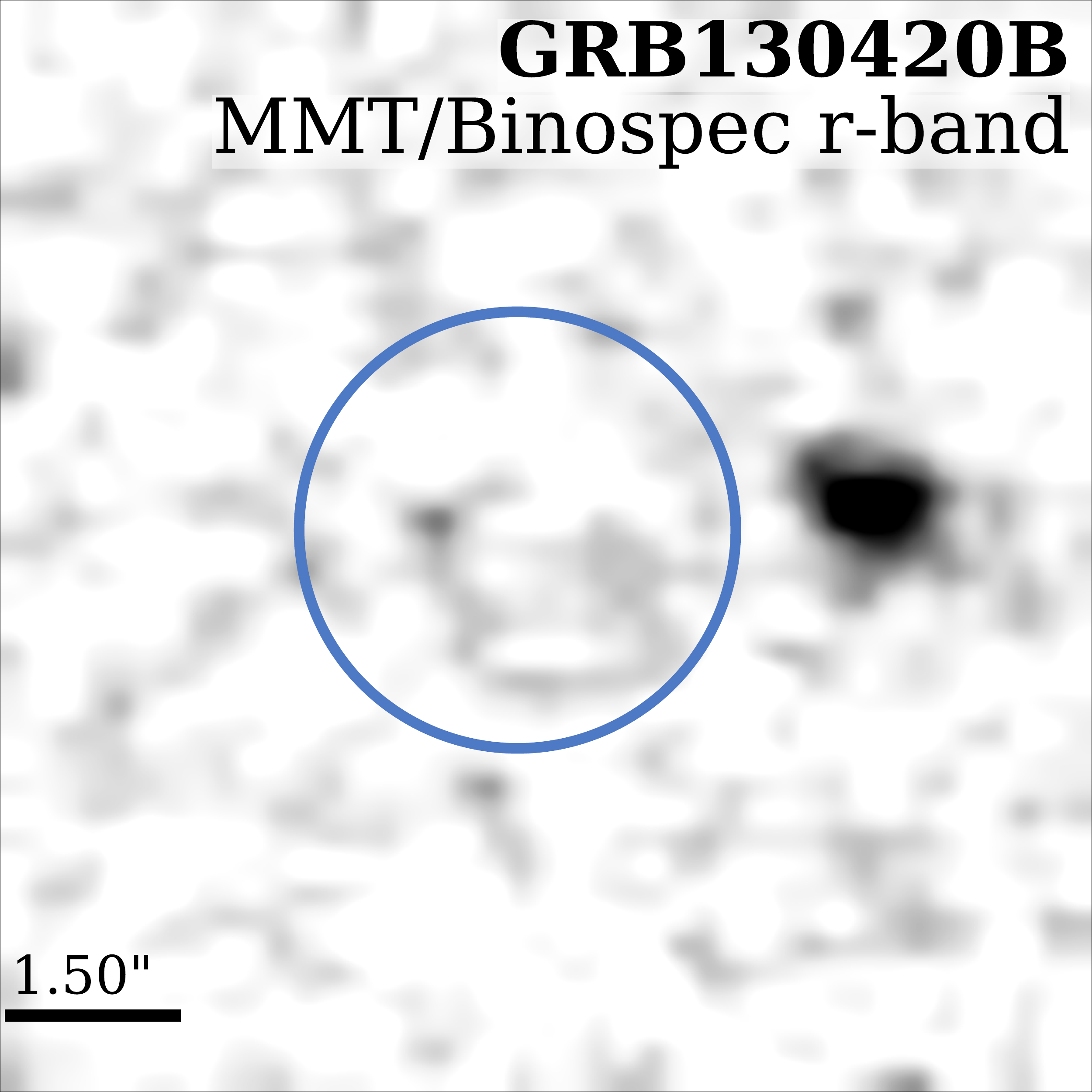}
\includegraphics[width=0.3\textwidth]{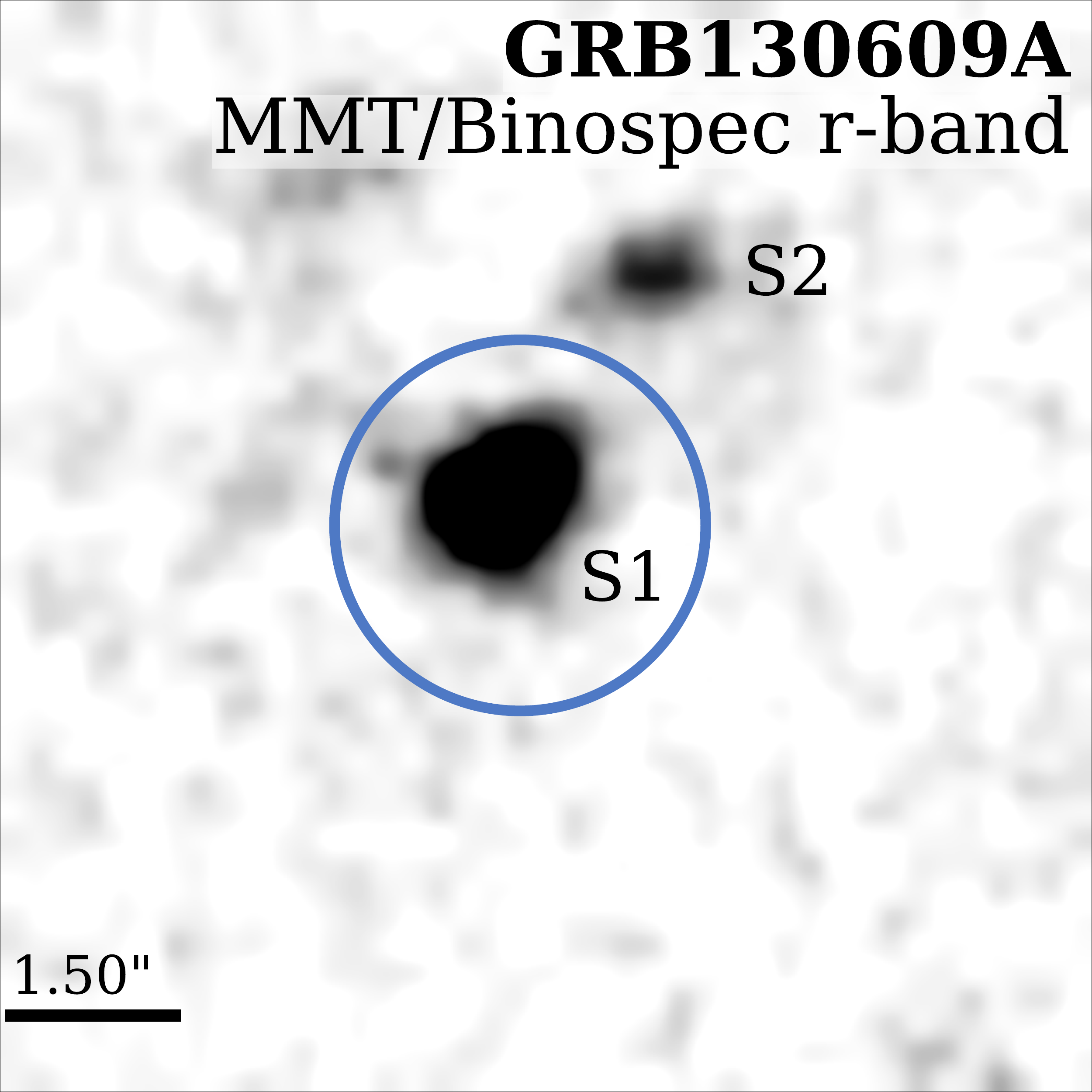}
\includegraphics[width=0.3\textwidth]{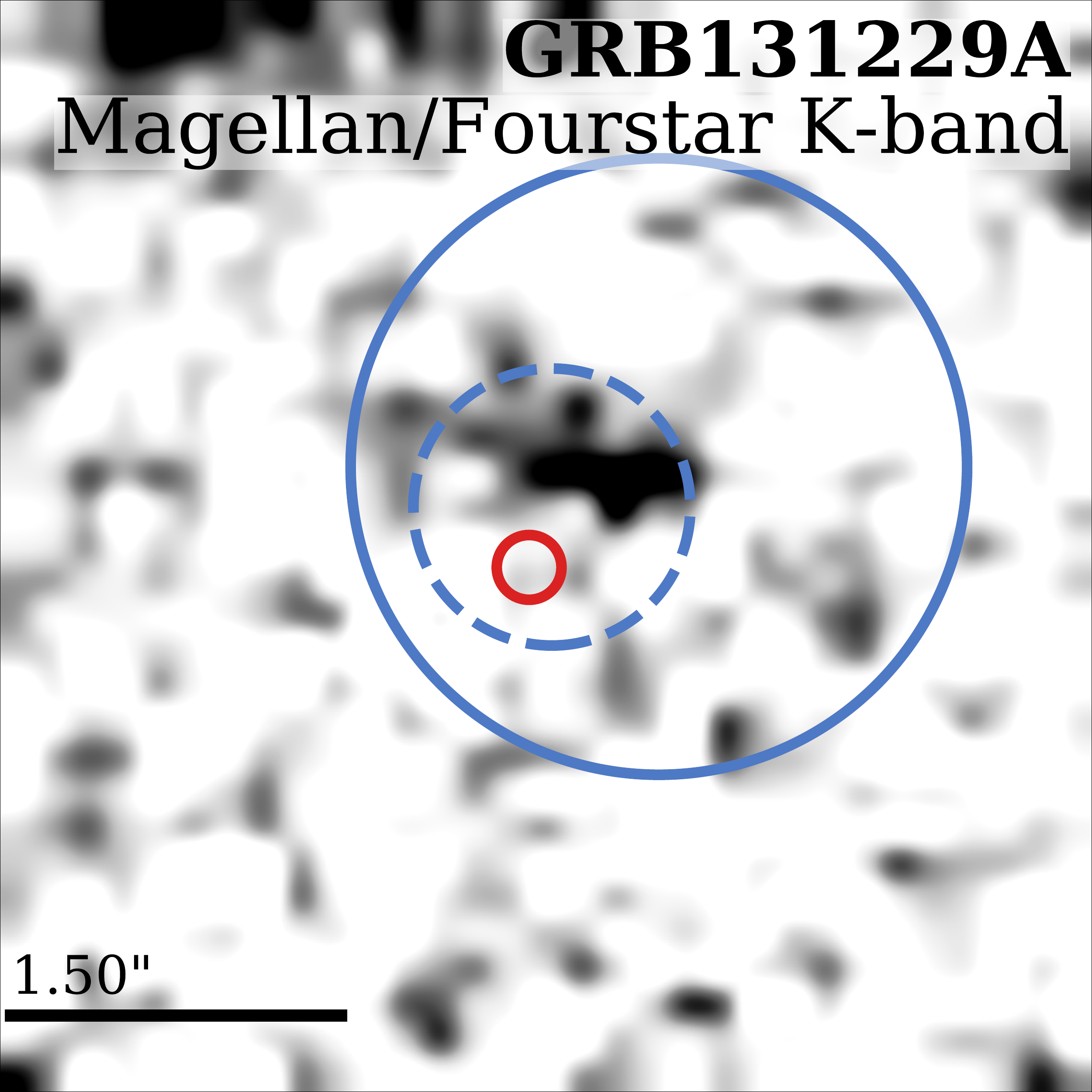}
\includegraphics[width=0.3\textwidth]{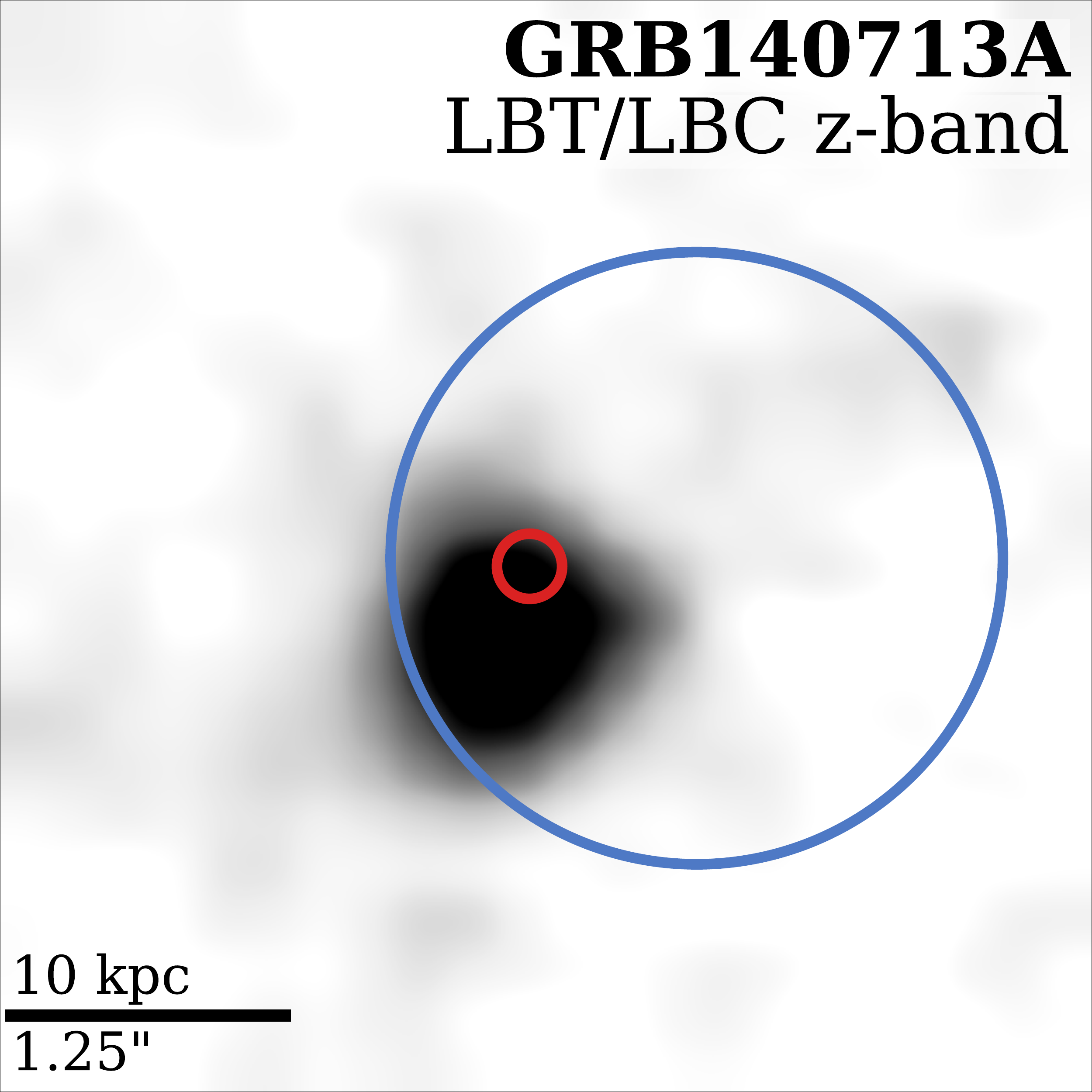}
\includegraphics[width=0.3\textwidth]{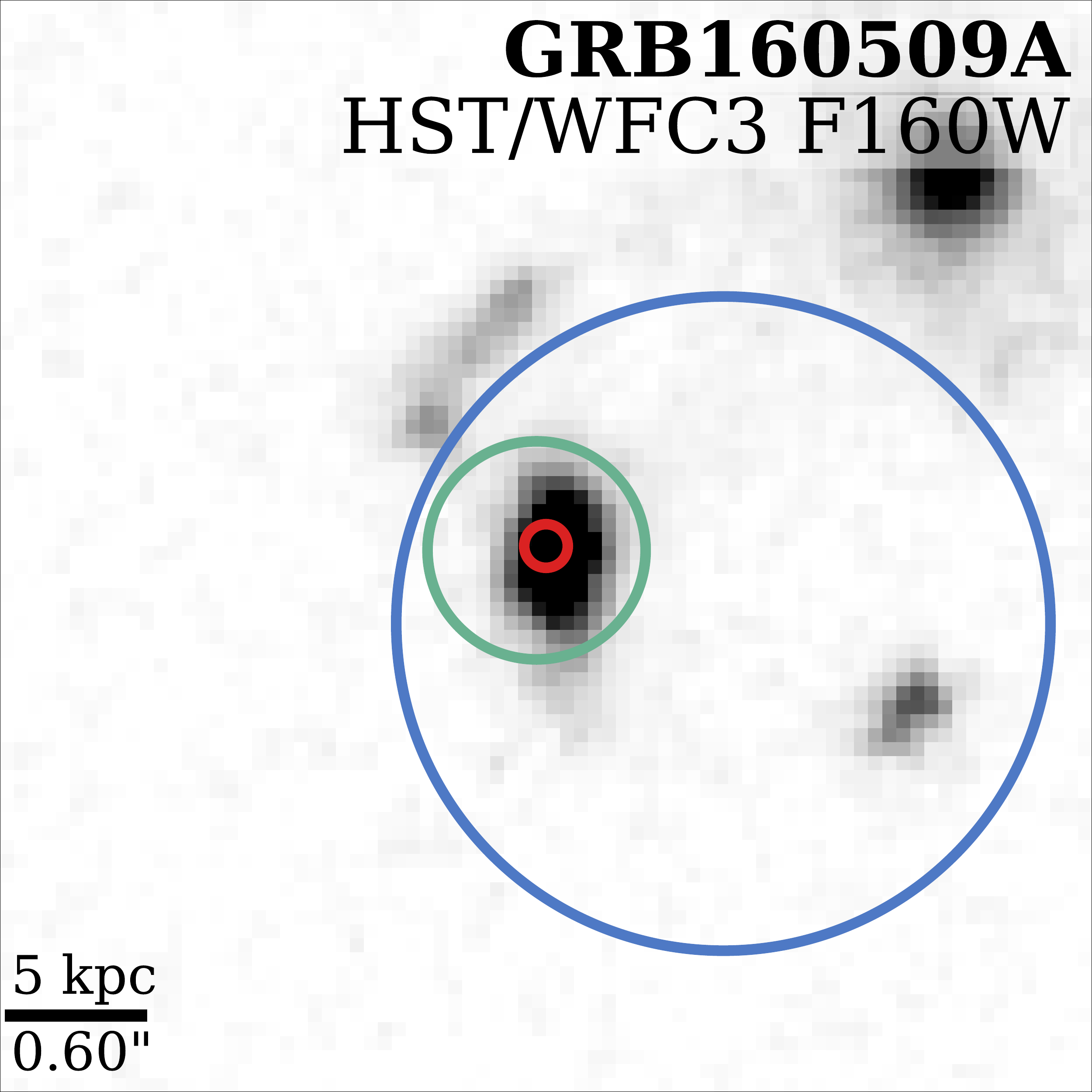}
\caption{Deep optical or NIR observations of the fields of GRBs\,130131A, 130420B, 130609A, 131229A, 140713A, and 160509A. In each panel, the blue solid-line circle indicates the XRT afterglow location (90\% confidence). When available, the red circle indicates the radio afterglow location ($1\sigma$), the green circle denotes the optical or NIR position ($1\sigma$), and the dashed blue circle indicates the {\it CXO} position. Images have been smoothed with a 3-pixel Gaussian, and the orientation is North up and East to the left. An angular scale is provided in all images, in addition to a physical scale when the redshift is known. For GRBs\,130420B and 130609A, we do not identify host galaxies.}
\label{fig:Host_Galaxy_Panel}
\end{figure*}

The full details of the X-ray, optical, near-infrared (NIR) and radio observations of the GRB afterglows and host galaxies, as well as their classifications as dark GRBs are presented in Appendix~\ref{appendix:Observations}. We show the fields and afterglow localizations of six dark GRBs in our sample in Figure~\ref{fig:Host_Galaxy_Panel}, and their afterglow detectabilities, as well as whether we model their afterglows and host galaxies, are summarized in Table~\ref{tab:Obs_summary}. 

%%%%% IMAGING FIGURE
%%%%% IMAGING FIGURE

%%%% Summary OBS TABLE
%%%%%%%%%%%%%%% TABLE
\tabletypesize{\normalsize}
\begin{deluxetable}{l|cccc}
\tablecolumns{4}
\tablewidth{0pc}
\tablecaption{Observation Summary
\label{tab:Obs_summary}}
\tablehead {
\colhead {GRB}                &
\colhead{Opt and/or NIR/Radio$^{a}$}           &
\colhead{Modeled$^{b}$}           &
\colhead {References}  
% \colhead {}    &
% \colhead{(J2000)}   &
% \colhead{(J2000)}   &
% \colhead{($\arcsec$)}   &
% \colhead{(s)}               &
% \colhead {($\times 10^{-7} {\rm erg}/{\rm cm}^{2}$)}          &
% \colhead {${\rm SFR}~[M_{\odot/~{\rm yr}}]$}           &
% \colhead {$t_{m}$ [Gyr]}          &
% \colhead {$A_{V}^{\rm host}$~[mag]}          
}
\startdata
110709B & N/Y & Y & 1, 2, 3\\
111215A & N/Y & Y & 1, 4, 5\\
130131A & Y/Y & Y & 5, 6 \\
130420B & N/N & N &  6 \\
130609A & N/N & N &  6 \\
131229A & N/Y$^{c}$ & Y$^{d}$ & 5, 6\\
140713A & N/Y & Y & 6, 7 \\
160509A & Y/Y & Y$^{e}$ & 6, 8, 9 \\
\enddata
\tablecomments{ $^{a}$ Whether or not the afterglow was detected in the optical/NIR or radio.\\
$^{b}$ Whether or not we model the afterglow and host galaxy of the burst in this paper \\
$^{c}$ We find a tentative 6.0~GHz detection for GRB\,131229A (see Section~\ref{sec:131229A_radio})
\\
$^{d}$ Due to the limited radio data, we present a simple analytical afterglow model for GRB\,131229A in Section~\ref{sec:131229A_BasicConsiderations}, and analytically derive afterglow properties. We model the host galaxy of GRB\,131229A in this work.
\\
$^{e}$ The X-ray to radio afterglow of GRB\,160509A was modeled within our modeling framework in \citet{Laskar2016_160509A_lab16}. We model the host galaxy of GRB\,160509A in this work.\\
{\bf References.} (1) \citet{Zauderer2013}; (2) \citet{Perley2016_I}; (3) \citet{Selsing2019}; (4) \citet{VanderHorst2015}; (5) \citet{Chrimes2019}; (6) This Work; (7) \citet{Higgins2019}; (8) \citet{Laskar2016_160509A_lab16}; (9) \citet{Kangas2020_160509A}} 
\end{deluxetable}
%%%%%%%%%%%%%%% TABLE
%%%% Summary OBS TABLE

\section{Afterglow Modeling}
\label{sec:afterglowmodeling}

We will consider the X-ray, optical, and radio afterglow light curves and spectral energy distributions (SEDs) of the GRBs in our sample in the context of synchrotron emission from the acceleration of electrons from a relativistic blast wave (i.e. \citealt{Sari1998, ChevalierLi2000, PanaitescuKumar2000, GS2002}). These electrons are accelerated to a non-thermal power law distribution, $ N(\gamma_{\rm e}) \propto \gamma_e^{-p}$. The afterglow SEDs can be described by power law segments which connect at three break frequencies (the self absorption frequency, $\nu_{\rm a}$, the characteristic frequency, $\nu_{\rm m}$, and the cooling frequency, $\nu_{\rm c}$) and the characteristic flux, $F_{\rm \nu,m}$ \citep{GS2002}. The SED and light curve temporal evolution are dependent on the following parameters: $p$, the isotropic-equivalent kinetic energy of the burst, $E_{K, \rm iso}$, the density of the environment\footnote{In a constant density interstellar medium (ISM) environment, $k = 0$ and $A = n_0 m_p$. In a wind environment, $k = 2$, $A = 5 \times 10^{11} \rm ~gm~cm^{-1} A_*$, normalized to a progenitor mass-loss rate of $A_* = 10^{-5}~M_{\odot} {\rm yr^{-1}}$ and wind velocity of $1000 {\rm ~km~s}^{-1}$; \citep{ChevalierLi2000}.}, $\rho = Ar^{-k}$, and the fractional energy density imparted on the electrons, $\epsilon_{\rm e}$, and the magnetic field, $\epsilon_{\rm B}$. 

In addition to the standard synchrotron frame work laid out in \citet{GS2002}, we also consider the effects of beaming on the GRB afterglow light curves. Collimation is expected to manifest itself as a jet break, which occurs when the angular size of the relativistic beam of the GRB jet approaches the value of the true opening angle of the jet ($\theta_{\rm jet}$). At the time that this jet break occurs, $t_{\rm jet}$, the observed light curve is predicted to steepen achromatically \citep{Rhoads1999, SariPiranHalpern1999}.
Determining $t_{\rm jet}$ from a GRB afterglow can therefore provide us with the opening angle of the jet, which allows for the correction of the energetics for beaming and in turn, we can derive the true kinetic ($E_{\rm K}$), $\gamma$-ray ($E_{\gamma}$), and total energies of the burst ($E_{\rm tot} = E_{\rm K} + E_{\gamma}$). Additionally, the evolution of the afterglow light curves at $\delta t > t_{\rm jet}$ (where $\delta t$ is the time after the {\it Swift}/Burst Alert Telescope trigger) is $F_\nu \propto t^{-p}$ when $\nu_{\rm obs} \geq \nu_{\rm m}$ \citep{SariPiranHalpern1999}, where $\nu_{\rm obs}$ is the observing frequency. Thus, identifying a jet break in the afterglow and measuring the post-break light curve slope provides an additional constraint on the value of $p$.

To properly model the X-ray afterglow, we include the effects of inverse Compton (IC) cooling in our model, which can lower the location of $\nu_{\rm c}$ (in comparison to the value predicted by a spherical blast-wave without IC cooling, e.g., \citealt{GS2002}) by a factor of $(1+Y)^2$, where $Y$ is the Compton $Y$-parameter (for the detailed explanation see: \citealt{SariEsin2001, Laskar2015_tl15}).

Prior to performing a full fit to the broad-band data, we first examine the SEDs and light curves of our afterglow observations and fit them using either a single power law function or a smoothly broken power law, given by 

\begin{equation}
    F_{\nu}(t) = F_b \left[ \frac{1}{2} \left(\frac{t}{t_{\rm b}} \right)^{-s \alpha_1} + \frac{1}{2} \left(\frac{t}{t_{\rm b}} \right)^{-s \alpha_2} \right]^{-1/s}, \label{eq:bpl_time}
\end{equation}

\begin{equation}
    F_{\nu}(\nu) = F_b \left[ \frac{1}{2} \left(\frac{\nu}{\nu_{\rm b}} \right)^{-s \beta_1} + \frac{1}{2} \left(\frac{\nu}{\nu_{\rm b}} \right)^{-s \beta_2} \right]^{-1/s} \label{eq:bpl_freq}
\end{equation}
\noindent where $F_b$ is the normalized flux of the break, $\nu_{\rm b}$ and $t_{\rm b}$ are the break frequency and time, respectively, $s$ is the break smoothness\footnote{Larger values of $s$ correspond to sharper breaks.}, and  $\alpha_1,~ \alpha_2$ and $\beta_1,~ \beta_2$ are the temporal or spectral slopes of the fits, respectively. We use the convention $F_\nu \propto t^\alpha \nu^\beta$ throughout.
Where data quality allows, we use Equation~\ref{eq:bpl_time} for light curves and  Equation~\ref{eq:bpl_freq} for SEDs.  When necessary, we use temporal power law fits to interpolate the flux to common times for our SED fitting. 

We use these basic broken power law considerations to place initial constraints on $p$, $t_{\rm jet}$, the nature of the burst environment (i.e. ISM vs. wind), and the location of the break frequencies with respect to our observing bands, when possible. We then model the GRB with the Markov Chain Monte Carlo (MCMC) modeling framework laid out in \cite{Laskar2014_tl14}, including IC effects. We also take into consideration the scattering effects of scintillation, which may cause variability on short timescales at GHz frequencies \citep{Rickett1990}. In the situations for which the burst environment cannot be constrained with initial considerations alone, we model the GRB afterglow with both an ISM and wind environment and choose the model that provides a better statistical fit. 

To account for any potential systematic uncertainties in flux calibration for data taken across different facilities, we include an uncertainty floor of 10\% on individual measurements prior to modeling. The free parameters for our model are $p$, $E_{K, \rm iso}$, $n_0$ in an ISM environment or $A_*$ for a wind environment,  $\epsilon_{\rm e}$, $\epsilon_{\rm B}$, and $t_{\rm jet}$. We fit each GRB with available broad-band afterglow data with the MCMC model, using 128 walkers for 10,000 steps, and discard the first $\sim 1\%$ of steps as ``burn in'', where the average likelihood across the chains have yet to reach a stable value. 

%%%%%%%%%%%%%%% TABLE
\tabletypesize{\normalsize}
\begin{deluxetable*}{c|ccccc}
\tablecolumns{5}
\tablewidth{0pc}
\tablecaption{Forward Shock Parameters
\label{tab:GAMMA_bestfit_stat}}
\tablehead {
\colhead {GRB}                &
\colhead{110709B}   &
\colhead{111215A}               &
\colhead {130131A }          &
\colhead {131229A$^{a}$}          &
\colhead {140713A}         \\
\colhead{Env.} &
\colhead{ISM} &
\colhead{Wind} &
\colhead{Wind} & 
\colhead{ISM} &
\colhead{Wind}
}
\startdata
$p$ &  $2.03$  &  $2.88$  &  $2.30$  &  --  &  $2.17$  \\
 &  $2.04^{+0.02}_{-0.02}$  &  $2.82^{+0.04}_{-0.06}$  &  $2.39^{+0.09}_{-0.07}$  &  $ 2.46^{+0.03}_{-0.03}$  &  $2.17^{+0.04}_{-0.03}$  \\
\hline
$E_{\rm K}$ &  $1.69 \times 10^{-1}$  &  $6.63 \times 10^{-1}$  &  $2.25 \times 10^{-3}$  &  --  &  $6.99 \times 10^{-3}$  \\
$(10^{52}~{\rm erg})$ &  $9.63^{+6.32}_{-4.5} \times 10^{-2}$  &  $4.99^{+0.94}_{-1.01} \times 10^{-1}$  &  $3.24^{+3.04}_{-1.26} \times 10^{-3}$  &  --  &  $7.71^{+1.63}_{-1.06} \times 10^{-3}$  \\
\hline
$A_*/n_{0}$ &  $3.85 \times 10^{-3}$  &  $1.88 \times 10^{-1}$  &  $4.0 \times 10^{-2}$  &  --  &  $6.25 \times 10^{-1}$  \\
$(-/{\rm cm}^{-3})$ &  $2.24^{+3.03}_{-1.28} \times 10^{-3}$  &  $1.55^{+0.19}_{-0.22} \times 10^{-1}$  &  $4.22^{+2.24}_{-1.22} \times 10^{-2}$  &  $\gtrsim 1.72 \times 10^{-5} \epsilon_{B}^{-5/3}$  &  $7.33^{+2.09}_{-1.18} \times 10^{-1}$  \\
\hline
$\epsilon_e$ &  $6.90 \times 10^{-1}$  &  $1.04 \times 10^{-1}$  &  $8.74 \times 10^{-1}$  &  --  &  $7.24 \times 10^{-1}$  \\
 &  $3.77^{+3.12}_{-1.74} \times 10^{-1}$  &  $9.47^{+0.59}_{-0.65} \times 10^{-2}$  &  $7.75^{+1.23}_{-1.59} \times 10^{-1}$  &  $\gtrsim 1.19 \times 10^{-2} \epsilon_{B}^{-1/3}$  &  $6.78^{+0.88}_{-0.89} \times 10^{-1}$  \\
\hline
$\epsilon_B$ &  $1.26 \times 10^{-3}$  &  $2.97 \times 10^{-3}$  &  $1.25 \times 10^{-1}$  &  --  &  $2.76 \times 10^{-1}$  \\
 &  $5.12^{+29.24}_{-3.91} \times 10^{-3}$  &  $6.11^{+4.81}_{-2.08} \times 10^{-3}$  &  $9.53^{+13.85}_{-6.98} \times 10^{-2}$  &  --  &  $2.79^{+0.92}_{-0.85} \times 10^{-1}$  \\
\hline
$t_{\rm jet}$ &  $1.38$  &  $26.09$  &  $3.49$  &  --  &  $3.61$  \\
$({\rm day})$ &  $1.39^{+0.11}_{-0.10}$  &  $25.36^{+4.04}_{-3.27}$  &  $4.89^{+8.04}_{-2.35}$  &  $\gtrsim 1.3$  &  $3.4^{+0.52}_{-0.42}$  \\
\hline
$\theta_{\rm jet}$ &  $1.65$  &  $2.60$  &  $8.17$  &  --  &  $21.27$  \\
$({\rm deg})$ &  $1.68^{+0.22}_{-0.19}$  &  $2.68^{+0.13}_{-0.12}$  &  $8.46^{+2.39}_{-1.36}$  &  $\gtrsim 1.32 \epsilon_{B}^{-1/4}$  &  $21.46^{+1.24}_{-1.07}$  \\
\hline
$A_{V, \rm GRB}$ &  $\gtrsim 3.9$ &  $\gtrsim 10.6$ &  $ \approx 2.2$ &  -- &  $\gtrsim 3.5$ \\
$({\rm mag})$ &  $\gtrsim 3.8^{b}$ &  $\gtrsim 10.6^{b}$ &  $2.3^{+0.1}_{-0.1}$ &  $\gtrsim 9.6$ &  $\gtrsim 3.5^{b}$ \\
\enddata
\tablecomments{
The top row for each parameter corresponds to the best fit forward shock value from our MCMC modeling. The bottom row for each parameter corresponds to the summary statistics from the marginalized posterior density functions (medians and 68\% credible intervals), except in the case of GRB\,131229A (see $^{a}$) \\
$^{a}$ Values derived from analytical arguments (see Section~\ref{sec:131229A_BasicConsiderations}) \\
$^{b}$ $A_{V, \rm GRB}$ value from afterglow model using median values \\
}
\end{deluxetable*}
%%%%%%%%%%%%%%% TABLE

\subsection{GRB\,110709B}
\label{sec:110709BAfterglowModeling}

%%%%% 110709B MODELING FIGURE
\begin{figure}
\centering
\includegraphics[width=0.45\textwidth]{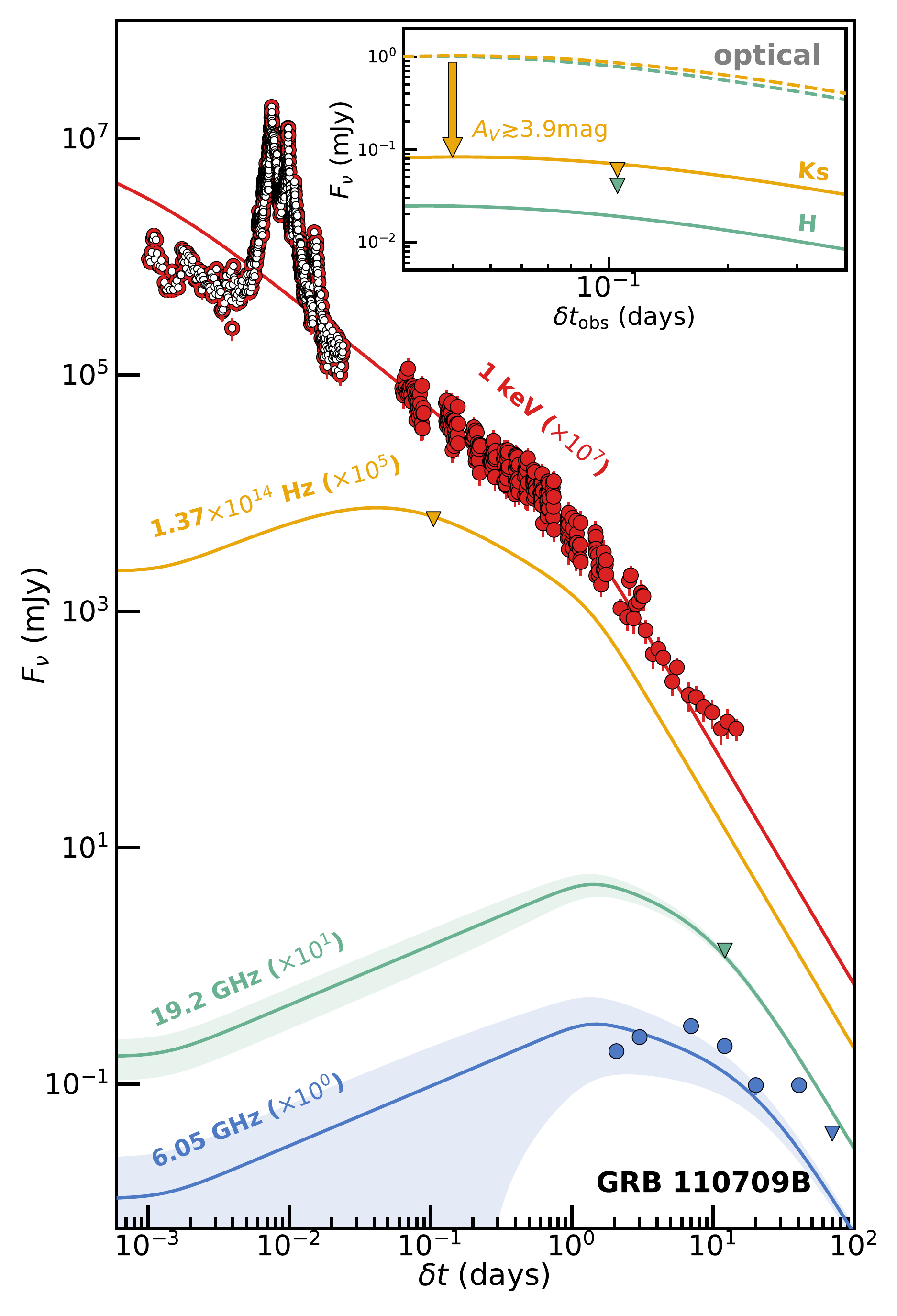}
\caption{X-ray, optical, and radio afterglow light curves of GRB\,110709B, together with the best-fit forward shock model in an ISM environment (lines). Circles represent detections, and triangles represent $3\sigma$ upper limits \citep{Zauderer2013}. Open symbols indicate data that are not included in the fit, and shaded regions represent variability due to scintillation. The inset shows the model $K_s$-band and $H$-band light curves (solid lines) as well as the non-extinguished models (dashed lines), with $A_{V, \rm GRB}\gtrsim 3.9$~mag necessary to be consistent with the upper limits.
} 
\label{fig:lc110709B}
\end{figure}
%%%%% 111215A MODELING FIGURE

We compile all available radio data of GRB\,110709B, along with optical upper limits and the {\it Swift} X-ray light curve (Section~\ref{sec:110709B_literature}), to model the afterglow. The data consists of a VLA 5.8~GHz light curve spanning $\delta t \approx 2.1-69.7~{\rm days}$, a VLA 21.8~GHz upper limit at $\delta t \approx 12.1~{\rm days}$, GROND optical/NIR ($g'r'i'z'JHK_{s}$) non-detections at $\delta t \approx 0.11~{\rm days}$, and the {\it Swift} X-ray afterglow light curve spanning $\delta t \approx 10^{-3}-10^{2}~{\rm days}$. 

The afterglow of GRB\,110709B was previously modeled by \citet{Zauderer2013}, and then by \citet{KangasFruchter2021}. \citet{Zauderer2013} did not include IC effects in their modeling, while \citet{KangasFruchter2021} did include IC effects, though they did not fit for $A_{V, \rm GRB}$, which is a key parameter of interest for our study. Here, we model GRB\,110709B to ensure consistency across our sample and to determine $A_{V, \rm GRB}$.

\subsubsection{Basic Considerations}
\label{sec:110709B_BasicConsiderations}
The X-ray afterglow light curve exhibits flaring activity until $\delta t \approx 0.02~{\rm days}$, and we only consider these data over $\delta t \approx 0.06 - 116~{\rm days}$ for our modeling. We created a time-sliced spectra
from the {\it Swift} online tool for the photon counting (PC) mode X-ray light curve, which found $\Gamma_{\rm X} = 2.06^{+0.04}_{-0.03} $ ($1 \sigma$, \citealt{Evans2009}), corresponding to $\beta_{\rm X} = -1.06^{+0.04}_{-0.03}$. We fit the X-ray light curve with a broken power law (Section \ref{sec:afterglowmodeling}; Eq \ref{eq:bpl_time}) characterized by $\alpha_{\rm X,1} = -0.78 \pm 0.05$ ($\alpha_{\rm X,1} = -0.86 \pm 0.03$) and $\alpha_{\rm X, 2} = -1.70 \pm 0.06$ ($\alpha_{\rm X, 2} = -1.58 \pm 0.04$), with the break occurring at $\delta t \approx 0.7 ~{\rm days}$ ($\delta t \approx 0.6 ~{\rm days}$) for $s = 2$ ($s = 8$). 

We first investigate the nature of the steepning in the X-ray light curve of GRB\,110709B. Such steepenings are often explained by either the passage of $\nu_c$ through the band, or a jet break. For the passage of $\nu_{\rm c}$ at $\delta t \approx 0.6~{\rm days}$, the expected change in temporal index is $\Delta \alpha = 0.25$, which is too shallow to explain the observed $\Delta \alpha \approx 0.7-0.9$. Thus, we attribute the steepening instead to a jet break.

We now use the X-ray spectral index and pre-break light curve to determine where $\nu_{\rm X}$ lies in relation to $\nu_{\rm c}$. If $\nu_{\rm m} < \nu_{\rm X} < \nu_{\rm c}$, the X-ray spectral index implies $p = 3.12 \pm 0.07$. However, we find that the measured value of $\alpha_{\rm X, 1}$ yields $p = 1.49 \pm 0.04$ in a wind environment or $p = 2.16 \pm 0.04$ in an ISM environment, neither of which are consistent with the value derived from $\beta_X$. On the other hand, for $\nu_{\rm m}, \nu_{\rm c} < \nu_{\rm X}$, we require $p = 2.12 \pm 0.07$ to match the X-ray spectral index and $p = 1.82 \pm 0.04$ (in both the wind and ISM environment) to match the light curve. The values of $p$ are in agreement to within $3 \sigma$, and we therefore conclude that $\nu_{\rm m} , \nu_{\rm c} < \nu_{\rm X}$ and $p \approx 1.8-2.1$. In this regime, the X-ray observations cannot be used to discriminate between an ISM and wind environment.

The expected temporal decay after a jet break is $\alpha = -p$, and our post-break X-ray light curve slope of $-1.7 \lesssim \alpha_{\rm X, 2} \lesssim -1.6$ is shallower than the expected $-2.1 \lesssim \alpha \lesssim -1.8$. However, the break time and slope are degenerate with the smoothness of the break, and a later break time is consistent with a steeper post break decline. This suggests that the break in the X-ray light curve is due to a jet break, whose onset occurs at $\delta t \gtrsim 0.6~{\rm days}$. Additionally, if $t_{\rm jet} \approx 0.6~{\rm days}$, then the $5.8$~GHz light curve, which does not have any observations prior to $t_{\rm jet}$, cannot be used to distinguish between the ISM and wind environment, as the behavior of the synchrotron model is the same regardless of environment after $t_{\rm jet}$.

In conclusion, the X-ray afterglow of GRB\,110709B is consistent with $\nu_{\rm m}, \nu_{\rm c} < \nu_{\rm X}$, $p \approx 1.8-2.1$, and $t_{\rm jet} \gtrsim 0.6~{\rm days}$. Additionally, the X-ray and radio light curves cannot be used to distinguish between the ISM and wind environment.

\subsubsection{MCMC Modeling}
\label{sec:110709B_MCMCModeling}

As we can not distinguish between the wind and ISM environments using preliminary analytical arguments, we therefore fit the GRB\,110709B afterglow data with both a wind and ISM environment, and choose the solution that provides a better statistical fit. 

We find the ISM environment model is marginally preferred by the data, with $\chi^2/{\rm d.o.f.} \approx 386/1289$ and likelihood ($L$) of $\approx 2092$, compared to the wind environment best fit model with $\chi^2/{\rm d.o.f.} \approx 434/1289$ and a lower $L$ of $\approx 2063$ (see Appendix~\ref{appendix:110709B} for wind model, provided for completeness). We present the best-fit (highest likelihood) ISM model in Figure~\ref{fig:lc110709B} and list the parameters  as well as the summary statistics from the marginalized posterior density functions (medians and 68\% credible intervals) in Table~\ref{tab:GAMMA_bestfit_stat}.

We find that $p \approx 2.02$ and $t_{\rm jet} \approx 1.4~{\rm days}$ (consistent with the arguments laid out in Section \ref{sec:110709B_BasicConsiderations}), resulting in $\theta_{\rm jet} \approx 1.4^\circ$. The SED remains in the slow cooling phase with a break frequency ordering of $\nu_{\rm sa} < \nu_{\rm m} < \nu_{\rm c}$ for the entirety of the afterglow observations considered. Our fit also confirms that $\nu_{\rm c} < \nu_{\rm X}$, as expected. For our best fit ISM model, the GROND optical/NIR limits imply $A_{V,\rm GRB} \gtrsim 3.94$~mag.

We comment here on a few notable discrepancies between the data and model light curves. The $5.8~{\rm GHz}$ model light curve under-predicts the last detection at $\delta t \approx 40~{\rm days}$ by $\approx 11.6 \sigma$. This discrepancy is too large to be reconciled with scintillation effects alone. Moreover, the X-ray model light curve under-predicts the data at $\delta t \gtrsim 7.6~{\rm days}$. This result is not unexpected, as we measure a shallower temporal index in the post-break X-ray light curve than would be expected for post jet-break behavior (see Section \ref{sec:110709B_BasicConsiderations}), and therefore we find an excess within our model. We first consider whether this excess flux in the X-ray afterglow is due to Klein Nishina (KN) effects. KN effects become important when $\widehat{\nu}_{m} \leq \nu_{\rm obs}$ (typically the $\nu_{\rm obs}$ is the X-ray frequency), where $\widehat{\nu}_{m} = \nu_{\rm m}(\widehat{\gamma}_{m})$, and $\widehat{\gamma}_{m}$ is the critical Lorentz factor \citep{Nakar2009}. Given our best fit parameters, we find that at $\delta t \approx 12.6~{\rm days}$, $ \nu_{\rm X} \ll \hat{\nu}_{\rm m} \approx 1.6 \times 10^{25}~{\rm Hz}$. Therefore, we conclude that KN effects are not causing the excess flux in the X-ray afterglow. We next consider whether IC emission is the cause of the observed X-ray excess. We calculate the flux of the IC spectra at $\delta t \approx 12.6~{\rm days}$ and find that the IC flux at $\nu_{\rm X}$ is $\approx 3.8 \times 10^{-10}~{\rm mJy}$, which is a factor $\approx 10^4$ times lower than the X-ray flux at that time. Therefore, we also conclude that IC effects are not contributing to the excess flux. 

Instead, the discrepancy between the X-ray and 5.8~GHz model light curves and observations could be reconciled with a slightly later jet break at $\delta t \approx 2.0~{\rm days}$. However, this would violate the 22~GHz upper limit at $\delta t \approx 12.1~{\rm days}$. In summary, there is not a natural explanation for the late-time excess emission in these bands, although fixing the time of the jet break to be later does not significantly affect the parameters of interest ($E_{\rm K}$, $n_0$, and $A_{V, \rm GRB}$), and they remain consistent with the fit presented above within errors.

Both \citet{Zauderer2013} and \citet{KangasFruchter2021} found the wind environment to best fit the afterglow of GRB\,110709B, whereas we found the ISM environment to best fit the afterglow of GRB\,110709B. The discussion of our wind environment model fit and comparison to the fits of \citet{Zauderer2013} and \citet{KangasFruchter2021} can be found in Appendix \ref{appendix:110709B}.

\subsection{GRB\,111215A}
\label{sec:111215AAfterglowModeling}

%%%%% 111215A MODELING FIGURE
\begin{figure*}
\centering
\includegraphics[width=\textwidth]{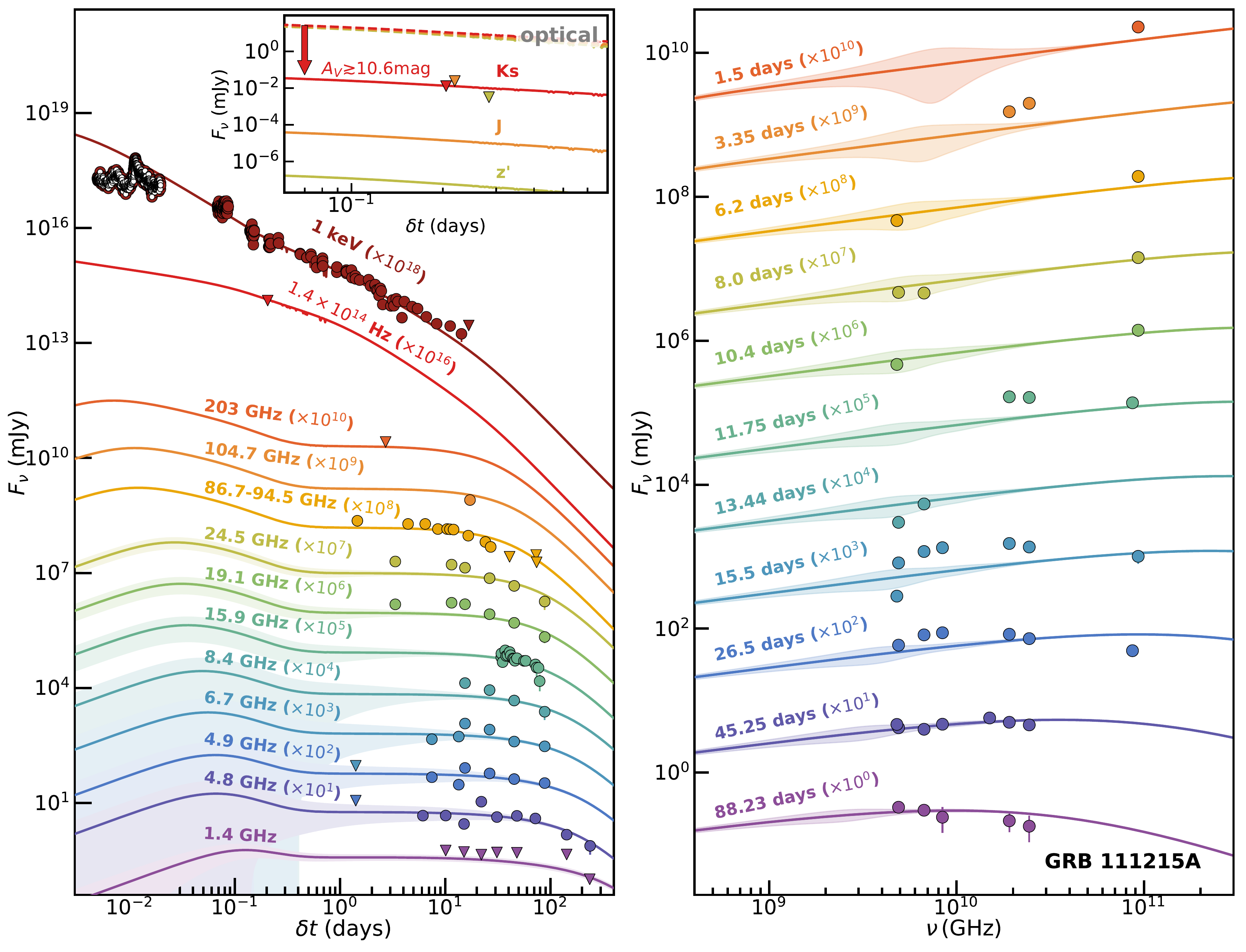}
\caption{{\it Left:} X-ray, optical, millimeter and radio afterglow light curves of GRB\,111215A, together with the best-fit forward shock model in a wind environment (lines). Circles represent literature data \citep{Zauderer2013, VanderHorst2015}, and triangles represent $3\sigma$ upper limits. Open symbols indicate data that are not included in the fit, and shaded regions represent variability due to scintillation. The inset shows the model $z'$-, $J$-, and $Ks$-band light curves (solid lines) as well as the non-extinguished models (dashed lines), indicating $A_{V, \rm GRB}\gtrsim 10.6$~mag to explain the upper limits.
{\it Right:} Radio and millimeter spectral energy distributions (SEDs) of the afterglow of GRB\,111215A from $\delta t \approx 1.5$-$88.23$ days, together with the best-fit forward shock model (lines). Overall the model provides a good match to the broad-band temporal and spectral evolution of the afterglow.} 
\label{fig:lcspec111215A}
\end{figure*}
%%%%% 111215A MODELING FIGURE

We compiled the radio and millimeter light curves of the afterglow of GRB\,111215A, along with NIR/optical upper limits and the {\it Swift} X-ray light curve for our afterglow modeling (Section \ref{sec:111215A_literature}). The radio afterglow of GRB\,111215A is one of the best sampled of any dark GRB, with 1.4~GHz non-detections spanning $\delta t \approx 10-237~{\rm days}$, 4.8, 4.9, and 6.7~GHz observations spanning $\delta t \approx 1.4-238~{\rm days}$ (all together C-band), 8.5~GHz observations spanning $\delta t \approx 15-88~{\rm days}$ (X-band), 15~GHz observations spanning $\delta t \approx 34-79~{\rm days}$, 19.1 and 24.5~GHz observations spanning $\delta t \approx 3-88~{\rm days}$ (all together K-band), 86.7, 93, 93.7, and 94.5~GHz observations spanning $\delta t \approx 1-74~{\rm days}$ (all together 3mm), 104.7~GHz detection at $\delta t \approx 2.7~{\rm days}$, and a 230~GHz upper limit at $\delta t \approx 17~{\rm days}$. The deepest optical and NIR upperlimits of GRB\,111215A span $\delta t \approx 0.2-0.3~{\rm days}$. The {\it Swift} X-ray afterglow of GRB\,111215A spans $\delta t \approx 4.9\times 10^{-3}-16.7~{\rm days}$. 

GRB\,111215A has previously been modeled by \citet{Zauderer2013} and \citet{VanderHorst2015}, though neither of them included IC effects. Additionally, \citet{KangasFruchter2021} modeled GRB\,111215A with the inclusion of IC effects, but they chose not to fit for $A_{V, \rm GRB}$. Like for GRB\,110709B, we choose to model GRB\,111215A in our modeling framework both for consistency across our sample, and to determine $A_{V, \rm GRB}$. 

\subsubsection{Basic Considerations}
\label{sec:111215A_basic}

We first consider the radio and millimeter light curves of GRB\,111215A to determine the location of the radio observations in relation to $\nu_{\rm sa}$ and $\nu_{\rm m}$. At $\delta t \approx 8.0~{\rm days}$ and $\delta t \approx 10.4~{\rm days}$, the spectral index between the C-band and 3mm afterglow is $\beta \approx 0.37-0.39$, suggesting the spectral ordering of $\nu_{\rm sa} <$~C-band~$< {\rm 3mm} < \nu_{\rm m}$ (expected $\beta = 1/3$). The 3mm light curve over $\delta t \approx 8.5-12.0~{\rm days}$ is roughly constant, with $F_{\nu,\rm 3mm} \approx 1.4~{\rm mJy}$, implying the burst occurred in a wind environment (expected $\alpha = 0$). We find the same conclusion from the C-band light curve, where from $\delta t \approx 6.1-10.1~{\rm days}$ the afterglow flux matches $F_{\nu} \propto t^0$. Thus the radio and millimeter afterglow of GRB\,111215A is consistent with a wind environment, and indicates $\nu_{\rm sa} <$~C-band~$< {\rm 3mm} < \nu_{\rm m}$ at least from $\delta t \approx 8.0-12~{\rm days}$. 

We now turn our attention to an interesting spectral phenomena in our radio and millimeter afterglow observations. As discussed, the spectrum between C-band and 3mm at $\delta t \approx 8.0$ and $10.4~{\rm days}$ indicates that $\nu_{\rm sa} <$~C-band~$< {\rm 3mm} < \nu_{\rm m}$ in a wind environment. This conclusion appears to still hold true for the spectrum at $\delta t \approx 45.25~{\rm days}$ ($\beta \approx 0.12$), as the passage of $\nu_{\rm m}$ through the millimeter and radio would result in a negative spectral slope. However, between $\delta t \approx 11.8-26.5~{\rm days}$, an additional bump in the spectrum peaking between X-band and K-band is inconsistent with the expected $\beta = 1/3$ (Figure~\ref{fig:lcspec111215A}). A possible explanation for this spectral feature is a reverse shock. However, the exploration of this possibility is outside the scope of this work.

We now consider the X-ray spectrum and light curve to determine the location of $\nu_{\rm c}$ and $p$. The X-ray afterglow of GRB\,111215A exhibits flaring until $\delta t \approx 0.02~{\rm days}$ (Figure~\ref{fig:lcspec111215A}). Thus, we only consider the X-ray afterglow for $\delta t \approx 0.02 -16.7~{\rm days}$, where we generate a PC-mode time-sliced spectra from the {\it Swift} online tool, which found $\Gamma_{\rm X} = 2.11 \pm 0.06$ ($1 \sigma$, \citealt{Evans2009}). The X-ray afterglow of GRB\,111215A is characterized by a spectral index $\beta_{\rm X} =  -1.11 \pm 0.06 $ and can be fit with a single temporal power law characterized by $\alpha_{\rm X} = -1.43 \pm 0.02$ (Figure~\ref{fig:lcspec111215A}). 

In a wind environment, the scenario of $\nu_{\rm m} < \nu_{\rm X} < \nu_{\rm c}$ $\beta_X$ implies $p = 3.22 ^{+0.12}_{-0.11}$, which is inconsistent with the $\alpha_X$ derived value of $p=2.25 \pm 0.03$, and we therefore rule out this spectral frequency ordering. However, in the case of $\nu_{\rm m}, \nu_{\rm c} < \nu_{\rm X}$, we require $p = 2.22 \pm 0.12$ to match the measured value of $\beta_X$, which is consistent within $3 \sigma$ to the $\alpha_X$ derived value of $p=2.58 \pm 0.03$. Therefore, we conclude that $\nu_{\rm m}, \nu_{\rm c} < \nu_{\rm X}$ and $p \approx 2.2-2.6$. Moreover, the X-ray light curve does not exhibit any break to $\delta t \approx 17~{\rm days}$, placing a lower limit on the time of the jet break to $t_{\rm jet} \gtrsim 17~{\rm days}$ (Figure~\ref{fig:lcspec111215A}).

In conclusion, the radio afterglow of GRB\,111215A is consistent with a wind environment where $\nu_{\rm sa} <$~C-band~$< {\rm 3mm} < \nu_{\rm m}$. Additionally, the X-ray afterglow is consistent with $\nu_{\rm m}, \nu_{\rm c} < \nu_{\rm X}$. We find a preliminary estimate of $p \approx 2.2-2.6$, and expect $t_{\rm jet} \gtrsim 17~{\rm days}$.

\subsubsection{MCMC Modeling}
\label{sec:111215A_MCMCModeling}

We fit the afterglow data of GRB\,111215A with a wind environment. We present the best fit parameters wind model in Figure \ref{fig:lcspec111215A} and list the parameters  as well as the summary statistics from the marginalized posterior density functions (medians and 68\% credible intervals) in Table~\ref{tab:GAMMA_bestfit_stat}.

For our best fit wind model, the optical/NIR limits imply $A_{V, \rm GRB} \gtrsim 10.6$~mag. Our model parameters are $p \approx 2.88$ and $t_{\rm jet} \approx 26.0$~days, resulting in a $\theta_{\rm jet} \approx 2.6^\circ$. The value of $p$ is higher than our initial prediction of $p \approx 2.2-2.6$ (see Section \ref{sec:111215A_basic}). We can reconcile the higher value of $p$ found in the full modeling compared to the simplistic calculations performed earlier by investigating the IC cooling effects for our best fit parameters. We find that the Compton $Y$-parameter decreases from $Y\approx 6.4$ at the time of the fast-to-slow cooling transition ($\delta t \approx 0.2~{\rm days}$) to $Y\approx 0.7$ at the time of the last X-ray detection ($\delta t \approx 14.3~{\rm days}$). This decrease in $Y$ results in a faster evolution of $\nu_{\rm c}$, and results in a shallower light curve at higher $p$.

Consistent with the likely existence of an additional component (Section~\ref{sec:111215A_basic}), our FS model under-predicts the light curves of the X- and K-band afterglow at $\delta t \lesssim 45~{\rm days}$, as well as the 3mm light curve at $\delta t \lesssim 8.0~{\rm days}$. 

Similar to the work presented here, \citet{KangasFruchter2021} modeled GRB\,111215A in a wind environment with IC effects included, finding similar parameter values to us, though our model most disagrees with \cite{KangasFruchter2021} on the placement of $t_{\rm jet}$, where they found $t_{\rm jet} \approx 10.8~{\rm days}$, driven mainly by the fit to the early data.

\subsection{GRB\,130131A}
\label{sec:130131AAfterglowModeling}

%%%%% 130131A MODELING FIGURE
\begin{figure*}
\centering
\includegraphics[width=\textwidth]{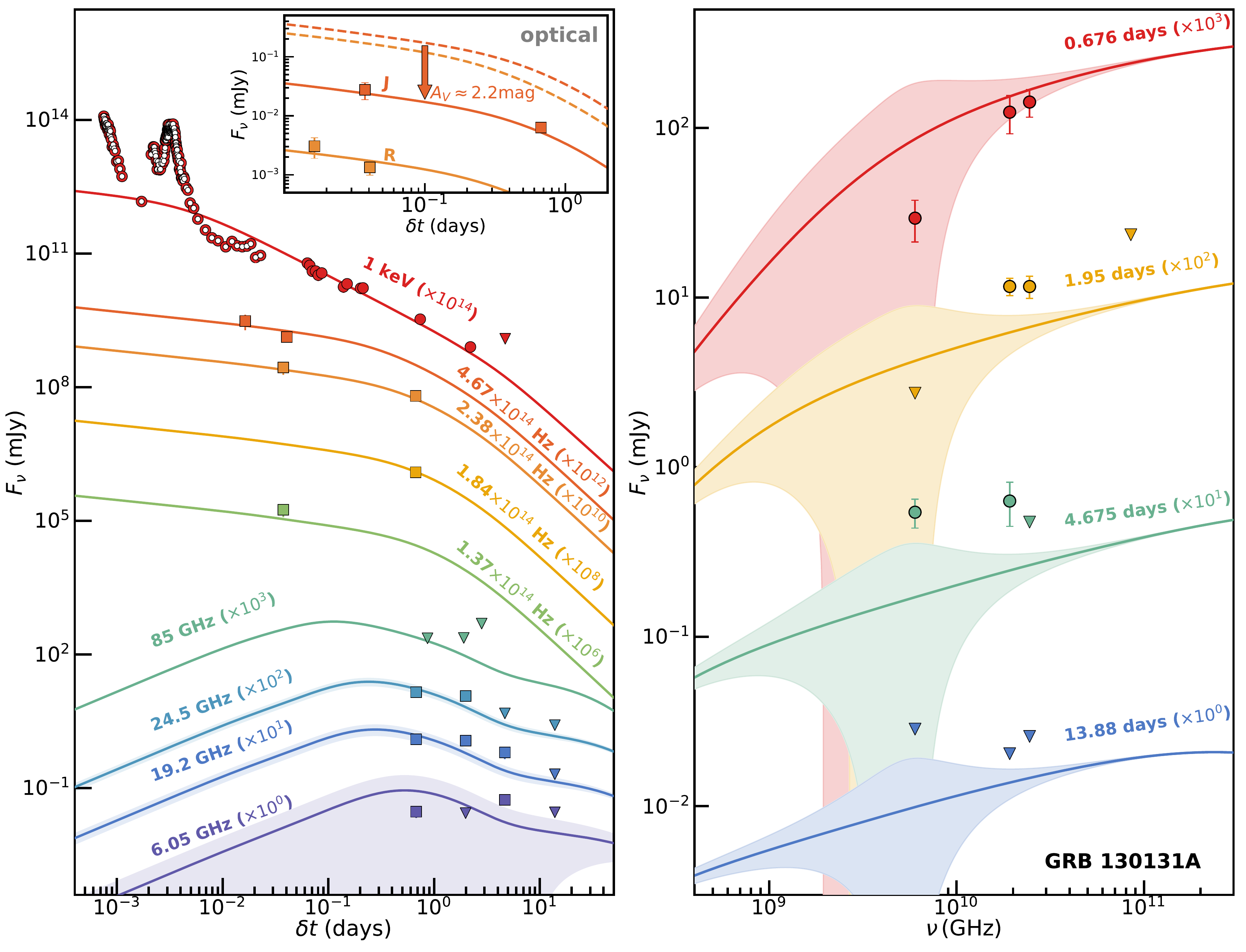}
\caption{{\it Left:} X-ray, optical, NIR, millimeter and radio afterglow light curves of GRB\,130131A, together with the best-fit forward shock model in a wind environment (lines). Squares represent data newly reported here, and triangles represent $3\sigma$ upper limits. Open symbols indicate data that are not included in the fit, and shaded regions represent variability due to scintillation. The inset shows the model $J$- and $R$-band light curves (solid lines) as well as the non-extinguished models (dashed lines), indicating $A_{V, \rm GRB} \approx 2.2$~mag to explain the detections.
{\it Right:} Radio and millimeter spectral energy distributions (SEDs) of the afterglow of GRB\,130131A from $\delta t \approx 0.7$-$13.9$ days, together with the best-fit forward shock model (lines). Overall the model provides a good match to the broad-band temporal and spectral evolution of the afterglow.} 
\label{fig:lcspec130131A}
\end{figure*}
%%%%% 130131A MODELING FIGURE

We compiled the radio light curves (6.0~GHz, 19.2~GHz, and 24.5~GHz, spanning $\delta t \approx 0.7-13.9~{\rm days}$) and millimeter light curve (85.5~GHz, spanning $\delta t \approx 0.9-2.8~{\rm days}$) of the afterglow of GRB\,130131A (Section~\ref{sec:130131Aradioobs}), along with the optical/NIR detections ($RJHK$, spanning $\delta t \approx 0.04-0.67~{\rm days}$) and the {\it Swift} X-ray light curve spanning $\delta t \approx 7.5\times 10^{-4}-4.7~{\rm days}$ (Section~\ref{sec:130131AXrayOpticalObs}) for our afterglow modeling.

\subsubsection{Basic Considerations}
\label{sec:130131ABasicConsideration}

To determine $p$ and the location of $\nu_{\rm c}$, we turn our attention to the X-ray afterglow. We ignore the X-ray flaring activity of GRB\,130131A ($\delta t \approx 0.08-2.28 \times 10^{-2}~{\rm days}$, Figure~\ref{fig:lcspec130131A}), and only consider the X-ray light curve and spectra from $\delta t \approx 0.02 - 4.7$~days. We generate a PC-mode time-sliced spectra from the {\it Swift} online tool, which found $\Gamma_{\rm X} = 2.4 ^{+0.3}_{-0.2}$ ($1 \sigma$, \citealt{Evans2009}). The X-ray spectra for this time range can be characterized with $\beta_{\rm X} = -1.4^{+0.3}_{-0.2}$ and the X-ray light curve for this time range can be fit with a single power law with $\alpha_{\rm X} = -1.14 \pm 0.04$. We first consider the scenario in which $\nu_{\rm m} < \nu_{\rm X} < \nu_{\rm c}$. In this regime, $\beta_{\rm X}$ would indicate $p = 3.8 \pm 0.5$. However, the temporal decline of the X-ray light curve would indicate $p = 1.85 \pm 0.06$ in the wind environment and $p = 2.52 \pm 0.06$ in an ISM environment. While the values of $p$ found for the ISM environment are consistent within $3 \sigma$, the nominal values of $p$ for the spectral and light curve analysis differ by $\approx 1.3$. On the other hand, for the regime $\nu_{\rm m}, \nu_{\rm c} < \nu_{\rm X}$, $\alpha_{\rm X} = -1.14 \pm 0.04$ implies $p = 2.19 \pm 0.06$, and $\beta_{\rm X} = -1.4^{+0.3}_{-0.2}$ implies $p = 2.8 \pm 0.5$. These values are more consistent ($2\sigma$), with a nominal value  difference of only $\approx 0.6$.  Therefore, the X-ray afterglow of GRB\,130131A indicates $\nu_{\rm m}, \nu_{\rm c} < \nu_{\rm X}$ and $p \approx 2.2-2.8$. Additionally, the X-ray light curve does not exhibit a break to $\delta t \approx 2.2~{\rm days}$, placing a lower limit on any jet break to $t_{\rm jet} \gtrsim 2.2~{\rm days}$. In this regime we cannot discriminate between the ISM and wind environments based on the X-ray afterglow alone.

We now consider whether the radio afterglow can provide additional constraints on the properties of GRB\,130131A. The radio light curves of GRB\,130131A are not well sampled, with only 4 epochs of observations (and 7 detections) between the 3 frequencies. Additionally, the 6.0~GHz light curve exhibits variability, with detections at $\delta t \approx 0.7$ and $\approx 4.7~{\rm days}$, interspersed with a non-detection at $\delta t \approx 2.0~{\rm days}$. Variability on short time scales at $\nu_{\rm obs} \lesssim 10~{\rm GHz}$ is likely attributable to scintillation \citep{Rickett1990}, and we therefore ignore the 6.0~GHz light curve in our analytical arguments (but include it, along with the anticipated scintillation effects, in our MCMC modeling). Turning our attention to the 19.2~GHz and 24.5~GHz observations, we note that the observations have a positive spectral slope of $\beta \approx 0.6$ at $\delta t \approx 0.7~{\rm days}$, and both the 19.2~GHz and 24.5~GHz light curves fade significantly at $\delta t > 2.0~{\rm days}$, requiring $\alpha \gtrsim -0.7$ to be consistent with the later non-detections. These spectral and temporal indices can be consistent with either a wind environment pre-jet break if $\nu_{\rm sa} < 19.2-24.5~{\rm GHz} < \nu_{\rm c} < \nu_{\rm m}$ (expected $\beta = 1/3$ and $\alpha = -2/3$), or an ISM or wind environment if $t_{\rm jet} \approx 2.0~{\rm days}$ and $\nu_{\rm sa} < 19.2-24.5~{\rm GHz} < \nu_{\rm m}$ (expected $\beta = 1/3$ and $\alpha = -1/3$). Therefore, we are unable discriminate between the ISM and wind environment with the 19.2~GHz and 24.5~GHz afterglow observations.

\subsubsection{MCMC Modeling}
\label{sec:130131A_MCMCModeling}

As we cannot distinguish between the wind and ISM environments from our afterglow observations of GRB\,130131A, we fit the data with both an ISM and wind environment. While the host galaxy of GRB\,130131A does not have a spectroscopically determined redshift (see Section~\ref{sec:130131Ahostobs}), we assume a photometric redshift of $z = 1.55$ for both fits,  based on host galaxy SED fitting (see Section~\ref{sec:Prospector}). We find our wind environment model is marginally preferred by the data, providing a $\chi^2/d.o.f. \approx 25/187$ and $ L \approx 137$, and we use this model for broader analysis in the rest of the paper. Comparatively, our ISM environment best fit model produced a $\chi^2/d.o.f. \approx 32/187$ and $ L  \approx 132$ (we provide the ISM fit in Appendix~\ref{appendix:130131A} for completeness). We present the best fit wind model in Figure~\ref{fig:lcspec130131A} and list the parameters  as well as the summary statistics from the marginalized posterior density functions (medians and 68\% credible intervals) in Table~\ref{tab:GAMMA_bestfit_stat}.

In confirmation of the arguments laid out in Section \ref{sec:130131ABasicConsideration}, we find $p \approx 2.3$, and $t_{\rm jet} \approx 3.5~{\rm days}$ from our best fit wind model. The spectrum remains in the fast cooling phase until $\delta t \approx 2.0~{\rm days}$, and $\nu_{\rm m}, \nu_{\rm c} < \nu_{\rm X}$ for the entirety of the X-ray afterglow that we consider. We find the optical/NIR afterglow detections are best fit with a line-of-sight extinction value of $A_{V, \rm GRB} \approx 2.2~{\rm mag}$.

The model matches the first epoch of 19.2~GHz and 24.5~GHz observations, but under-predicts the light curves at $\delta t \gtrsim 2.0~{\rm days}$ by $2-4 \sigma$. This discrepancy can be reconciled somewhat by setting $t_{\rm jet} \gtrsim 14~{\rm days}$ (past the time of all of our afterglow observations), but the model parameters of interest ($E_{\rm K}$, $n_0$, and $A_{V, \rm GRB}$) are not significantly affected by this change. As suggested in Section~\ref{sec:130131ABasicConsideration}, we expect strong scintillation at 6.0~GHz based on our model, consistent with the large variability observed in the 6.0~GHz light curve. 

\subsection{GRB\,131229A}
\label{sec:131229A_BasicConsiderations}

%%%%% 131229A MODELING FIGURE
\begin{figure*}
\centering
\includegraphics[width=\textwidth]{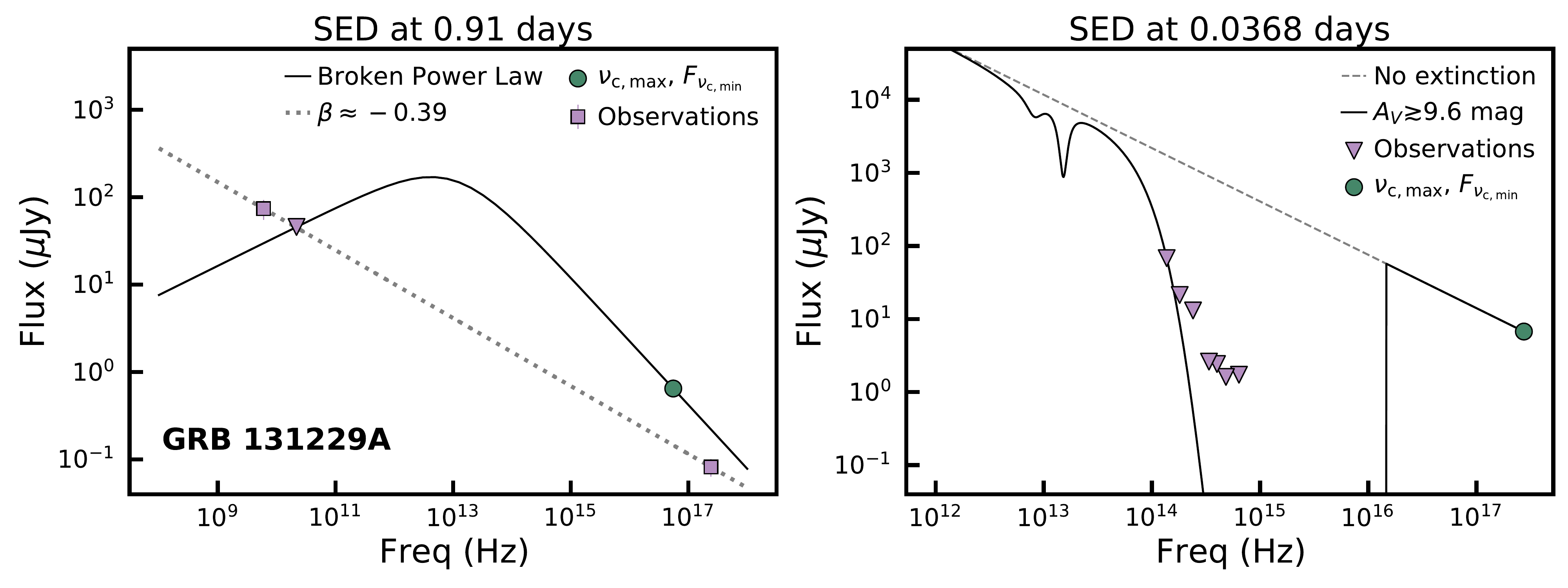}
\caption{{\it Left}: Radio to X-ray SED of GRB\,131229A at 0.91 days, with observations (purple) and estimated $F_{\nu_{\rm c, min}}$ at $\nu_{\rm c, max}$ (green). Black solid line indicates the broken power law fit between the 21.8~GHz upper limit and $F_{\nu_{\rm c}}$, where the break indicates $\nu_{\rm m} \approx 10^{13}~{\rm Hz}$. Dashed line indicates the extrapolation from $F_{\nu_{\rm c, min}}$ to radio frequencies.
{\it Right}: Optical to X-ray SED of GRB\,131229A at 0.0368 days, with optical/NIR upperlimits (purple) and estimated $F_{\nu_{\rm c, min}}$ at $\nu_{\rm c, max}$ (green). Dashed grey line represents the extrapolation from $F_{\nu_{\rm c, max}}$ assuming $\nu_{\rm m} < {\rm NIR/optical} < \nu_{\rm c}$ with no extinction ($A_{V, \rm GRB} = 0~{\rm mag}$). Black solid line represents the extinction curve with the minimum extinction required to be consistent with the upper limits ($A_{V, \rm GRB} \gtrsim 9.6~{\rm mag}$).
} 
\label{fig:SED131229A}
\end{figure*}
%%%%% 131229A MODELING FIGURE

The radio and millimeter afterglow of GRB\,131229A consists of single epoch observations at 6.0, 21.8, and 93~GHz, optical at $\delta t\approx 0.9-1.0~{\rm days}$ (Section~\ref{sec:131229A_radio}). The optical afterglow of GRB\,131229A was not detected, with the deepest upper limits at $\delta t \approx 0.03~{\rm days}$. The {\it Swift} X-ray afterglow spans $\delta t \approx 1.2\times 10^{-3}-1.3~{\rm days}$ (Section~\ref{sec:131229AXray_and_optical}).

We first investigate the X-ray light curve of GRB\,131229A to determine $p$, the location of $\nu_{\rm c}$, and place limits on $t_{\rm jet}$. The combined windowed timing (WT) mode and PC-mode X-ray light curve can be characterized by a broken power law (Equation~\ref{eq:bpl_time}), with $\alpha_1 \approx -1.0 $ and $\alpha_2 \approx -1.4$. As the break in the light curve occurs between the WT-mode and PC-mode ($\delta t \approx 0.39-4.8 \times 10^{-2}~{\rm days}$), we investigate each mode separately. The WT-mode X-ray light curve of GRB\,131229A is characterized by a single power law of $\alpha_{\rm WT} = -1.03 \pm 0.04$, where as the PC-mode X-ray light curve of GRB\,131229A is characterized by a single power law of $\alpha_{\rm PC} = -1.36 \pm 0.03$. The change in temporal index of $\Delta \alpha$ (see Section~\ref{sec:110709B_BasicConsiderations}) is $\approx 0.34$, indicating this temporal break may be the passage of $\nu_{\rm c}$ through the X-ray band (expected $\Delta \alpha = 0.25$).

If the break in the X-ray light curve is indeed the passage of $\nu_{\rm c}$, the derived $p$ from the temporal and spectral indices should be consistent before and after the break. We create a WT-mode time-sliced spectra from the {\it Swift} online tool, and find the WT-mode ($\delta t \approx (1.2-3.9) \times 10^{-3}~{\rm days}$) photon index to be $\Gamma_{\rm WT} = 1.84^{+0.05}_{-0.05}$ ($1 \sigma$, \citealt{Evans2009}), corresponding to a WT-mode spectral index of $\beta_{\rm WT} = -0.84^{+0.05}_{-0.05}$. The hardness of this spectrum implies $\nu_{\rm m} < \nu_{\rm X} < \nu_{\rm c}$, and we derive $p = 2.68^{+0.09}_{-0.09}$, assuming $\beta_{\rm WT} = (1-p)/2$. In a wind environment, the measured value of $\alpha_{\rm WT}$ yields $p = 1.71^{+0.05}_{-0.05}$, which is inconsistent with our $\beta_{\rm WT}$ derived $p$, and we therefore rule out a wind environment. On the other hand, in an ISM environment, the measured value of $\alpha_{\rm WT}$ yields $p = 2.37^{+0.05}_{-0.05}$, which is consistent with the $\beta_{\rm WT}$ derived $p$ within $3 \sigma$. We find a mean weighted WT-mode $p$ of $p_{\rm WT} = 2.44 \pm 0.04$. Past the break ($\delta t \approx 0.05-1.32~{\rm days}$), we create a PC-mode time-sliced spectra from the {\it Swift} online tool, which finds the PC-mode photon index to be $\Gamma_{\rm PC} = 2.14^{+0.11}_{-0.10}$ ($1 \sigma$, \citealt{Evans2009}), corresponding to a PC-mode spectral index of $\beta_{\rm PC} = -1.14^{+0.11}_{-0.10}$. Assuming $\nu_{\rm m}, \nu_{\rm c} < \nu_{\rm X}$, we find $\beta_{\rm PC}$ yields $p = 2.28^{+0.21}_{-0.20}$, consistent with $p_{\rm WT}$. Furthermore, $\alpha_{\rm PC}$ yields $p = 2.49^{+0.04}_{-0.04}$, once again consistent with $p_{\rm WT}$, and we find a mean weighted PC-mode $p$ of $p_{\rm PC} = 2.48 \pm 0.04$. We therefore conclude that an ISM environment is preferred for GRB\,131229A, and that $\nu_{\rm c}$ passes through the X-ray band between $\delta t \approx (0.39-4.8) \times 10^{-2}~{\rm days}$, yielding an overall mean weighted $p$ of $p = 2.46 \pm 0.03$. Furthermore, the X-ray light curve does not exhibit any additional steepening, placing a limit of $t_{\rm jet} \gtrsim 1.3~{\rm days}$.

We next examine the radio to X-ray SED at the time of the radio observations ($\delta t \approx 0.91~{\rm days}$) to place constraints on the location of $\nu_{\rm m}$. The shallow radio to X-ray spectral index at $\delta t \approx 0.91~{\rm days}$ ($\beta_{\rm radio-X} \approx -0.39$, dotted line, Figure ~\ref{fig:SED131229A}) is inconsistent with a single, optically thin power law spectrum with index $\beta=(1-p)/2\approx-0.73$ extending from the radio to the X-rays, and instead requires a spectral peak in between.

In the regime of $6.0~{\rm GHz} < 21.8~{\rm GHz} < \nu_{\rm m} < \nu_{\rm c}$, we expect a positive spectral index of $\beta = 1/3$ for the radio band. However, the observed limit of $\beta \lesssim -0.38$ between the 6.0~GHz and 21.8~GHz observations is inconsistent with this expectation. The discrepancy between the tentative detection at 6.0~GHz and the upper limit at 21.8~GHz may be explained by a variety of factors, such as the 6.0~GHz counterpart being unrelated to the FS afterglow of GRB\,131229A, or due to scintillation effects which can cause variability on short timescales at $\nu_{\rm obs} \lesssim 10~{\rm GHz}$ \citep{Rickett1990}. We instead utilize the 21.8~GHz non-detection to place constraints on the FS emission.

We use the 21.8~GHz non-detection and the X-ray light curve to place constraints on the location of $\nu_{\rm c}$ and $\nu_{\rm m}$.
Based on our analytical arguments, the latest time $\nu_{\rm c}$ can reasonably pass through the X-ray band is at the start of the PC-mode X-ray light curve ($\delta t \approx 4.8 \times 10^{-2}~{\rm days}$), and therefore we assume $\nu_{\rm c} \approx \nu_{\rm X}$ at this time. Scaling $\nu_{\rm c}$ to the time of the radio observations ($\nu_{\rm c} \propto t^{-1/2}$), we find the maximum value of $\nu_{\rm c}$ to be $\nu_{\rm c, max} \approx 5.5\times 10^{16}~{\rm Hz}$ at $\delta t \approx 0.91~{\rm days}$ (resulting in $F_{\nu, \rm c, min} \approx 0.65~\mu{\rm Jy}$). We fit a broken power law (Equation~\ref{eq:bpl_freq}) with the 21.8~GHz non-detection and $F_{\nu_{\rm c, min}}$, fixing $\beta_1 = 1/3$, $\beta_2 = (1-p)/2 \approx -0.73$, and $s = 1.84-0.40p \approx 0.856$. We find $\nu_{\rm m} \gtrsim 10^{13}~{\rm Hz}$ (solid line, Figure~\ref{fig:SED131229A}). 

Our X-ray analysis provides measured values of the synchrotron flux above $\nu_{\rm c}$, and our radio analysis provides a constraint on the synchrotron flux between $\nu_{\rm sa}$ and $\nu_{\rm m}$ (Table~1 in \citealt{GS2002}). Combined with our constraint on $\nu_{\rm c}$ (Table~2 in \citealt{GS2002}), we place limits on the energy, density, and microphysics of the system. Assuming a redshift of $z = 1.04$ (see Section~\ref{sec:Prospector}), we find $E_{\rm K, iso}\lesssim 3.21 \times 10^{1} \epsilon_{B}^{1/3} 10^{52}~{\rm erg} $, $n_0 \gtrsim 1.72 \times 10^{-5} \epsilon_{B}^{-5/3} {\rm cm}^{-3}$, and $\epsilon_{e} \gtrsim 1.19 \times 10^{-2} \epsilon_{B}^{-1/3}$. Using these constraints, along with our constraint on $t_{\rm jet}$, we place constraints on $\theta_{\rm jet}$ and find $\theta_{\rm jet} \gtrsim 1.32 \epsilon_{B}^{-1/4}(^{\circ})$ \citep{SariPiranHalpern1999}.

We next place constraints on $A_{V, \rm GRB}$ of GRB\,131229A. Assuming $\nu_{\rm m} < {\rm NIR/optical} < \nu_{\rm c}$, and using our assumption of $\nu_{\rm c} \approx \nu_{\rm X}$ at $\delta t \approx 4.8\times 10^{-2}~{\rm days}$, we interpolate the SED between $\nu_{\rm m}$ and $\nu_{\rm c}$ at the time of the most constraining optical/NIR upper-limits ($\delta t \approx 0.0368~{\rm days}$). We find that the extinction necessary to be consistent with the optical/NIR upper limits is $A_{V, \rm GRB} \gtrsim 9.6~{\rm mag}$ (see Figure~\ref{fig:SED131229A}).

In conclusion, we find for GRB\,131229A $p = 2.46 \pm 0.03$, $t_{\rm jet} \gtrsim 1.3~{\rm days}$, and $A_{V, \rm GRB} \gtrsim 9.6~{\rm mag}$. We also place constraints on the isotropic energy and circumburst density of GRB\,131229A of $E_{\rm K, iso}/10^{52}~{\rm erg} \lesssim 3.21 \times 10^{1} \epsilon_{B}^{1/3}$ and $n_0 \gtrsim 1.72 \times 10^{-5} \epsilon_{B}^{-5/3}$, leading to further constraints on $\theta_{\rm jet}$% and $E_{\rm K}$
. With limited radio observations, it is difficult to derive meaningful constraints on the afterglow properties of GRB\,131229A within our MCMC modeling framework, and therefore we instead report our analytically derived values in Table~\ref{tab:GAMMA_bestfit_stat}.

\subsection{GRB\,140713A}
\label{sec:140713AAfterglowModeling}

%%%%% 140713A MODELING FIGURE
\begin{figure*}
\centering
\includegraphics[width=\textwidth]{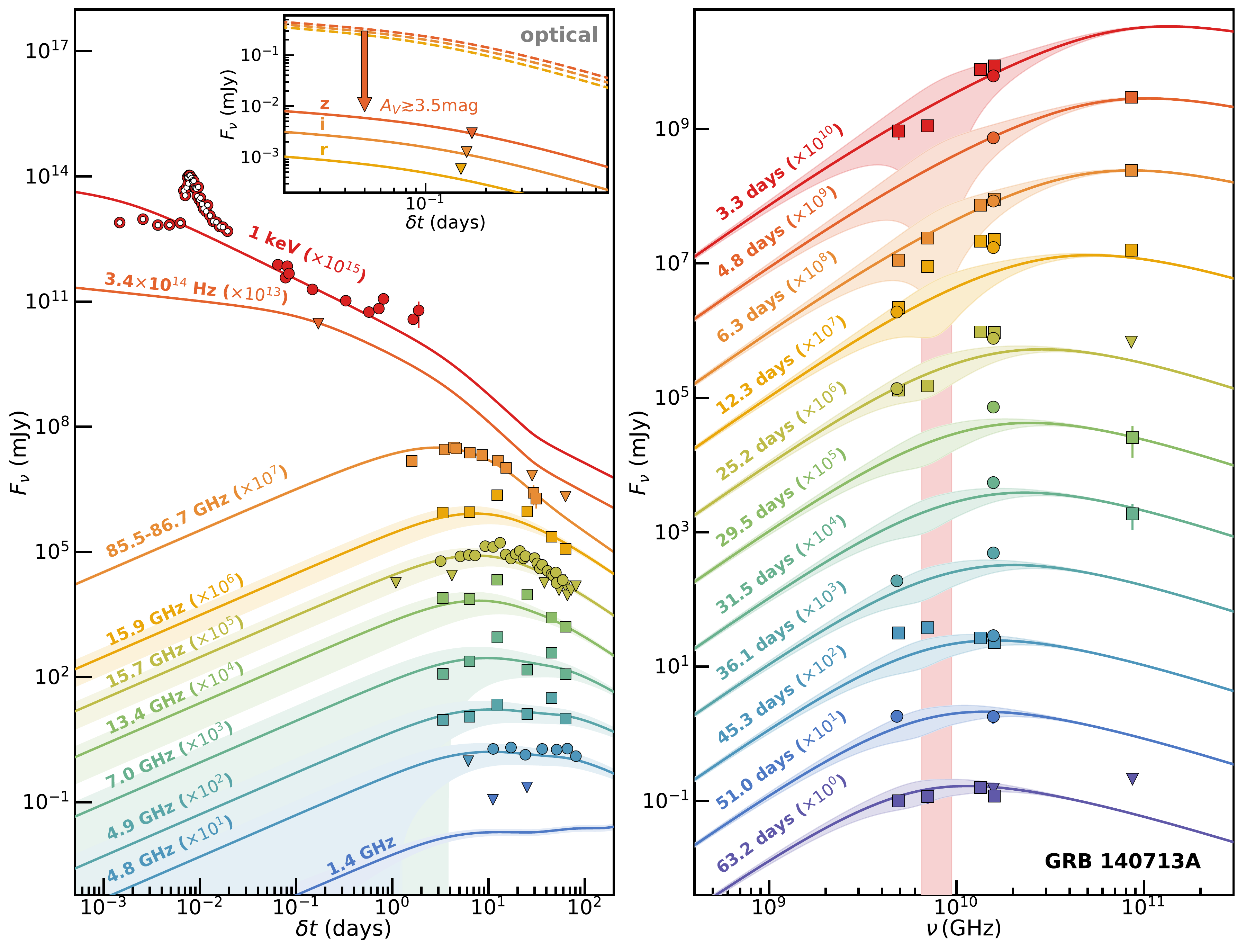}
\caption{{\it Left:} X-ray, optical, millimeter and radio afterglow light curves of GRB\,140713A, together with the best-fit forward shock model in a wind environment (lines). Squares represent data newly reported here, circles represent literature data (\cite{Higgins2019}), and triangles represent $3\sigma$ upper limits. Open symbols indicate data that are not included in the fit, and shaded regions represent variability due to scintillation. The inset shows the model $r$-, $i$-, and $z$-band light curves (solid lines) as well as the non-extinguished models (dashed lines), indicating $A_{V, \rm GRB}\gtrsim3.5$~mag to explain the upper limits. 
{\it Right:} Radio and millimeter spectral energy distributions (SEDs) of the afterglow of GRB\,140713A from $\delta t \approx 3.0$-$31.5$ days, together with the best-fit forward shock model (lines). Overall the model provides a good match to the broad-band temporal and spectral evolution of the afterglow.} 

\label{fig:lcspec140713A}
\end{figure*}
%%%%% 140713A MODELING FIGURE

We compiled the radio and millimeter light curves of the afterglow of GRB\,140713A (Sec \ref{sec:140713Aradioobs}), with WSRT and GMRT upper limits at 1.4 GHz spanning $\delta t \approx 11.1-25.1 ~{\rm days}$, WSRT observations at 4.8 GHz spanning as well as VLA observations at 4.9 and 7.0 GHz $\delta t \approx 3.3-81.0 ~{\rm days}$ (all together, C-band), VLA observations at 13.4 and 15.9 GHz, as well as AMI observations at 15.7 GHz spanning $\delta t \approx 0.1-81.0~{\rm days}$ (all together, K$_u$-band), and CARMA observations at 85.5~GHz and PdBI observations at 86.7 GHz spanning $\delta t \approx 1.5-63.1~{\rm days}$ (all together, 3mm). In combination with the radio and millimeter light curves, we include the NOT optical limits at $\delta t \approx 0.2~{\rm days}$ and the {\it Swift} X-ray light curve spanning $\delta t \approx 1.5\times 10^{-3}-1.9~{\rm days}$ in our afterglow modelling (Sec \ref{sec:140713AXrayOpticalObs}). 

The afterglow of GRB\,140713A has been previously modeled by \citet{Higgins2019}, though they did not include IC effects. We choose to model GRB\,140713A in our modeling framework both for consistency across the bursts in our sample, and because we are introducing new VLA and 3mm afterglow observations that have not previously been modeled.

\subsubsection{Basic Considerations}
\label{sec:140713A_BasicConsiderations}

We observe a steep decline in the 3mm light curve, with $\alpha \approx -2.2$ at $\delta t \gtrsim 12~{\rm days}$. We cannot reconcile this steep decline with the standard synchrotron model of a spherical blast-wave, as the light curve slopes predicted by the standard model are too shallow. We therefore conclude that this decline is caused by a jet break, and that $t_{\rm jet} \lesssim 12$ days. Furthermore, this decline indicates that $p \approx 2.2$, $\nu_{\rm m} \approx$ 3mm, and $F_{\rm \nu, m} \approx 1.7$ mJy at $\delta t \approx 12$ days (see Section~\ref{sec:afterglowmodeling}). 

Assuming $\nu_{\rm m} \approx$ 3mm at $\delta t \approx 12$ days, and evolving this break frequency and $F_{\rm \nu, m}$ forward in time ($\nu_{\rm m} \propto t^{-2}$, $F_{\rm \nu, m} \propto t^{-1}$ post jet break, \citealt{SariPiranHalpern1999}) from 12 days, we find that $\nu_{\rm m} \approx {\rm K}_{u}$-band at $\delta t \approx 28$ days, with a characteristic flux of $F_{\rm \nu, m} \approx 0.76$ mJy. Indeed, we can fit the 15.7 GHz light curve with a steep power law of $\alpha \approx -2.0$ at $\delta t \gtrsim 27$ days, and find the flux at 27 days to be $\approx 0.78$ mJy. Therefore, the K$_{u}$-band data corroborates that $t_{\rm jet} \lesssim 12$ days, and we conclude that $p \approx 2.0-2.2$. With $t_{\rm jet} \lesssim 12~{\rm days}$, the majority of our radio and millimeter afterglow observations are taken at $\delta t \gtrsim t_{\rm jet}$, and we are unable to discriminate between an ISM and wind environment using these observations. 

We now determine the location of the X-rays in relation to $\nu_{\rm c}$. We ignore the X-ray flare ($\delta t \approx (0.15-1.9) \times 10^{-2}~{\rm days}$, see Figure~\ref{fig:lcspec140713A}) of GRB\,140713A, and as such we only consider the X-ray light curve and spectra from $\delta t \approx 0.06 - 1.88~{\rm days}$ in the synchrotron framework to determine the location of the X-rays in relation to $\nu_{\rm c}$. We create a time-sliced PC-mode spectra from the {\it Swift} online tool, which finds the X-ray photon index to be $\Gamma_{\rm X} = 2.0 \pm 0.2$ (1$\sigma$, \citealt{Evans2009}). The X-ray spectra for this time range is characterized by $\beta_{\rm X} = -1.0 \pm 0.2$ and the X-ray light curve for this time range can be fit with a single power law with $\alpha_X = -0.9 \pm 0.1$. 

For $p \approx 2.1$, derived from the identification of a jet break in the radio afterglow, we would expect a spectral index of $\beta_{\rm X} = (1-p)/2 \approx -0.55$ if $\nu_{\rm m} < \nu_{\rm X} < \nu_{\rm c}$ and $\beta_{\rm X} = -p/2 \approx -1.05$ if $\nu_{\rm m},\nu_{\rm c} < \nu_{\rm X}$. The measured value of $\beta_{\rm X} = -1.0 \pm 0.2$ is more consistent with the latter case, within $1 \sigma$, where as the former case is only consistent within $3 \sigma$. In the regime of $\nu_{\rm m},\nu_{\rm c} < \nu_{\rm X}$, we would expect $\alpha_{\rm X} = (2-3p)/4 \approx -1.1$, consistent with our measured value of $\alpha_{\rm X} \approx -0.9$ within $2 \sigma$. Therefore, we conclude $\nu_{\rm m}, \nu_{\rm c} < \nu_{\rm X}$, and note that in this regime we can not discriminate between the ISM and wind environment with the X-ray light curve.

In conclusion, the radio and X-ray afterglow of GRB\,140713A is consistent with $p \approx 2.0-2.2$, $t_{\rm jet} \lesssim 12~{\rm days}$, and $\nu_{\rm m},\nu_{\rm c} < \nu_{\rm X}$. Neither the radio nor X-ray observations allow us to analytically distinguish between the ISM and wind environment. 

\subsubsection{MCMC Modeling}
\label{sec:140713A_MCMCModeling}

As we cannot distinguish between the wind and ISM environments from our afterglow observations of GRB\,140713A, we fit the data with both an ISM and wind environment. We find that our wind environment model better fits the data, with the best fit model having a reduced $\chi^2 \approx 114/102$ and $L \approx 79$. Comparatively, our ISM environment best fit model produced a reduced $\chi^2 \approx 272/102$ and $L \approx 11$ (for completeness, we present the best fit ISM model in in Appendix \ref{appendix:140713A}). We present the best fit wind model in Figure \ref{fig:lcspec140713A} and list the parameters as well as the summary statistics from the marginalized posterior density functions (medians and 68\% credible intervals) in Table~\ref{tab:GAMMA_bestfit_stat}.

In confirmation of the arguments laid out in Section~\ref{sec:140713A_BasicConsiderations}, the parameters of our best-fit model are $p \approx 2.17$ and $t_{\rm jet} \approx 3.6$~days, resulting in a $\theta_{\rm jet} \approx 21^\circ$. 
Additionally, the SED remains in the fast cooling phase until $\delta t \approx 35$~days, and the ordering of the break frequencies at $\delta t \approx 1.65$~days is $\nu_{\rm c} < \nu_{\rm m} < \nu_{\rm X}$. For our best fit wind model, the NOT optical limits imply $A_{V,\rm GRB} \gtrsim 3.52~{\rm mag}$.

%%%% Prospector Parameters Table
%%%%%%%%%%%%%%% TABLE
\tabletypesize{\normalsize}
\begin{deluxetable*}{l|cccccc}
\tablecolumns{6}
\tablewidth{0pc}
\tablecaption{Host Galaxy Properties
\label{tab:hostgal_properties}}
\tablehead {
\colhead {GRB Name}                &
\colhead{$z$}           &
\colhead{$\log_{10}(M/M_\odot)$}           &
\colhead {$\log_{10}(Z/Z_\odot)$}           &
\colhead {${\rm SFR}~[M_{\odot/~{\rm yr}}]$}           &
\colhead {$t_{m}$ [Gyr]}          &
\colhead {$A_{V}^{\rm host}$~[mag]}          
}
\startdata
110709B & $2.109$  &  $9.06^{+0.18}_{-0.29}$ & $-0.63^{+0.33}_{-0.24}$ & $2.73^{+2.07}_{-1.02}$ & $0.40^{+0.45}_{-0.27}$ & $0.61^{+0.23}_{-0.22}$ \\
111215A & $2.012$  & $9.98^{+0.30}_{-0.15}$ & $0.15^{+0.03}_{-0.04}$ & $50.83^{+18.27}_{-23.38}$ & $0.20^{+0.60}_{-0.10}$ & $0.83^{+0.09}_{-0.24}$ \\
130131A & $1.55^{+0.02}_{-0.04}$  & $9.62^{+0.20}_{-0.20}$  & $0.10^{+0.06}_{-0.21}$ & $78.74^{+34.44}_{-18.51}$ & $0.05^{+0.06}_{-0.03}$ & $1.08^{+0.05}_{-0.04}$ \\
131229A & $1.04^{+0.32}_{-0.39}$  & $9.50^{+0.27}_{-0.34}$  & $-0.50^{+0.40}_{-0.30}$ & $1.46^{+5.02}_{-1.45}$ & $1.29^{+1.78}_{-0.87}$ & $1.20^{+0.83}_{-0.66}$ \\
140713A &  $0.935$  & $9.90^{+0.03}_{-0.04}$ & $0.12^{+0.04}_{-0.05}$ & $3.97^{+0.40}_{-0.46}$ & $2.18^{+0.21}_{-0.29}$ &  $0.31^{+0.06}_{-0.06}$\\
160509A &  $1.17$  & $10.30$-$10.86$ & $-0.08$-$0.17$ & $52.14$-$277.17$ & $0.07$-$1.10$ &  $1.60$-$4.66$\\
\enddata
\tablecomments{Host galaxy properties derived by \texttt{Prospector}.}
\end{deluxetable*}
%%%%%%%%%%%%%%% TABLE
%%%% Prospector Parameters Table

The X-ray model light curve under-predicts the data at $\delta t \gtrsim 0.7$~days. This is not unexpected, as we found in Section~\ref{sec:140713A_BasicConsiderations} that the X-ray temporal slope was shallower than the expected slope for $p \approx 2.1$. We first consider whether this excess flux in the X-ray light curve is due to KN effects, which become important when $\nu_{\rm X} \geq \widehat{\nu}_{m}$. Given our best fit parameters, we find that at $\delta t \approx 1.65$~days, $\widehat{\nu}_{m} \approx 1.7 \times 10^{21}~{\rm Hz} \gg \nu_{\rm X}$. Therefore, we conclude that KN effects are not causing the excess flux in the X-ray light curve. We next consider whether IC effects are the cause of the excess flux in the X-ray light curve. We calculate the flux of the IC spectra  at $\nu_{\rm X}$ at $\delta t \approx 1.65$~days, and find that the IC flux at $\nu_{\rm X}$ is $\approx 4.7 \times 10^{-7}$~mJy, $\approx 80$ times smaller than the X-ray flux at that time. Therefore, IC effects cannot account for the excess X-ray flux, although we note that such excess emission has been seen in other events \citep{FongBM2014, MarguttiGL2015, LaskarBC2018_GRBsI_lbc18, LaskarvES2019_lves19}.

\citet{Higgins2019} have previously modeled the afterglow of GRB\,140713A. Their model allowed for $p < 2$, and they found a value of $p \approx 1.85$, smaller, but not far off from our value of $p \approx 2.17$. Their other parameters are also similar to ours, with the biggest difference in our model being that we identify a jet break in the 3mm light curve, and therefore find $t_{\rm jet} \approx 3.61~$days, where as they predict $t_{\rm jet} \approx 25-30~$days.

\section{Host Galaxy Modeling}
\label{sec:Prospector}

To model the stellar population properties of the host galaxies, we use the stellar population inference code \texttt{Prospector} \citep{Leja_2017}. \texttt{Prospector} determines properties such as total mass formed, age of the galaxy at the time of observation ($t_\text{age}$), optical depth of old and young stars, stellar metallicity ($Z_*$), the star formation history, and redshift using the available photometric and/or spectroscopic data for each host.  We apply a nested sampling fitting routine with \texttt{dynesty} \citep{Speagle2020} to the observational data of each host to produce posterior distributions in each property. Model SEDs are built using \texttt{Python-fsps} (Flexible Stellar population synthesis; \citealt{Conroy2009,Conroy2010}). Unless the redshift of a host is known, we allow redshift to be a sampled parameter. For hosts with spectra, we fit their spectral continuum with a 10$^{\text{th}}$ order Chebyshev polynomial and add a gas-phase metallicity and gas ionization parameter to accurately fit the nebular emission lines. We also assume a Chabrier initial mass function (IMF) \citep{Chabrier2003}, Milky-Way Dust Extinction Law \citep{Cardelli1989}, and a parametric delayed-$\tau$ SFH ($\text{SFH} \propto t*e^{-t/\tau}$), where the $\tau$ is a sampled parameter in the \texttt{Prospector} fitting. Furthermore, we apply the Gallazzi 2005 Mass-Metallicity relation \citep{Gallazzi2005} and a 2:1 ratio in the dust attenuation between old and young stars respectively, as stellar populations are noticed to follow this trend \citep{cab+20,pkb+14}. The total dust attenuation in optical depth is converted to a $V$-band magnitude, and hence forth referred to as $A_{V, \rm Host}$. We follow the methods in \cite{Nugent2020} to determine the mass-weighted age $t_m$, stellar mass ($M_*$), and star formation rate (SFR).

\begin{figure*}
\centering
\includegraphics[width=\textwidth]{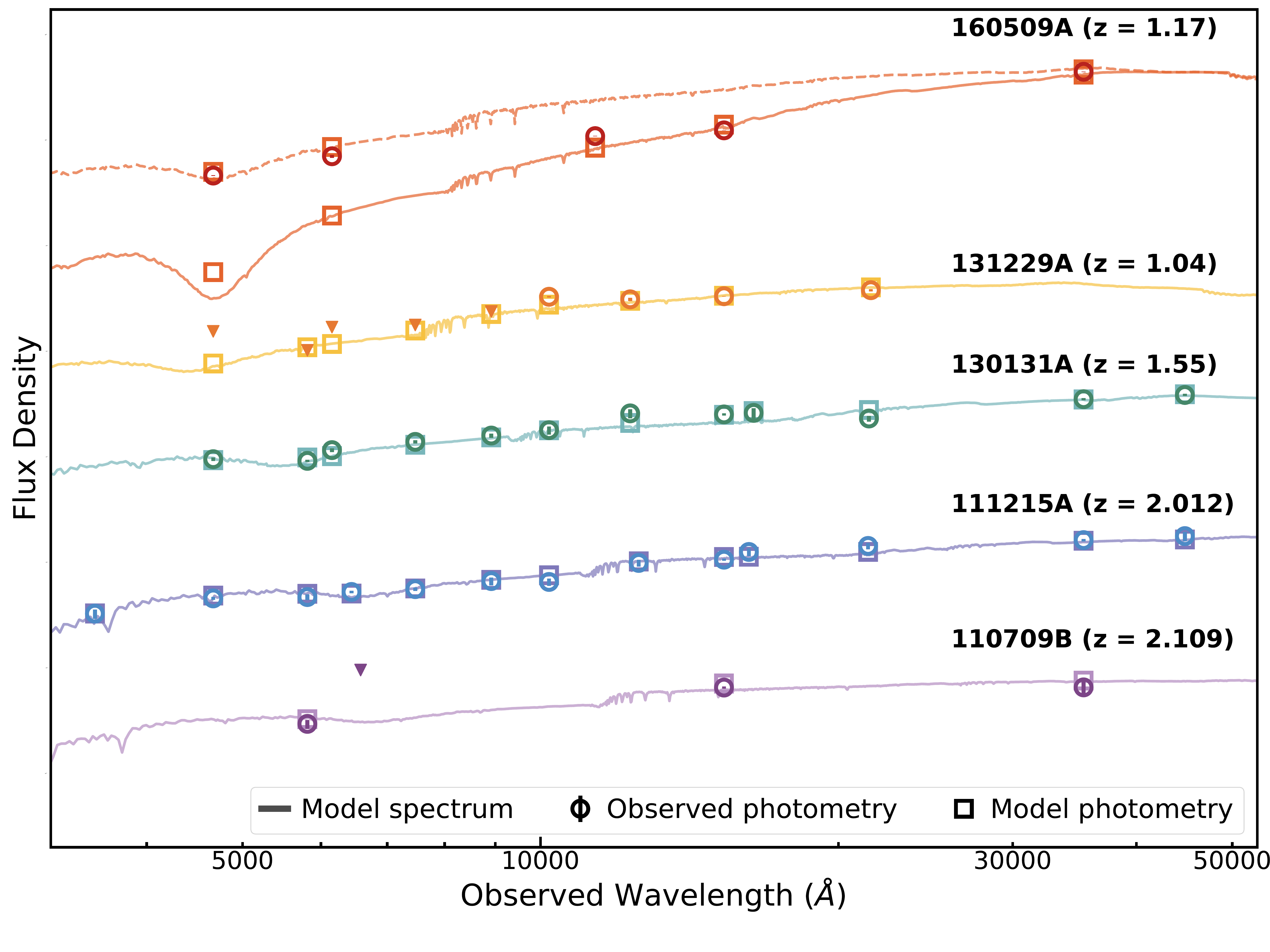}
\caption{The observed host photometry (circles) of the dark GRBs modeled in this paper (GRBs\,110709B, 111215A, 130131A, 131229A, and 160509A), as well as the \texttt{Prospector} model photometry (squares) and best-fit model spectrum (lines). For GRB\,160509A, a second \texttt{Prospector} model is plotted that does not include the HST/WFC3 photometry of the host (dashed line, see Section~\ref{sec:Prospector} for details). The GRB SEDs are arbitrarily scaled in flux for clarity.}
\label{fig:Prospector_phot_only}
\end{figure*}

We model the host galaxy of GRB 110709B with photometry from \citet{Zauderer2013} and \citet{Selsing2019} with a fixed $z=2.109$ \citep{Perley2016_I, Selsing2019}. We find that the stellar population has dust extinction $A_{V, \rm Host} = 0.61^{+0.26}_{-0.25}$~mag (Table~\ref{tab:hostgal_properties}). Though the \texttt{Prospector} SED model fits the photometric data well, the model is based on only three detections and one limit (Figure~\ref{fig:Prospector_phot_only}). 

We fit the host galaxy of GRB\,111215A with photometry from \citet{VanderHorst2015}, corrected for Galactic extinction in the direction of the burst at a fixed the redshift of $z=2.012$ \citep{VanderHorst2015, Chrimes2019}. We find the host has dust attenuation $A_{V, \rm Host} = 0.91^{+0.06}_{-0.08}$~mag. Our host galaxy properties are similar to those found by \citet{VanderHorst2015}.

For the host of GRB\,130131A, the redshift is unknown, although the spectrum indicates $1.3 < z < 4$ (see Section \ref{sec:130131Ahostobs}). Thus, we leave redshift as a free parameter with a flat prior of $1.3<z<4$. We find a photometric redshift of $z=1.55^{+0.01}_{-0.05}$, and $A_{V, \rm Host} = 1.14^{+0.06}_{-0.04}$~mag. 

\begin{figure*}
\centering
\includegraphics[width=0.8\textwidth]{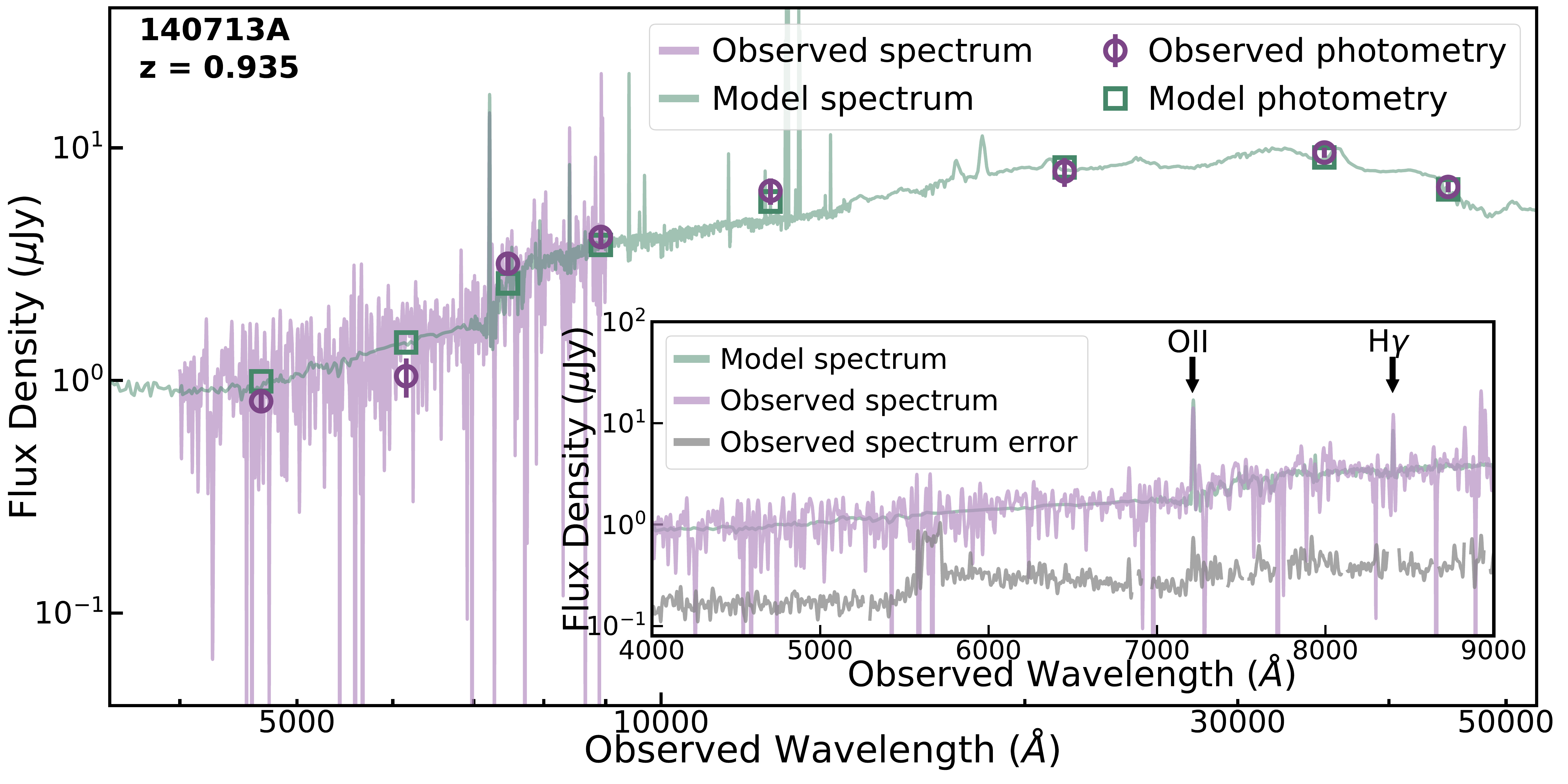}
\caption{The $grizJK$ and Spitzer photometry (purple circles) of GRB140713A compared to the \texttt{Prospector}-produced model spectrum (purple line) and photometry (green squares) at $z=0.935$. The model is consistent with the observed photometry across the wide wavelength range. The inset shows a zoom-in of the observed LRIS spectrum (purple line), as well as the error spectrum (grey line), compared to the model spectrum.
}
\label{fig:140713A_Prospector_SED}
\end{figure*}

For the host of GRB\,131229A, the redshift is unknown, although the deep optical limits ($\gtrsim 23~-\gtrsim 26 ~{\rm mag}$) indicates a redshift of $z \gtrsim 1-1.5$ (see Section~\ref{sec:131229Ahostobs}). Thus, we leave redshift as a free parameter with a flat prior of $0.1<z<3$. We find a photometric redshift of $z=1.04^{+0.28}_{-0.36}$, and $A_{V, \rm Host} = 1.17^{+0.72}_{-0.66}$~mag.

We jointly fit the photometric and spectroscopic data (Table \ref{tab:host}) of the host of GRB 140713A. We find that the stellar population has a low $A_{V, \rm Host} = 0.21^{+0.04}_{-0.04}$~mag. We also find that the photometry and spectrum of the host are overall well-fit by the $\texttt{Prospector}$ SED model, especially the [OII] ($\lambda 3727 \AA$) and H$\gamma$ spectral line strengths (Figure \ref{fig:140713A_Prospector_SED}). Our host galaxy properties are similar to those found by \citet{Higgins2019}.

Finally, we fit the host galaxy of GRB\,160509A, with a redshift of $z = 1.17$ \citep{Laskar2016_160509A_lab16, Kangas2020_160509A}. When we fit the full host galaxy SED of Keck/LRIS, HST/WFC3, and {\it Spitzer} photometry, our \texttt{Prospector} model over-predicts the Keck/LRIS photometry by an order of magnitude. This may indicate that the Keck/LRIS observations are dominated by afterglow contribution, in contradiction with \citet{Laskar2016_160509A_lab16}. However, this \texttt{Prospector} model also finds a high $A_{V, \rm Host}$ of $\approx 4.7~{\rm mag}$, much higher than what is expected for the normal dark GRB host population (i.e. \citealt{Perley2013_DarkGRB_AV}). This high value of $A_{V, \rm Host}$ is driven by the color between the HST/WFC3 and {\it Spitzer} bands, and thus we also fit the host of GRB\,160509A with only the Keck/LRIS and {\it Spitzer} photometry. This method results in a more expected  $A_{V, \rm Host} \approx 1.6~{\rm mag}$, but the model under-predicts the HST/WFC3 photometry by a factor of $\approx 5$. Without further observations of the host galaxy of GRB\,160509A, including re-observing at later times in $g'$- and $r'$-band, we cannot conclusively determine which fit is correct, and as such we quote the ranges for both fits in Table~\ref{tab:hostgal_properties} and present both SEDs in Figure~\ref{fig:Prospector_phot_only}. 

The full details of the best-fit properties for all of the host galaxies of our sample can be found in Table \ref{tab:hostgal_properties} and the model fits are displayed in Figure~\ref{fig:Prospector_phot_only} and Figure~\ref{fig:140713A_Prospector_SED}.

\section{Discussion}
\label{sec:Discussion}

We have presented multi-wavelength observations and modeling of the afterglows of five dark GRBs (GRBs\,110709B, 111215A, 130131A, 131229A, 140713A), as well as SED modeling of the host galaxies of six dark GRBs with \texttt{Prospector} (GRBs\,110709B, 111215A, 130131A, 131229A, 140713A, 160509A). We have classified two additional long GRBs as dark (GRBs\,130420B, 130609A) and presented their radio and millimeter observations. However, for these two events, there is insufficient afterglow and host galaxy follow-up to allow for more in-depth modeling. For the purposes of this discussion, we group bursts that have been classified as dark (either explicitly, or satisfying $\beta_{\rm OX} < 0.5$, i.e. \citealt{Cenko2009_darkGRBs, Melandri2012, Kruhler2012_TOUGH, Rossi2012, LittleJohns2015}) and/or ``dusty" (i.e. \citealt{Kruhler2015, Perley2016_I}) and refer to both groups as ``dark". Equipped with this larger sample, including the new bursts presented here with particularly high extinction site-lines,  we now investigate whether dark GRBs differ from the broader long GRB population in terms of their $\gamma$-ray, afterglow, and host properties. In our comparisons we will refer to any long GRB not classified as dark as a ``typical'' GRB, for brevity. 
 
\subsection{The $\gamma$-ray and afterglow properties of Dark GRBs}

\begin{figure}
\centering
\includegraphics[width=0.48\textwidth]{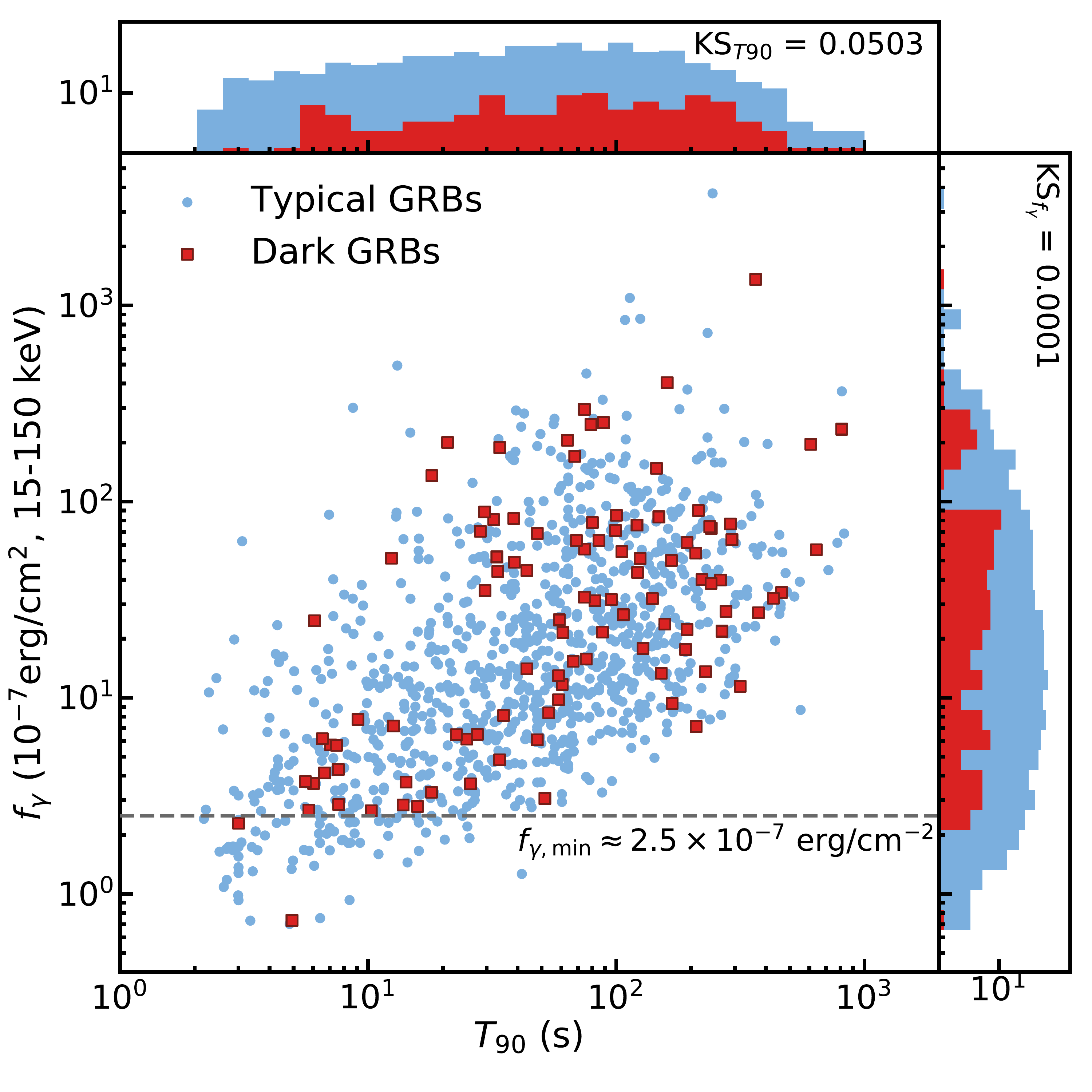}
\caption{
{\it Swift}/BAT fluence ($f_\gamma$) in the 15-150~keV band versus duration ($T_{90}$) of 976~long GRBs (blue points) and 103~dark GRBs (red squares), as derived by \citet{Lien2016}. Dark GRB classification is based on the either explicit classification, or satisfying $\beta_{\rm OX} < 0.5$ (\citealt{Cenko2009_darkGRBs, Kruhler2011, Kruhler2012_TOUGH, Melandri2012, Rossi2012, Perley2013_DarkGRB_AV, Hunt2014, Schady2014, Jeong2014, LittleJohns2015, Perley2016_I, Chrimes2019}, This Work). The dashed horizontal line represents the theoretical minimum $f_{\gamma,\rm min} \approx 2.5 \times 10^{-7}~{\rm erg/cm}^2$, below which the X-ray (and therefore optical) afterglow is typically too faint to be classified as dark with ground based rapid optical follow-up that reaches limits of $R > 24~{\rm mag}$. A simple 2 sample Kolmogorov-Smirnov (KS) test from the \texttt{scipy.stats} package which tests the null hypothesis that dark GRBs are drawn from the same population as the typical GRB population, finds $p = 0.0001$ for the $f_\gamma$ distribution and $p = 0.0503$ for the $T_{90}$ distribution, where a value of $p < 0.05$ rejects the null hypothesis.
} 
\label{fig:Fluence_vs_T90}
\end{figure}

To determine whether dark GRBs differ from the typical GRB population, we first examine their $\gamma$-ray properties: fluence ($f_\gamma$, 15-150 keV band) and duration ($T_{90}$), as even the dark GRBs with sparse afterglow data often have uniformly derived $\gamma$-ray properties from {\it Swift}/BAT. The exception to this is GRB\,160509A, which was instead discovered by the {\it Fermi} Large Area Telescope (LAT; \citealt{2016GCN_160509A_discovery}). We plot these properties from the catalog in \citet{Lien2016} in Figure \ref{fig:Fluence_vs_T90} for 103 dark GRBs (as classified by \citealt{Jakobsson2004, CastroTirado2007, Cenko2009_darkGRBs, vanderHorst2009, Kruhler2011, Kruhler2012_TOUGH, Zauderer2013, Perley2013_DarkGRB_AV, Hunt2014, Chrimes2019}, and This Work) as red points, and all other typical GRBs as blue points. We find that there is a notable lack of dark GRBs in the parameter space corresponding to low fluence ($f_\gamma \lesssim 2 \times 10^{-7}~{\rm erg/cm}^2$) and short duration ($2~{\rm s} < T_{90} \lesssim 5~{\rm s}$). For each parameter, we test the null hypothesis that the the dark GRB population is drawn from the same distribution as the typical GRB population using a two-sample Kolmogorov-Smirnov (KS) test from the \texttt{scipy.stats} package, where a value of $p < 0.05$ rejects the null hypothesis. We obtain $p = 0.0001$ for the $f_\gamma$ distribution and $p = 0.0503$ for the $T_{90}$ distribution, implying that dark GRBs are not drawn from the typical GRB population in terms of their fluence, but may be in terms of their duration.

%%%%% Fluence vs T90 FIGURE
%%%%% Fluence vs T90 FIGURE

One possible explanation for the lack of dark GRBs in the low $f_{\gamma}$ and short $T_{90}$ parameter space may be an observational bias in missing dark GRBs with low fluence\footnote{$f_\gamma$ and $T_{90}$ are correlated ($f_\gamma \propto T_{90}^{-1.11}$; \citealt{Balazs2004}) \citep{vanderHorst2009}. Therefore we focus our discussion on $f_\gamma$.}, as these bursts have been shown to have systematically fainter X-ray afterglows, and in turn fainter optical afterglows than the rest of the population \citep{Gehrels2008}. Thus, such bursts may have afterglows which are fainter than the sensitivity threshold of optical afterglow searches, preventing accurate classification of these lower $f_{\gamma}$ events as dark. To quantify this effect, we use the \citet{Jakobsson2004} darkness classification of $\beta_{\rm OX} < 0.5$, and assume prompt optical observations ($\delta t \approx 0.1~{\rm days}$) of $R > 24~{\rm mag}$, which represents normal GRB follow-up capabilities. The minimum X-ray flux at 0.1~days needed to accurately classify a burst as dark is $F_{\rm X, min} \approx 4.1 \times 10^{-2}~\mu{\rm Jy}$. We extrapolate $F_{\rm X, min}$ to $\delta t=11$~hr assuming $F_{\rm X} \propto t^{-1}$ (see \citealt{Nousek2006, Zhang2006, Evans2009}), and use the derived $F_{\rm X,{\rm 11~hr}}-f_\gamma$ relation of \citet{Gehrels2008} to calculate the minimum fluence of $f_{\gamma,{\rm min}} \approx 2.5 \times 10^{-7}~{\rm erg}~{\rm cm}^{-2}$ necessary to produce an X-ray afterglow bright enough to accurately classify a GRB as dark. This limit lies just below the majority of dark GRBs with the lowest $f_{\gamma}$ (Figure~\ref{fig:Fluence_vs_T90}). Therefore, it is plausible that the lack of low $f_\gamma$ dark bursts is due to an observational bias, as opposed to an intrinsic effect. Additionally, as $f_\gamma$ and $T_{90}$ are correlated \citep{Balazs2004}, this also provides a natural explanation for the lack of observed dark bursts at $ 2~{\rm s} < T_{90} < 5~{\rm s}$. If we exclude bursts with $f_\gamma < 2.5 \times 10^{-7}~{\rm erg}~{\rm cm}^{-2}$ then the dark GRB population does become more statistically similar to the long GRB population ($p = 0.0002$ for the $f_\gamma$ distribution, $p = 0.0734$ for the $T_{90}$ distribution). Finally, we note that there is not a complete catalogue of all dark GRBs, and we may be missing a significant fraction of the dark GRB population. This is made apparent when one considers that the estimated fraction of dark GRBs with respect to all long GRBs is 10-50\% \citep{Jakobsson2004, Cenko2009_darkGRBs, Fynbo2009, Greiner2011, Melandri2012, Perley2013_DarkGRB_AV}, whereas only $\sim 10\%$ of the long GRBs in Figure \ref{fig:Fluence_vs_T90} have been classified as dark.  

We next explore the dark GRB inferred burst explosion properties (e.g., kinetic energies and opening angles) to investigate whether dark GRBs differ from the typical population. As most dark GRBs do not have extensive broadband afterglow modeling, we focus on the six dark GRBs (GRBs\,110709B, 111215A, 130131A, 131229A, 140713A, 160509A) in our sample which are uniformly modeled. We find that the dark GRB sample spans a wide range of beaming-corrected kinetic energies ($E_{\rm K} \approx 0.01-6.6\times10^{51}~{\rm erg}$). Compared to the values for other long GRBs with afterglow modeling \citep{Panaitescu2002_pk02, Price2002_pbr02, Yost2003_yhsf03, Frail2005_fsk05, Chandra2008_ccf08, Cenko2010_cfh10, Cenko2011_cfh11, Laskar2013_lbz13, Laskar2014_tl14, Laskar2015_tl15,     Alexander2017_alb17, Tanvir2018_nt18,  LaskarBC2018_GRBsI_lbc18, LaskarBM2018_GRBsII_lbm18, LaskarAB2018_lab18, LaskarvES2019_lves19}, we find that the kinetic energies of the dark GRB sample are consistent with those of the typical GRB sample, $E_{\rm K} \approx 0.01-36.0\times10^{51}~{\rm erg}$. 
Additionally, our dark GRB sample spans a wide range of jet opening angles ($\theta_{\rm jet} \approx 1.6-21^\circ$), with a distribution again consistent with that of the typical GRB population ($\theta_{\rm jet} \approx 1.1-50^\circ$). There is no clear evidence that dark GRBs are distinct from the typical GRB population in afterglow properties. Our conclusions do not change if we broaden our sample to include the small sample of long GRBs with $A_{V, \rm GRB} > 1~{\rm mag}$ that have not been classified as dark (GRBs\,980329 and 980703; \citealt{Yost2003_yhsf03}, and GRB\,011121; \citealt{Price2002_pbr02}).

\subsection{The origin of the dust along the line-of-sight}
\label{sec:Discussion_Origin_of_dust}
We next consider whether our dark GRB sample has different environmental properties than typical GRBs. In practice, the higher line-of-sight extinctions derived for dark GRBs could be a result of local environment (probed by the afterglow), larger structures such as star-forming regions, global host galaxy dust distributions, or a combination of all three. An investigation into the global and local environmental properties of long GRBs will help determine the cause of the extinction of dark GRBs. 

\begin{figure}
\centering
\includegraphics[width=0.48\textwidth]{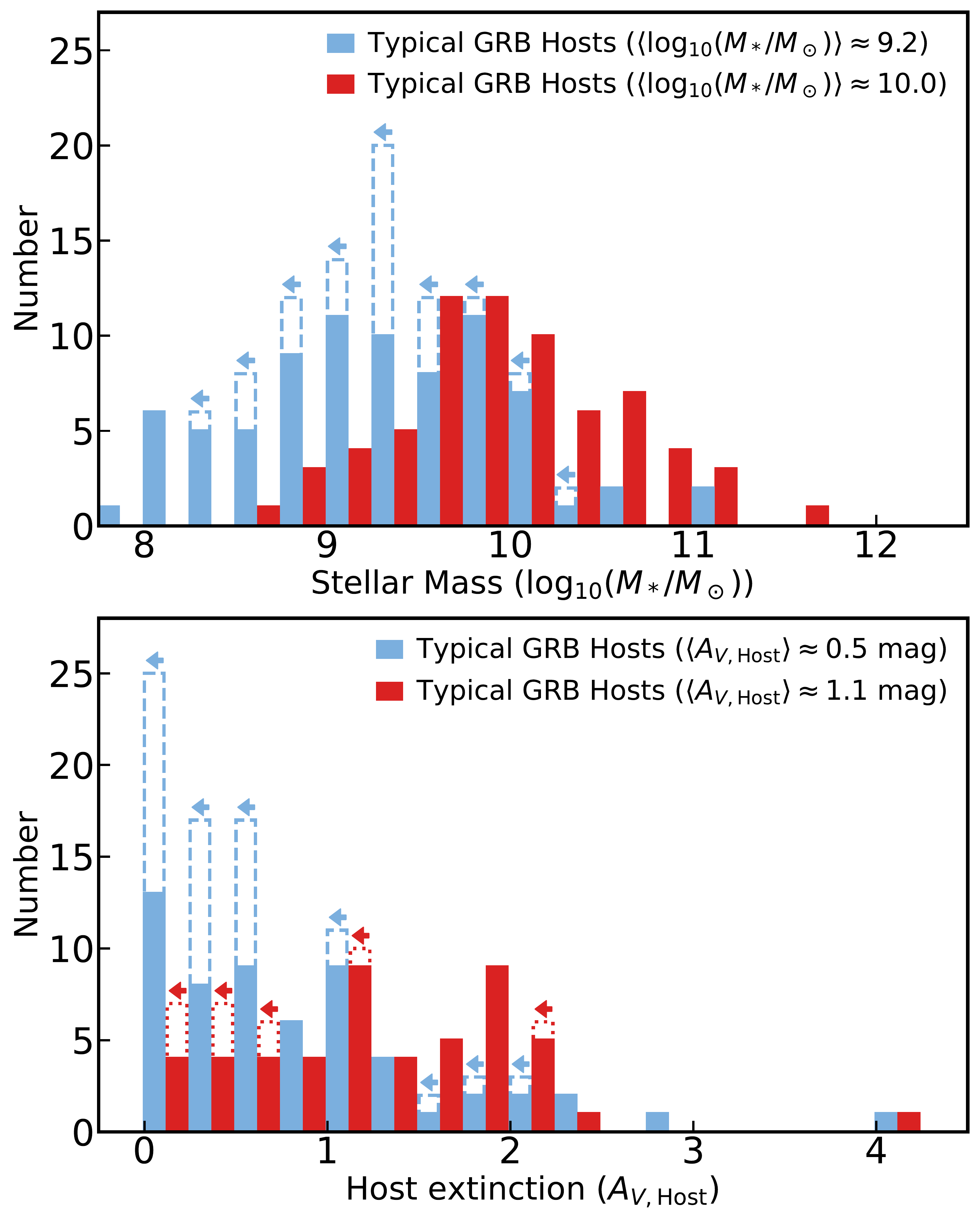}
\caption{
{\it Top:} Distribution of inferred stellar masses ($\log_{10}(M_*/M_\odot)$) of the hosts of 82 typical GRBs (blue) and the hosts of 68 dark GRBs (red), gathered from the literature (\citealt{Savaglio2009, Liebler&Berger2010, Perley2013_DarkGRB_AV, Hunt2014, Piranomonte2015, Perley2016_II, Japelj2016_LGRB_Bat6_II, Palmerio2019_LGRB_Bat6_III}, This Work). Available upper limits are denoted by dashed, unfilled distribution with leftward arrow. The hosts of the dark GRB population are generally more massive than the typical GRB population (median $\log_{10}(M_*/M_\odot) \approx 9.9$ for the hosts of dark GRB population, vs median $\log_{10}(M_*/M_\odot) \approx 9.2$ for the typical GRB population)
{\it Bottom:} Distribution of extinction, $A_{V, \rm Host}$, of 92 typical GRB hosts (blue) and 60 dark GRB hosts (red), gathered from the literature (\citealt{Savaglio2009, Perley2013_DarkGRB_AV, Hunt2014, Kruhler2015, Piranomonte2015, Japelj2016_LGRB_Bat6_II}, This Work), where upper limits are denoted by dashed, unfilled distribution with leftward arrow.
The hosts of the dark GRB population are generally more dusty than the typical GRB population (median $A_{V, \rm Host} \approx 1.1~{\rm mag}$ for the hosts of dark GRB population, vs median $A_{V, \rm Host} \approx 0.5~{\rm mag}$ for the typical GRB population)
} 
\label{fig:Mass_AV_hist}
\end{figure}

First, we investigate the global host properties of dark GRBs, to examine how their hosts differ from the typical GRB host population. We gather a sample of 150 GRB hosts that have inferred stellar mass measurements ($M_*/M_\odot$, \citealt{Savaglio2009, Liebler&Berger2010, Perley2013_DarkGRB_AV, Hunt2014, Piranomonte2015, Perley2016_II, Japelj2016_LGRB_Bat6_II, Palmerio2019_LGRB_Bat6_III}, This Work), and plot the $\log(M_*/M_\odot)$ distribution of the typical and dark GRB host populations  in Figure \ref{fig:Mass_AV_hist}. We find that the host galaxies of dark GRBs tend to be more massive (median $\log(M_*/M_\odot) \approx 9.9$) than typical GRB host galaxies (median $\log(M_*/M_\odot) \approx 9.2$), in alignment with previous results based on smaller samples \citep{Kruhler2011, Perley2013_DarkGRB_AV}. Moreover, we find that only $\sim 16\%$ (12/82, excluding upper limits) of the hosts of the typical GRB population are more massive than the median host mass of the dark GRB population. A natural explanation could be that high-mass galaxies have a larger number of obscured sight lines as they overall have larger dust contents (e.g. \citealt{Santini2014, Calura2017}) which would result in high-$A_{\rm V}$, dark GRBs in these types of hosts \citep{Perley2013_DarkGRB_AV}. 

%%%%% Mass and A_V,Host Histogram FIGURE

To further this point, we gather a sample of 152 GRBs that have inferred global host extinctions ($A_{V, \rm Host}$, \citealt{Savaglio2009, Perley2013_DarkGRB_AV, Hunt2014, Kruhler2015, Piranomonte2015, Japelj2016_LGRB_Bat6_II}, This Work), and plot the $A_{V, \rm Host}$ distribution of the typical GRB population and the dark GRB population in Figure \ref{fig:Mass_AV_hist}. Indeed, we find that dark GRBs typically occur in dustier host galaxies (median $A_{V, \rm Host} \approx 1.1~{\rm mag}$) compared to typical GRBs (median $A_{V, \rm Host} \approx 0.5~{\rm mag}$) (see also \citealt{Perley2013_DarkGRB_AV}). This trend, combined with the tendency of dark GRBs to occur in more massive galaxies, implies that the mass of dark GRB host galaxies is linked to a dustier host galaxy overall. Indeed, there are known positive correlations between stellar mass and dust mass (e.g. \citealt{Santini2014, Calura2017}), that support this idea. The trend of dark GRBs originating in massive, dusty hosts indicates that the high line-of-sight extinction of dark GRBs is, at least in part, attributed to global host extinction.

\begin{figure}
\centering
\includegraphics[width=0.48\textwidth]{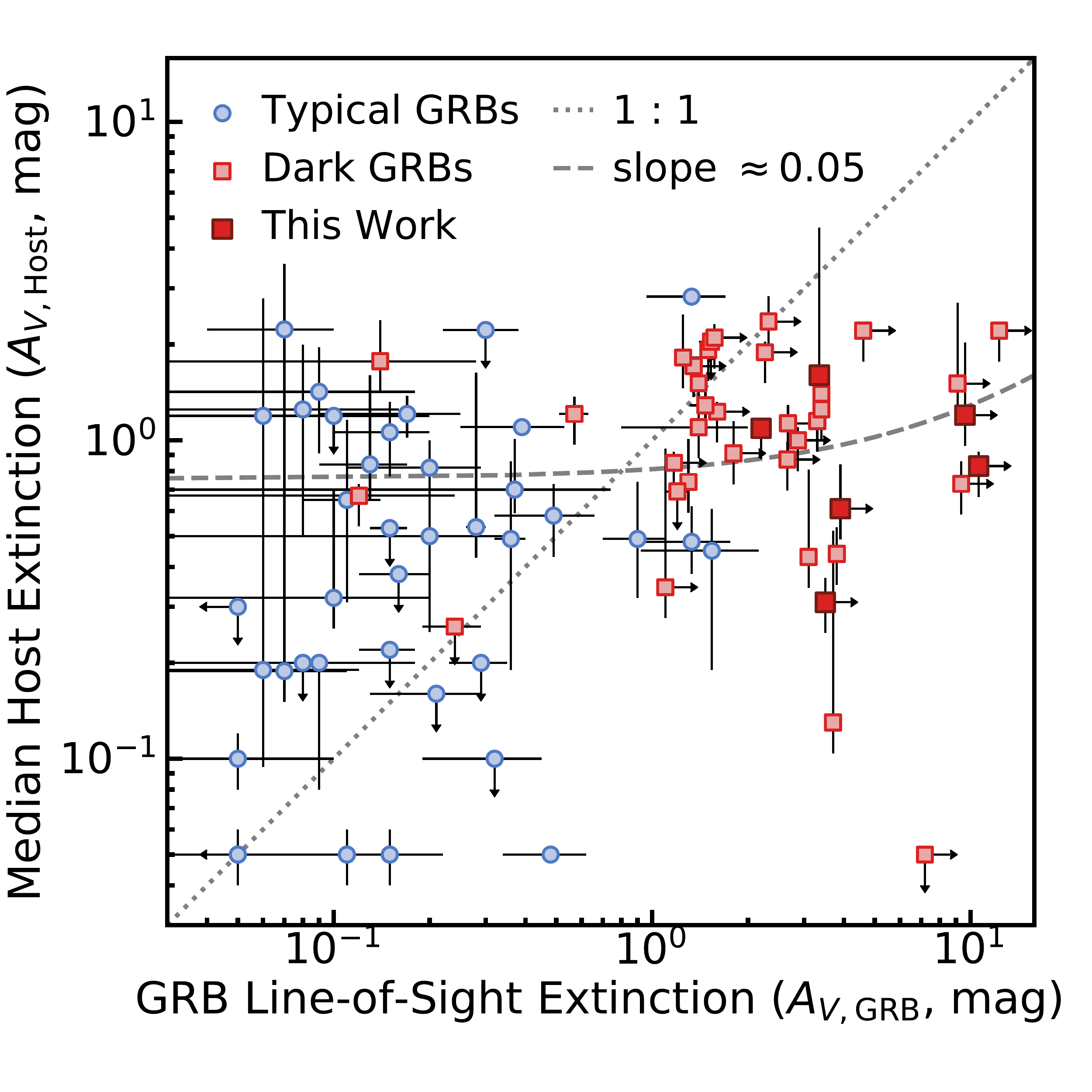}
\caption{
Median Host Extinction ($A_{V, \rm Host}$) versus GRB line-of-sight extinction ($A_{V,\rm GRB}$) for the dark (light red squares) and typical, optically-bright (blue circles) long GRB populations, 
with $A_{V,\rm GRB}$ from \citet{Kann2006, Kann2010,Zafar2011, Greiner2011, Kruhler2011, Perley2013_DarkGRB_AV} and $A_{V,\rm Host}$ from \citet{Savaglio2009, Perley2013_DarkGRB_AV, Kruhler2015, Japelj2016_LGRB_Bat6_II,Kruhler2011, Perley2013_DarkGRB_AV}.
Our sample of six dark GRBs with $A_{V, \rm Host}$ and $A_{V, \rm GRB}$ measurements are represented by red squares with dark outlines.
When necessary, we convert $E_{B-V}$ values to $A_{V, \rm Host}$ using the convention $R = E_{B-V}/A_V = 3.14$ \citep{Schultz1975}.
To be consistent with Figure~15 from \citet{Perley2013_DarkGRB_AV}, for host galaxies with upper limits on their median host extinction, we plot the upper limit at the upper bound ($+ 1 \sigma$) of the allowed $A_{V, \rm Host}$ range. Additionally, we choose the $A_{V, \rm GRB}$ derived from the SMC dust fit from Table 1 of \citet{Kann2006} and Table 3 of \citet{Kann2010}, both to be consistent with Figure~15 from \citet{Perley2013_DarkGRB_AV} and to be consistent with our sample, for which we also use an SMC dust extinction law. For clarity, for GRBs with $A_{V, \rm Host}$ or $A_{V, \rm GRB} \approx 0.0~{\rm mag}$, we plot their upper bounds at $0.05~{\rm mag}$. The dotted line denotes a $1:1$ relation between median host extinction and GRB line-of-sight extinction. The dashed line denotes the correlation found between $A_{V, \rm Host}$-$A_{V, \rm GRB}$ pairs, with a slope of $\approx 0.05$ (see Section \ref{sec:Discussion_Origin_of_dust}). These lines are meant to guide the readers eye.
} 
\label{fig:AV_Host_vs_GRB}
\end{figure}

We next investigate whether the dust extinction of dark GRBs as inferred from their afterglows is linked to the dust content of the host galaxy by determining whether $A_{V, \rm GRB}$ is directly linked to the $A_{V, \rm Host}$. \citet{Perley2013_DarkGRB_AV} suggested that the majority of long GRBs have $A_{V, \rm GRB}$ values within a factor of $\sim\!2-3$ of their host $A_{V, \rm Host}$, following a near $1:1$ relation between $A_{V, \rm GRB}$ and $A_{V, \rm Host}$. This correlation between $A_{V, \rm GRB}$ and $A_{V, \rm Host}$ could occur, for instance, if host galaxies of GRBs have uniform dust distributions, resulting in a homogenous screen of dust in the hosts' diffuse ISM (see \citealt{Perley2013_DarkGRB_AV}). To test this, we gather available $A_{V, \rm GRB}$ measurements from the literature for typical \citep{Kann2006, Kann2010, LiangLi2010, Covino2013, Zafar2011, Greiner2011, LittleJohns2015} and dark GRBs (\citealt{Kann2010, LiangLi2010, Kruhler2011, Greiner2011, Zafar2011, Perley2013_DarkGRB_AV, Covino2013, LittleJohns2015, Laskar2016_160509A_lab16}, This Work) and compare them to their corresponding $A_{V, \rm Host}$ (\citealt{Cenko2009_darkGRBs, Savaglio2009, Kruhler2011, Perley2013_DarkGRB_AV, Perley2015, Kruhler2015, Piranomonte2015, Vergani2015_LGRB_Bat6_I, Japelj2016_LGRB_Bat6_II}, This Work). We plot the $A_{V, \rm Host}$-$A_{V, \rm GRB}$ pairs in Figure \ref{fig:AV_Host_vs_GRB}, and find that the entire long GRB population spans a wide range of $A_{V, \rm Host}$ ($\sim 0-4.7~{\rm mag}$) and $A_{V, \rm GRB}$ ($\sim 0$ to $\gtrsim 12.3~{\rm mag}$).

To test whether there is a relation between $A_{V, \rm Host}$ and $A_{V, \rm GRB}$, we randomly sample $A_{V, \rm Host}$-$A_{V, \rm GRB}$ pairs from the long GRB sample for which there are published $A_{V, \rm Host}$ and $A_{V, \rm GRB}$ values, taking into account error bars and upper/lower limits,\footnote{For cases in which there are quoted error bars for an $A_V$ value (either $A_{V, \rm Host}$ or $A_{V, \rm GRB}$), we randomly sample using an asymmetric Gaussian, with the $A_V$ value as the mean, $\mu$, and the upper and lower errors on $A_V$ as the standard deviation, $\sigma$, for either side of the Gaussian, respectively. For cases in which there is an upper limit quoted for an $A_V$ value, we randomly sample using a tophat function from 0 to the $A_V$ value. For GRB\,160509A we assume a tophat function for $A_{V, \rm Host}$ from 1.60-4.66~{mag} (See Section~\ref{sec:Prospector}) Finally, for cases in which there is a lower limit quoted for an $A_{V, \rm GRB}$ value, we randomly sample from a half Gaussian distribution with the $A_{V, \rm GRB}$ lower limit as $\mu$, and a $5\sigma$ maximum $A_{V, \rm GRB}$ of $25~{\rm mag}$, such that $\sigma = (25-\mu)/5$.} and fit a line to the randomly drawn sample using \texttt{curve\char`_fit} from the \texttt{scipy} package. We repeat this process $1000$ times and produce a distribution of line slopes that are fit by the $A_{V, \rm Host}$-$A_{V, \rm GRB}$ pairs. We find that our distribution of $A_{V, \rm Host}$-$A_{V, \rm GRB}$ slopes has a median value of $\approx 0.05$, a nearly flat correlation instead of a $1:1$ relation. Moreover, within the dark GRB sample alone, we also find only weak correlation between $A_{V, \rm Host}$-$A_{V, \rm GRB}$ (median slope of $\approx 0.03$). The weak correlation between $A_{V, \rm Host}$ and $A_{V, \rm GRB}$ could indicate that either that the high line-of-sight extinction is caused by dust extinction from the extremely local ($\sim$parsec) environment of the dark GRB, or that the host galaxies of long GRBs have patchy, rather than a uniform, dust distributions. In this latter case, the dust extinction driving the high $A_{V, \rm GRB}$ is a geometrical line-of-sight effect that is probabilistic in nature.

%%%%% A_V Host vs GRB FIGURE
%%%%% A_V Host vs GRB FIGURE

To investigate the contribution of the extremely local environment of the GRB (within the blast wave radius $\sim 0.2-50~{\rm parsec}$) to the high line-of-sight extinctions, we compare the circumburst densities as inferred from the afterglow. The naive expectation is that if the dust that is providing obscuration of the afterglow originates on $\sim$parsec scales, dark GRBs will trace environments with higher inferred densities. The sample of six dark GRBs we model in this paper spans a wide range of densities ($n_0 \sim 10^{-3}-10^{1}~{\rm cm^{-3}}$; in the cases of wind environments we calculate $n_0$ at $10^{17}~{\rm cm}$). Compared to typical long GRBs \citep{Panaitescu2002_pk02, Price2002_pbr02, Yost2003_yhsf03, Frail2005_fsk05, Chandra2008_ccf08, Cenko2010_cfh10, Cenko2011_cfh11, Laskar2013_lbz13, Laskar2014_tl14, Laskar2015_tl15,     Alexander2017_alb17, Tanvir2018_nt18,  LaskarBC2018_GRBsI_lbc18, LaskarBM2018_GRBsII_lbm18, LaskarAB2018_lab18, LaskarvES2019_lves19}, the densities of the dark GRBs in our sample fall well within the bounds of the typical GRB sample as a whole ($n_0 \sim 10^{-5}-10^{3}~{\rm cm}^{-3}$). Additionally, we find that $A_{V, \rm GRB}$ is not correlated with the circumburst density. Indeed, the long GRBs with the highest measured densities (GRBs\,050904 and 120404A, $n_0 \sim 3.5-6.3 \times 10^{2}~{\rm cm}^{-3}$) have low inferred $A_{V, \rm GRB}$ values ($\lesssim 0.05-0.13~{\rm mag}$; \citealt{Laskar2014_tl14, Laskar2015_tl15}). This implies that the higher inferred line-of-site extinction for dark GRBs is not a result of the extremely local environment ($\sim 0.2-50~{\rm pc}$).

In summary, we find that dark GRBs tend to occur in more massive, dustier host galaxies than typical GRBs. Additionally, the observed dust obscuration of dark GRBs favors a patchy dust distribution over a uniform one in host galaxies, as $A_{V, \rm GRB}$ is only weakly correlated to $A_{V, \rm Host}$. Furthermore, the dust obscuration of dark GRBs is not purely a result of the extremely local ($\sim$parsec) environment of the GRB, as $A_{V, \rm GRB}$ is not correlated to the circumburst density. The combination of high dust content and a patchy dust distribution results in a higher probability of any given line-of-sight to intersect a patch of dust, leading to a high $A_{V, \rm GRB}$ and dark GRB. The combination of high line-of-sight extinctions, patchy dust distributions, and association of long GRBs with star-foming galaxies, make dark GRBs exciting probes of obscured star formation. 

\subsection{SFR and Radio Limits}
\label{sec:SFR_Radio_Limits}

%%%%% SFR vs SFR FIGURE
\begin{figure*}
\centering
\includegraphics[width=0.7\textwidth]{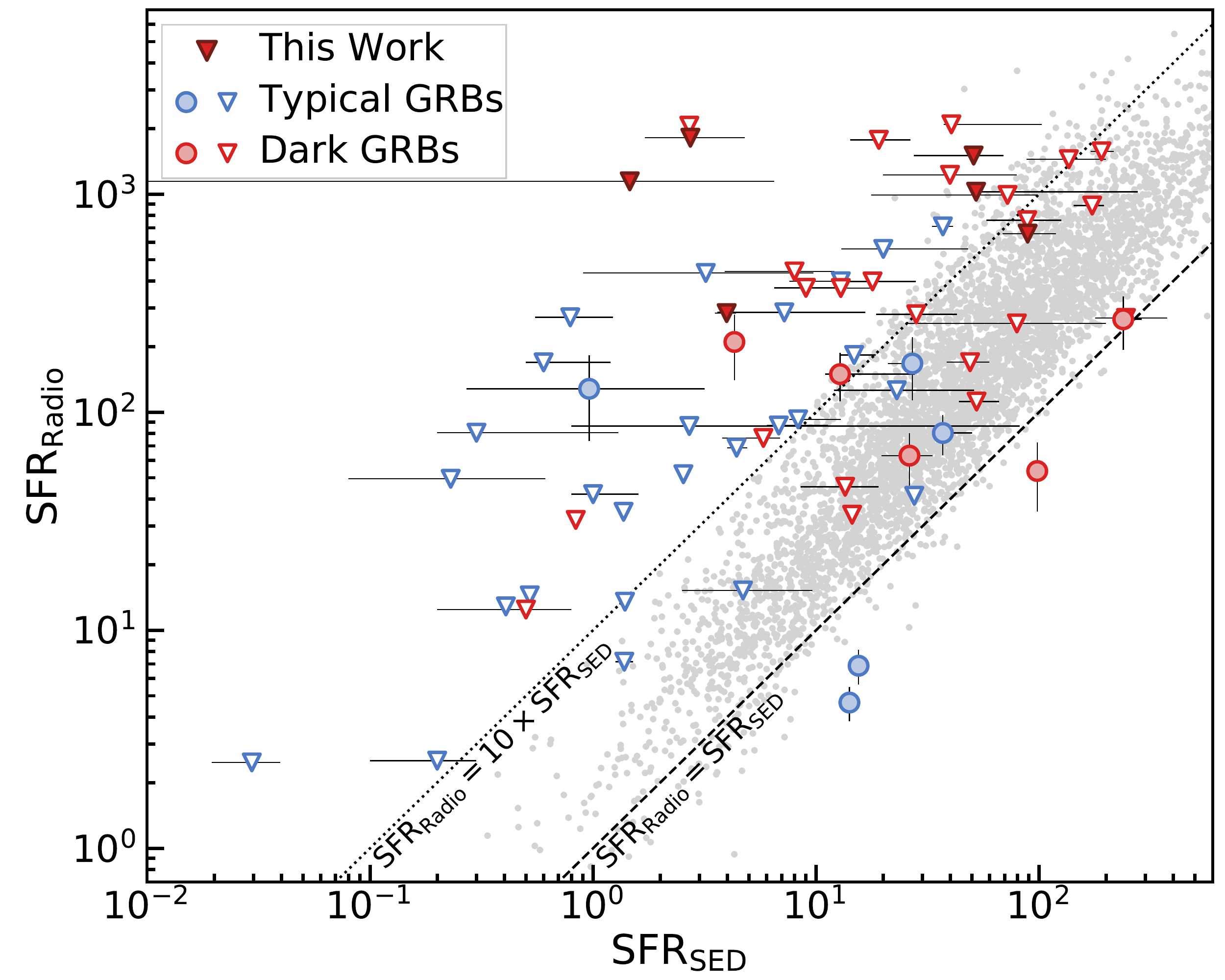}
\caption{
Radio SFR (${\rm SFR}_{\rm Radio}$) versus SED derived SFR (${\rm SFR}_{\rm SED}$) for typical GRB host galaxies (blue) and dark GRB host galaxies (red). Triangles correspond to radio upper limits. ${\rm SFR}_{\rm Radio}$ derived from radio observations in the literature \citep{PerleyPerley2013_RadioSFR, Stanway2014, Perley2015, Greiner2016, Perley2017, Gatkine2020, Eftekhari2021} using the ${\rm SFR}_{\rm Radio}$ relation in \citet{Greiner2016} and assuming a spectral index of $-0.75$. ${\rm SFR}_{\rm SED}$ is taken from the literature \citep{Perley2013_DarkGRB_AV, Perley2015} and supplemented with UV derived SFR or SFR tracers such as $H_\alpha$ \citep{Hunt2014, Kruhler2015}. Our sample of 6 dark GRBs are shown as filled in red triangles, with ${\rm SFR}_{\rm Radio}$ from the most constraining radio afterglow upper limits and ${\rm SFR}_{\rm SED}$ from our \texttt{Prospector} host galaxy modeling.
Also shown are a sample of star forming galaxies for comparable redshifts $z \approx 0.03 - 3.4$ from the VLA-COSMOS source catalog (grey, \citealt{VLA_COSMOS_II_2017}). The dashed and dotted line indicate ${\rm SFR}_{\rm Radio} = {\rm SFR}_{\rm SED}$ and ${\rm SFR}_{\rm Radio} = 10 \times {\rm SFR}_{\rm SED}$, respectively. %The top panel shows the distribution of ${\rm SFR}_{\rm SED}$ for the typical GRB (blue) and dark GRB host (red) populations that are plotted. 
} 
\label{fig:SFR_radio_vs_SFR_SED}
\end{figure*}

We next place limits on the amount of obscured star formation (SF) occurring in the host galaxies of the typical GRBs and dark GRBs. Long GRBs are inherently linked to SF due to their massive star progenitors and therefore their association with star-forming galaxies \citep{Djorgovski1998, Christensen2004, Japelj2016_LGRB_Bat6_II, Palmerio2019_LGRB_Bat6_III}. Additionally, massive, dusty, galaxies such as those that host dark GRBs often have higher SFRs \citep{Santini2014, Calura2017}, which has been corroborated by dark GRB host studies (e.g. \citealt{Perley2013_DarkGRB_AV}), and patchy dust distributions within dark GRB hosts may increase the probability of ongoing SF in the host to be obscured \citep{PerleyPerley2013_RadioSFR}. 
This obscured SF may result in radio bright host galaxies, as the dust becomes transparent at radio wavelengths and reveals the true SFR (${\rm SFR}_{\rm Radio}$), whereas the SFR measured from UV/optical (e.g. emission line) diagnostics and from stellar population synthesis modeling of broad-band, galaxy-integrated photometry is only sensitive to the SFR unobscured by dust  \citep{PerleyPerley2013_RadioSFR}. The search for radio emission from typical GRB host galaxies and dark GRB hosts is of interest as the results help answer the question of whether long GRBs are biased or unbiased tracers of SF across the universe (i.e. \citealt{PerleyPerley2013_RadioSFR, Perley2015, Gatkine2020}). 

Several studies have been conducted to search for radio emission from long GRB host galaxies \citep{PerleyPerley2013_RadioSFR, Stanway2014, Perley2015, Greiner2016, Perley2017, Gatkine2020, Eftekhari2021}, with a handful ($\sim 10$) of successful detections. These studies are not necessarily representative of the overall long GRB host population, as some focused exclusively on dark GRBs, some focused on typical GRBs at high redshift ($z > 2.5$), and few took an unbiased approach to all long GRB hosts.
As there is no large, unbiased, survey of radio observations of long GRB host galaxies, we must rely on the these smaller studies, despite their potential biases in sample selection. We collect the radio fluxes from these studies to provide a broad view of the currently available observations, as well as to make comparisons to our sample of six dark GRBs. We calculate the ${\rm SFR}_{\rm Radio}$ of these observations using the relation in \citet{Greiner2016} (assuming a spectral index of $-0.75$), and plot them against the SED-derived SFR (${\rm SFR}_{\rm SED}$, \citealt{Perley2013_DarkGRB_AV, Perley2015}, supplementing with UV-derived SFR or SFR tracers such as ${\rm H}\alpha$, ${\rm H}\beta$, and ${\rm O}{\rm II}$  from \citealt{Hunt2014, Kruhler2015}), in Figure \ref{fig:SFR_radio_vs_SFR_SED}. For comparison, we also plot a sample of star-forming field galaxies at comparable redshifts ($z \approx 0.03 - 3.4$) from the VLA-COSMOS source catalog \citep{VLA_COSMOS_II_2017}. 

The VLA-COSMOS sources typically have SFR ratios, $R_{\rm SFR} = {\rm SFR}_{\rm Radio}/{\rm SFR}_{\rm SED}$, on the order of $\approx 1-10$. We define galaxies with ``significant'' obscured SF as galaxies that satisfy $R_{\rm SFR} \gg 10$. Of the $\sim 80$ long GRB host galaxies that have been observed at radio wavelengths, only $\sim 10$ have unambiguous host detections \citep{PerleyPerley2013_RadioSFR, Stanway2014, Perley2015, Perley2017}. Of the $\sim 10$ detected long GRB host galaxies, only 3 display significant amounts of obscured SF, and the majority of radio detected long GRB hosts have $R_{\rm SFR}$ within the range we would expect compared to other star forming galaxies ($R_{\rm SFR} \approx 1-10$, i.e. the VLA-COSMOS sources). 

To assess the detectability of such obscured SF in the hosts of the six dark GRBs in our sample, we now place upper limits on the ${\rm SFR}_{\rm radio}$ of the host galaxies. We consider the most constraining afterglow upper limits of our sample (\citealt{Zauderer2013, VanderHorst2015, 140713A_GMRT_Radio_upperlimit, Laskar2016_160509A_lab16}, This Work), and we place limits of ${R}_{\rm SFR} \lesssim 8-700$ (also plotted in Figure~\ref{fig:SFR_radio_vs_SFR_SED}). With the afterglow radio limits, we are unable to rule out significant amounts of obscured SF for all bursts in our sample except GRB\,130131A. We also searched the VLASS \citep{VLASS} for radio emission at 3~GHz at the positions of the bursts, but found the non-detections to be less constraining than the radio afterglow upper limits; this holds true even for the projected total sensitivity of VLASS of the combined 3 epochs (${R}_{\rm SFR} \lesssim 30-2000$).

With the next generation VLA (ngVLA), we would be sensitive to radio emission at the level of $\sim 0.75~\mu{\rm Jy}$ with one hour of observation at 2.4~GHz \citep{ngVLA}\footnote{https://ngvla.nrao.edu/page/performance}. This radio flux would correspond to ${\rm SFR}_{\rm Radio} \gtrsim 0.02-16~M_\odot/{\rm yr}$
for $z = 0.1-2.0$. With the ngVLA, we would be able to detect at least 5 of the 6 dark GRB hosts in our sample (with the exception of GRB\,110709B, whose ${\rm SFR}_{\rm SED}$ is $\approx 7 \times$ lower than that of the ngVLA limit at the host redshift), assuming ${R}_{\rm SFR} \approx 1$ (corresponding to unobscured SF). If we consider an unbiased sample of long GRBs, such as the {\it Swift}/BAT6 sample \citep{Salvaterra2012_bat6}, we can calculate the expected detection fraction of the long GRB host galaxies by the ngVLA. Of the 24 {\it Swift}/BAT6 host galaxies at $0.1 < z < 2 $ with ${\rm SFR}_{\rm SED}$ (or lower limits on ${\rm SFR}_{\rm SED}$) presented in \citet{Japelj2016_LGRB_Bat6_II} and  \citet{Palmerio2019_LGRB_Bat6_III}, we calculate a detection rate with the ngVLA of $\approx 50\%$, assuming $R_{\rm SFR} \approx 1$. A detection rate much larger than these estimates would indicate that long GRB hosts have some amount of obscured star formation, and exact measurements of ${\rm SFR}_{\rm Radio}$ would determine the fraction of long GRB host galaxies that house significant obscured SF.

\section{Conclusions}
\label{sec:Conclusions}

We have newly classified 2 long GRBs as dark (GRB\,130420B and GRB\,160509A) and presented VLA, CARMA, and PdBI afterglow observations of five dark GRBs (GRB\,130131A, 130420B, 130609A, 131229A, and 140713A). We uniformly modeled the radio to X-ray afterglow of five dark GRBs with VLA detections (GRB\,110709B, 111215A, 130131A, 131229A, and 140713A), using our afterglow modeling software that incorporates effects due to jet breaks, scintillation, and IC cooling, and include one dark GRB from the literature which was modeled using the same method and software. The radio detections allowed us to determine the environment and break frequencies of the synchrotron afterglow, in turn constraining the burst energetics, circumburst density, and geometries. Additionally, we fit the host galaxies of 6 dark GRBs (GRB\,110709B, 111215A, 130131A, 131229A, 140713A and 160509A) using \texttt{Prospector}, and present photometric redshifts for 2 of the dark GRBs (GRB\,130131A at $z \approx 1.55$ and GRB\,131229A at $z \approx 1.04$). We come to the following conclusions:
\begin{itemize}
    \item Dark GRBs are not distinct from typical long GRBs in terms of duration, burst kinetic energy, jet opening angle, or circumburst density. However, they are statistically distinct from typical long GRBs in terms of fluence, though this distinction may be attributed to observational biases and inconsistent classification of dark GRBs.
    \item Our sample of six uniformly modeled dark GRBs with VLA detections have line-of-sight extinction values of $A_{V, \rm GRB} \approx 2.2- \gtrsim10.6~{\rm mag}$, demonstrating the importance of radio observations in revealing GRBs with heavily dust-obscured sightlines. These values are $\gtrsim 0.7$ to $\gtrsim 12.8$ times greater than their median host galaxy values $A_{V, \rm Host} \approx 0.3-4.7~{\rm mag}$. 
    \item While dark GRBs do occur in dustier and more massive galaxies than typical long GRBs, the line-of-sight extinction is not strongly correlated to the median host extinction, nor to the circumburst density. This indicates that the origin of the dust along the line-of-sight is due to a clumpy, rather than uniform, dust distribution within the host galaxy. This also disfavors a dust origin from the extremely local ($\sim{\rm parsec}$) environment of the burst. 
    \item Targeted radio searches with $\sim \mu{\rm Jy}$ sensitivity (e.g. the ngVLA) should be capable of detecting $\approx 50\%$ of long GRB host galaxies at $0.1 < z < 2$, where a higher detection rate and exact flux measurements will determine the amount of obscured SF within long GRB hosts. 
\end{itemize}

Our work demonstrates the unique power of rapid-response radio observations with the VLA in uncovering the most obscured GRBs via their afterglows. This is especially important given that these events by definition have extinguished optical emission. Additionally, observations of dark GRB environments, from parsec to kiloparsec scales, lends insight on the distribution of dust and star formation in the galaxies which give rise to these relatively rare transients. Looking forward, next generation radio facilities, in conjunction with UV-optical observations, can be leveraged to determine the degree of obscured star formation for a large population of GRB environments across redshift.

\section{Acknowledgements}

G.S. acknowledges for this work was provided by the NSF through Student Observing Support award SOSP20B-001 from the NRAO. The Fong Group at Northwestern acknowledges support by the National Science Foundation under grant Nos. AST-1814782, AST-1909358 and CAREER grant No. AST-2047919. W.F. gratefully acknowledges support by the David and Lucile Packard Foundation. AJCT acknowledges support from the Spanish Ministry Project PID2020-118491GB-I00, Junta de Andaluc\'ia Project P20\_01068 and the "Center of Excellence Severo Ochoa" award for the Instituto de Astrofísica de Andalucía (SEV-2017-0709)

The National Radio Astronomy Observatory is a facility of the National Science Foundation operated under cooperative agreement by Associated Universities, Inc. This work made use of data supplied by the UK Swift Science Data Centre at the University of Leicester. The National Radio Astronomy Observatory is a facility of the National Science Foundation operated under cooperative agreement by Associated Universities, Inc. This work is based on observations carried out under project number S14DD004 %[XXX-YY]
with the IRAM NOEMA Interferometer. IRAM is supported by INSU/CNRS (France), MPG (Germany) and IGN (Spain). This research was supported in part through the computational resources and staff contributions provided for the Quest high performance computing facility at Northwestern University which is jointly supported by the Office of the Provost, the Office for Research, and Northwestern University Information Technology. W. M. Keck Observatory and MMT Observatory access was supported by Northwestern University and the Center for Interdisciplinary Exploration and Research in Astrophysics (CIERA). Some of the data presented herein were obtained at the W. M. Keck Observatory, which is operated as a scientific partnership among the California Institute of Technology, the University of California and the National Aeronautics and Space Administration. The Observatory was made possible by the generous financial support of the W. M. Keck Foundation. The authors wish to recognize and acknowledge the very significant cultural role and reverence that the summit of Maunakea has always had within the indigenous Hawaiian community. We are most fortunate to have the opportunity to conduct observations from this mountain. Some observations reported here were obtained at the MMT Observatory, a joint facility of the University of Arizona and the Smithsonian Institution. The United Kingdom Infrared Telescope (UKIRT) was supported by NASA and operated under an agreement among the University of Hawaii, the University of Arizona, and Lockheed Martin Advanced Technology Center; operations are enabled through the cooperation of the East Asian Observatory. We thank the Cambridge Astronomical Survey Unit (CASU) for processing the WFCAM data and the WFCAM Science Archive (WSA) for making the data available. This paper includes data gathered with the 6.5 meter Magellan Telescopes located at Las Campanas Observatory, Chile. The LBT is an international collaboration among institutions in the United States, Italy and Germany. The LBT Corporation partners are: The University of Arizona on behalf of the Arizona university system; Istituto Nazionale di Astrofisica, Italy;  LBT Beteiligungsgesellschaft, Germany, representing the Max Planck Society, the Astrophysical Institute Potsdam, and Heidelberg University; The Ohio State University; The Research Corporation, on behalf of The University of Notre Dame, University of Minnesota and University of Virginia. IRAF is distributed by the National Optical Astronomy Observatory, which is operated by the Association of Universities for Research in Astronomy (AURA) under a cooperative agreement with the National Science Foundation. This publication makes use of data products from the Two Micron All Sky Survey, which is a joint project of the University of Massachusetts and the Infrared Processing and Analysis Center/California Institute of Technology, funded by the National Aeronautics and Space Administration and the National Science Foundation.

\facilities{VLA, CARMA, PdBI, Swift, Keck:LRIS, Keck:MOSFIRE, LBT:LBC, MMT:MMIRS, MMT:Binospec, UKIRT:WFCAM, Magellan:LDSS3, Magellan:Fourstar, Spitzer:IRAC}
\software{CASA \citep{CASA}, Miriad \citep{Miriad}, pwkit \citep{Pwkit}, scipy \citep{2020SciPy-NMeth}, matplotlib \citep{matplotlib}, pandas \citep{pandas}, numpy \citep{numpy}, Prospector \citep{Leja_2017}}

\clearpage

% \begin{appendices}
\appendix
\restartappendixnumbering
%%%% Xray property TABLE
%%%%%%%%%%%%%%% TABLE
\tabletypesize{\normalsize}
\begin{deluxetable*}{l|cccccc}
\tablecolumns{6}
\tablewidth{0pc}
\tablecaption{{\it Swift}/XRT Properties
\label{tab:Xray_properties}}
\tablehead {
\colhead {GRB}                &
\colhead{R.A.}           &
\colhead{Dec.}           &
\colhead{90\% Error}           &
\colhead{$T_{90}$ (15-350 keV)}           &
\colhead {$f_{\gamma}$ (15-150 keV)}      \\
\colhead {}    &
\colhead{(J2000)}   &
\colhead{(J2000)}   &
\colhead{($\arcsec$)}   &
\colhead{(s)}               &
\colhead {($\times 10^{-7} {\rm erg}/{\rm cm}^{2}$)}          &
% \colhead {${\rm SFR}~[M_{\odot/~{\rm yr}}]$}           &
% \colhead {$t_{m}$ [Gyr]}          &
% \colhead {$A_{V}^{\rm host}$~[mag]}          
}
\startdata
130131A & \ra{11}{24}{30.31} & \dec{+48}{04}{32.9} & 1.4 & 51.52 & $3.1 \pm 0.6$ \\
130420B & \ra{12}{12}{30.79} & \dec{+54}{23}{26.3} & 2.0 & 12.64 & $7.2 \pm 0.5$ \\
130609A & \ra{10}{10}{40.44} & \dec{+24}{07}{56.6} & 1.7 & 7.06 & $5.7 \pm 0.4$ \\
131229A & \ra{05}{40}{55.62} & \dec{-04}{23}{46.8} & 1.4  & 48.0 & $68.9 \pm 1.5$ \\
140713A & \ra{18}{44}{25.41} & \dec{+59}{38}{00.5} & 1.4 & 6.02 & $3.7 \pm 0.3$ \\
\enddata
\tablecomments{{\it Swift}/XRT Properties from \citet{Evans2009, Lien2016}.} 
\end{deluxetable*}
%%%%%%%%%%%%%%% TABLE
%%%% Xray proptery TABLE

\section{Observations}
\label{appendix:Observations}

Here we present the X-ray to radio afterglow observations of the 8 dark GRBs in our sample, as well as their host galaxy observations. For the five dark GRBs with no previously published VLA observations, we summarize the Neil Gehrels {\it Swift} Observatory ({\it Swift}) X-ray Telescope (XRT) properties in Table~\ref{tab:Xray_properties}. Unless otherwise stated, all VLA data were manually reduced using standard procedures with the Common Astronomy Software Applications (CASA, \citealt{CASA}), and all Combined Array for Research in Millimeter Astronomy (CARMA) data were manually reduced using standard procedures with the \texttt{Miriad} software package \citep{Miriad}. For VLA and CARMA observations, we measure the flux density and position of the afterglow using the \texttt{imtool} program under the \texttt{pwkit} package, which fits the afterglow to a point source \citep{2017ascl.soft04001W}. The radio observations, including the configuration, gain, bandpass, and flux, calibrators, are summarized in Table~\ref{tab:obs}.

\subsection{GRB\,130131A}

\subsubsection{Swift and Optical Observations}
\label{sec:130131AXrayOpticalObs}

GRB\,130131A was discovered by the Burst Alert Telescope (BAT) on-board {\it Swift} on 2013 January 31.58 \citep{2013GCN.14156SwiftDet}. The XRT started observations of GRB\,130131A at $\delta t = 58.5$~s (where $\delta t$ is the time after BAT trigger), finding an uncatalogued X-ray source within the BAT position \citep{2013GCN.14156SwiftDet, 2013GCN.14160XRTposition}.

An uncatalogued, fading, optical/near-infrared (NIR) source was found within the XRT error circle at $\delta t \approx 0.02- 0.04$~days in $R$-, $J$-, and $K$-band, and was determined to be the optical/NIR afterglow after the source faded by $\gtrsim 2$~mag in $K-$band by $\delta t \approx 0.85~$days \citep{2013GCN.14182OpticallyDark, 2013GCN.14157AfterglowCandidate, 2013GCN.14175AfterglowConfirmed}. We initiated MMT SAO Widefield InfraRed Camera (SWIRC) observations at $\delta t \approx 0.67$ days in $J$- and $H$-band and detected a source coincident with the optical/NIR afterglow  \citep{2013GCN.14182OpticallyDark, 2013GCN.14157AfterglowCandidate, 2013GCN.14175AfterglowConfirmed} in both bands. For photometric calibration, we use sources in the field in common with the 2MASS catalog, and perform aperture photometry using IRAF. We find afterglow magnitudes of $ 21.9 \pm 0.1$~mag in $J$-band and $21.2 \pm 0.1$~mag in $H$-band at a position of R.A.=\ra{11}{24}{30.35} and Dec=\dec{+48}{04}{33.08} with a positional uncertainty of $0.10\arcsec$.

Interpolating the XRT light curve to the times of the $R$-band observations, we find $\beta_{\rm OX}\approx-0.2$ at 0.016~days and 0.041~days, meeting the \citet{Jakobsson2004} criterion and confirming the darkness classification first asserted by \citet{2013GCN.14182OpticallyDark}. We find $\beta_{\rm X} = -1.4 ^{+0.3}_{-0.2}$, and therefore, $\beta_{\rm OX} - \beta_{\rm X} > 0.5$ for the $R$-band detections, indicating that GRB\,130131A also meets the \citet{vanderHorst2009} darkness criterion.

\subsubsection{Radio Afterglow Discovery}
\label{sec:130131Aradioobs}

%%%% AFTERGLOW OBS TABLE
%%%%%%%%%%%%%%% TABLE
\startlongtable
\begin{deluxetable*}{lccccccc}
\tabletypesize{\normalsize}
\tablecolumns{8}
\tablewidth{0pt}
\tablecaption{Radio Afterglow Observations of Dark GRBs
\label{tab:obs}}
\tablehead{
\colhead{GRB} &
\colhead{Facility} &
\colhead{Config.} &
\colhead {Mid-time}	 &
\colhead {$\delta t$} &
\colhead{Gain/Band-pass/Flux} &
\colhead{$\nu$} &
\colhead {$F_{\nu}$} \\
\colhead {} 		&
\colhead {} 		&
\colhead {} 		&
\colhead {(UT)}	&
\colhead {(d)}	 &
\colhead {(Calibrators)} 		&
\colhead {(GHz)}	 &
\colhead {($\mu$Jy)}	
}
\startdata
130131A & VLA & D & 2013 Feb 1.27 & 0.68 & J1146+3958/3C286/3C286 & 6.0 & 29.4 $\pm$ 8.1 \\
 & & & & & J1153+4931/3C286/3C286 & 19.2 & 124.0 $\pm$ 31.6\\
 & & & & & & 24.5 & 142.3 $\pm$ 26.5 \\
 & & & 2013 Feb 2.59 & 1.99 & J1146+3958/3C286/3C286 & 6.0 & $< 27.3^{a}$ \\
 & & & & & J1153+4931/3C286/3C286 & 19.2 & 116.3 $\pm$ 13.7\\
 & & & & & & 24.5 & 116.2 $\pm$ 17.3\\
 & & & 2013 Feb 5.28 & 4.68 & J1146+3958/3C286/3C286 & 6.0 & 54.2 $\pm$ 10.5\\
 & & & & & J1153+4931/3C286/3C286 & 19.2 & 63.1 $\pm$ 18.4\\
 & & & & & & 24.5 & $<$ 47.4\\
 & & & 2013 Feb 14.47 & 13.88 & J1146+3958/3C286/3C286 & 6.0 & $<$ 28.5\\
 & & & & & J1153+4931/3C286/3C286 & 19.2 & $<$ 20.4 \\
 & & & & & & 24.5 & $< 25.8$\\
 & CARMA & B & 2013 Feb 1.45 & 0.67 & 1153+495/0927+390/Jupiter & 85.5 & $<$ 230\\
 & & & 2013 Feb 2.49 & 1.91 & 1153+495/3C273/3C273 & 85.5 & $<$ 235 \\
 & & & 2013 Feb 3.40 & 2.82 & & 85.5 & $<$ 491 \\
 \hline
130420B & VLA & D & 2013 Apr 23.32 & 2.78 & J1219+4829/3C286/3C286 & 6.0 & $< 22.8$ \\
 & & & 2013 Apr 23.32 & 2.78 & & 21.8 & $< 34.8$ \\
 & CARMA & C & 2013 Apr 23.14 & 2.60 & 1153+495/0927+39/3C84 & 85.5 & $< 436$ \\
 \hline
130609A & VLA & C & 2013 Jun 9.9 & 0.78 & J0956+2515/3C147/3C147 & 6.0 & $< 28.5$ \\
 & & & 2013 Jun 9.9 & 0.78 & & 21.8 & $< 38.4$ \\
 & CARMA & D & 2013 Jun 10.15 & 1.02 & 0956+252/0927+390/0854+201 & 85.5 & $< 341$\\
 \hline
131229A & VLA & B & 2013 Dec 30.18 & 0.91 & J0541-0541/J0541-0541/J0541-0541 & 6.0 & $74 \pm 19$ \\
 & & & 2013 Dec 30.18 & 0.91 & & 21.8 & $< 45.6$ \\
 \hline
140713A & VLA & D & 2014 Jul 17.12 & 3.34 & J1927+6117/3C286/3C286 & 13.4 & 779.4 $\pm$ 35.2 \\
 & & & & & & 15.9 & 872.6 $\pm$ 37.3 \\
 & & & 2014 Jul 17.14 & 3.36 & & 4.9 & 94.3 $\pm$ 24.3 \\
 & & & & & & 7.0 & 119.5 $\pm$ 18.4 \\
 & & & 2014 Jul 20.11 & 6.33 & & 13.4 & 739.1 $\pm$ 47.0 \\
 & & & & & & 15.9 & 901.9 $\pm$ 73.3 \\
 & & & 2014 Jul 20.13 & 6.35 & & 4.9 & 111.8 $\pm$ 20.3 \\
 & & & & & & 7.0 & 237.5 $\pm$ 15.1 \\
 & & & 2014 Jul 26.08 & 12.30 & & 13.4 & 2149.6 $\pm$ 38.6 \\
 & & & & & & 15.9 & 2274.0 $\pm$ 35.0 \\
 & & & 2014 Jul 26.1 & 12.32 & & 4.9 & 219.5 $\pm$ 18.8 \\
 & & & & & & 7.0 & 897.8 $\pm$ 21.9 \\
 & & & 2014 Aug 8.05 & 25.27 & & 13.4 & 950.3 $\pm$ 22.0 \\
 & & & & & & 15.9 & 931.5 $\pm$ 21.9 \\
 & & & 2014 Aug 8.07 & 25.29 & & 4.9 & 130.2 $\pm$ 27.9 \\
 & & & & & & 7.0 & 150.0 $\pm$ 15.2 \\
 & & & 2014 Aug 28.04 & 45.26 & & 13.4 & 267.0 $\pm$ 16.0 \\
 & & & & & & 15.9 & 229.6 $\pm$ 14.7 \\
 & & & 2014 Aug 28.06 & 45.28 & & 4.9 & 315.6 $\pm$ 14.2 \\
 & & & & & & 7.0 & 379.1 $\pm$ 39.7 \\
 & & & 2014 Sep 14.97 & 63.19 & & 13.4 & 159.1 $\pm$ 14.3 \\
 & & & & & & 15.9 & 118.2 $\pm$ 19.7 \\
 & & & 2014 Sep 14.99 & 63.21 & & 4.9 & 100.5 $\pm$ 13.1 \\
 & & & & & & 7.0 & 116.5 $\pm$ 26.8 \\
 & PdBI & D & 2014 Jul 18.16 & 4.38 & --/--/MWC349 & 86.7 & 3180.0 $\pm$ 90.0\\
 & & & 2014 Jul 28.97 & 15.19 & & 86.7 & 1030.0 $\pm$ 110.0\\
 & & & 2014 Aug 12.08 & 29.3 & & 86.7 & 260.0 $\pm$ 130.0\\
 & & & 2014 Aug 14.04 & 31.26 & & 86.7 & 190.0 $\pm$ 80.0\\
 & & & 2014 Sep 14.85 & 63.07 & & 86.7 & $< 210.0$\\
 & CARMA & E & 2014 Jul 15.37 & 1.59 & 1849+670/1927+739/MWC349 & 85.5 & 1506.8 $\pm$ 255.0\\
 & & & 2014 Jul 17.28 & 3.50 & & 85.5 & 3863.9 $\pm$ 297.2\\
 & & & 2014 Jul 18.41 & 4.63 & & 85.5 & 3008.0 $\pm$ 302.3\\
 & & & 2014 Jul 20.18 & 6.40 & 1849+670/1512-090/MWC349 & 85.5 & 2396.6 $\pm$ 332.3\\
 & & & 2014 Jul 22.37 & 8.59 & 1849+670/1927+739/MWC349 & 85.5 & 2072.9 $\pm$ 215.0\\
 & & & 2014 Jul 26.32 & 12.54 & & 85.5 & 1529.3 $\pm$ 198.6\\
 & & & 2014 Aug 11.14 & 28.36 & 1849+670/3C279/MWC349 & 85.5 & $<$ 670.2\\
\enddata
\tablecomments{All listed data are new to this work. Our modeling is supplemented by radio data from \cite{2013GCN.15638_131229A_CARMA, 2014GCN.16641GMRT, Anderson2018, Higgins2019} \\
$^{a}$ Upper limits correspond to 3$\sigma$. \\}
\end{deluxetable*} 

We initiated a series of four observations of GRB\,130131A starting on 2013 February 1.28 UT with the VLA under Program 13A-046 (PI: E. Berger) at 6.0~GHz and 21.8~GHz (side-band frequencies of 19.2~GHz and 24.5~GHz). We detect a clear source within the XRT error circle with $3.6\sigma$ significance of $\approx 29.4~\mu{\rm Jy}$ at 6.0~GHz (the information presented here supersedes the information from a preliminary reduction reported in \citealt{2013GCN.14171VLADet}). We took further VLA 6.0~GHz and 21.8~GHz observations until 2013 February 14.47 UT. The first 3 epochs of observations were taken with 8-bit sampling, while we used 3-bit sampling for the final 21.8~GHz observations to improve sensitivity. Due to the increased complexity of the 3-bit observations, we used the CASA VLA pipeline (version 2020.1.0.36) for the reduction \citep{CASA}. The details of the VLA observations are listed in Table~\ref{tab:obs}. Using the observation with the highest signal to noise (the second 21.8~GHz epoch),  we  derive  a  position  for  the  radio afterglow of R.A.=\ra{11}{24}{30.401} ($\pm 0.739"$) and Dec=\dec{+48}{04}{33.15} ($\pm 0.357\arcsec$).

We initiated a series of three observations of GRB\,130131A starting on 2013 February 1.45 UT using CARMA under program number c0999 (PI: A. Zauderer). We observed the field at a mean frequency of 85.5~GHz and do not detect a source to a $3 \sigma$ limit of $\lesssim 230~\mu{\rm Jy}$ (the information presented here supercedes the marginal detection reported in \citealt{2013GCN.14172CARMAObs}). We took further observations until 2013 February 3.40 UT, with no significant detection. The details of the CARMA observations are listed in Table~\ref{tab:obs}.

\subsubsection{Host Galaxy Observations}
\label{sec:130131Ahostobs}

The faint host galaxy of GRB\,130131A was first reported in \citet{Chrimes2019}, within the XRT position \citep{Evans2009}. We initiated $griz-$band observations of the host galaxy on 2015 December 3 UT with the Large Binocular Camera (LBC) on the Large Binocular Telescope (LBT/LBC) atop Mount Graham, Arizona. We reduced these images using standard routines in the IRAF/{\tt ccdproc} package \citep{iraf1,iraf2}. We applied bias and flat-field corrections and co-added the dithered images in each filter. We performed absolute astrometry with IRAF/{\tt ccmap} and {\tt ccsetwcs} using sources in common with the SDSS DR12 catalog \citep{SDSSDR12}. Using {\tt SExtractor} \citep{SExtractor}, we measure a position of the host: R.A.=\ra{11}{24}{30.317} and Dec =$+$\dec{48}{04}{33.32}. The radio and NIR afterglows are at projected offsets of 0.857$\arcsec$ and 0.401$\arcsec$ from the host, respectively (Figure~\ref{fig:Host_Galaxy_Panel}).

%%%% HOST GALAXY OBS TABLE
%%%%%%%%%%%%%%% TABLE

\tabletypesize{\normalsize}
\begin{deluxetable*}{lcccccccc}%[!t]
\tablecolumns{8}
\tablewidth{0pc}
\tablecaption{Host Galaxy Observations
\label{tab:host}}
\tablehead{
\colhead{GRB} &
\colhead{Date} &
\colhead {Telescope/Instrument}	 &
\colhead {Filter} &
\colhead{Exposure Time} &
\colhead {Magnitude} &
\colhead{$A_{\lambda}$} &
\colhead {Reference} \\
\colhead {}	&
\colhead {(UT)} 		&
\colhead {}	&
\colhead { }	 &
\colhead {(s)}	 &
\colhead {(AB)}	&
\colhead {(AB)}	&
\colhead {}	&
}
\startdata
130131A & 2015-12-03 & LBT/LBC & $g$ & $8 \times 300$ & $24.01 \pm 0.076$ & $0.052$ & This work \\
        & 2014-10-09 & HST/WFC3 & F606W &  1101 & $24.089 \pm 0.037$ & 0.039  & \citet{Chrimes2019}\\
        & 2015-12-03 & LBT/LBC & $r$ & $6 \times 300$ & $23.58 \pm 0.085$ & $0.036$ &  This Work\\
        & 2015-12-03 & LBT/LBC & $i$ & $12 \times 150$ & $23.19 \pm 0.075$ & $0.027$ &  \\
        & 2015-12-03 & LBT/LBC & $z$ & $16 \times 150$ & $22.89 \pm 0.097$ & $0.020$ &   \\
        & 2020-12-26 & MMT/MMIRS & $Y$ &  $20\times60$ & $22.633 \pm 0.178$ & 0.017 & \\
        & 2016-03-18 & UKIRT/WFCAM & $J$ & $36 \times 60$ & $21.839 \pm 0.191$ & $0.011$ \\
        & 2016-03-18 & UKIRT/WFCAM & $H$ & $36 \times 60$ & $21.817 \pm 0.226$ & $0.007$ \\
        & 2021-01-08 & MMT/MMIRS & $K$ &  $20\times60$ & $22.0821 \pm 0.130$ & 0.005 & \\
        & 2014-10-09 & HST/WFC3 & F160W & 1059 & $21.889 \pm 0.022$ & 0.008 & \citet{Chrimes2019}\\
        & 2016-07-17 & Spitzer/IRAC & 3.6$\mu$m & 36 $\times$ 100 & $21.167 \pm 0.01 $ & 0.002 & This work \\
        & 2016-07-17 & Spitzer/IRAC & 4.5$\mu$m & 36 $\times$ 100 & $20.968 \pm 0.01$ & - \\
\hline 
130420B & 2020-11-23 & MMT/Binospec & $r$ & $12 \times 120$ & $> 24.2$ & $0.038$ & This work \\
\hline 
130609A & 2020-11-20 & MMT/Binospec & $r$ & $62 \times 30$ & \nodata$^{\dagger}$ & $0.074$ & This work \\
 & 2020-12-01 & MMT/MMIRS & $J$ & $30 \times 60$ & \nodata$^{\dagger}$ & $0.023
$ &  \\
\hline 
131229A & 2014-10-22 & LBT/LBC & $g$ & $2 \times 300$ & $> 24.9$ & $0.892$ & This
work \\
  & 2014-08-14 & HST/WFC3 & $F606W$ & $1125$ & $> 25.8$ & $0.672$ & \citet{Chrimes2019}\\
 & 2014-10-22 & LBT/LBC & $r$ & $5 \times 300$ & $> 24.7$ & $0.617$ &  This Work\\
 & 2014-10-22 & LBT/LBC & $i$ & $10 \times 150$ & $> 24.6$ & $0.459
$ &  \\
 & 2014-10-22 & LBT/LBC & $z$ & $4 \times 150$ & $> 23.2$ & $0.341$ &  \\
 & 2013-12-29 & Magellan/LDSS3 & $z$ & $7 \times 180$ & $> 23.96$ & $0.341$ & \citet{2013GCN.15637_131229A_Magellan}$^{a}$ \\
 & 2020-01-08 & MMT/MMIRS & $Y$ & $90 \times 60$ & $23.260 \pm
0.192$ & $0.294$ & This Work \\
 & 2015-03-29 & Magellan/Fourstar & $J$ & $22 \times 61$ & $23.382 \pm 0.381 $ & $0.191$ &  \\
 & 2014-08-14 & HST/WFC3 & $F160W$ & $1209$ & $23.235 \pm 0.077$ & $0.138$ & \cite{Chrimes2019} \\
 & 2015-03-30 & Magellan/Fourstar & $K_s$ & $198 \times 11.6$ & $22.949 \pm  0.285$ & $0.082$ & This work \\
\hline
140713A & 2014-10-23 & LBT/LBC & $g$ & 300 $\times$ 3& 24.12 $\pm$ 0.09 & 0.16 &This work\\
 & 2014-10-23 & LBT/LBC & $r$ & 300 $\times$ 4 & 23.85 $\pm$ 0.19 & 0.11 &\\
 & 2014-10-23 & LBT/LBC & $i$ & 150 $\times$ 8 & 22.65 $\pm$ 0.09 & 0.08 &\\
 & 2014-10-23 & LBT/LBC & $z$ & 150 $\times$ 6& 22.36 $\pm$ 0.12 & 0.06 &\\
 & 2018-10-18 & Keck/MOSFIRE & $J$ & 60 $\times$ 27 & 21.93 $\pm$ 0.16 & 0.04 &\\
 %& 2019-09-20 & MMT/MMIRS & $H$ & 13 $\times$ 180 & 22.38 $\pm$ 0.24 & 0.02 &\\
 & 2018-10-18 & Keck/MOSFIRE & K$_s$  & 60 $\times$ 29 & 21.69 $\pm$ 0.18 & 0.02 &\\
 & 2016-11-08 & Spitzer/IRAC & 3.6$\mu$m & 36 $\times$ 100 & $21.45 \pm 0.05 $& 0.01 &\cite{Higgins2019}\\
 & 2016-11-08 & Spitzer/IRAC & 4.5$\mu$m & 36 $\times$ 100 & $21.82 \pm 0.05$ & - \\
\hline
160509A & 2016-06-07 & Keck/LRIS & $g'$ & 3 $\times$ 330 & 24.434 $\pm$ 0.12 & 0.956 & \citet{Laskar2016_160509A_lab16}\\
 & 2016-06-07 & Keck/LRIS & $r'$ & 3 $\times$ 300 & 23.519 $\pm$ 0.35 & 0.661 & \\
  & 2017-07-05 & HST/WFC3 & $F110W$ & 2697 & 22.565 $\pm$ 0.03 & 0.255 & This Work \\
 & 2017-07-05 & HST/WFC3 & $F160W$ & 2797 & 22.292 $\pm$ 0.03 & 0.148 & \\
 & 2017-11-05 & Spitzer/IRAC & 3.6$\mu$m & 32 $\times$ 93.6 & $19.5172 \pm 0.041$ & 0.047 & \\
\enddata
\tablecomments{All values are in AB magnitudes and are corrected for the Galactic extinction, $A_{\lambda}$, in the direction of the burst \citep{sf11}. \\
$^{a}$ We use this $z$-band limit in our host galaxy modeling as it is the most constraining for this burst (see Section~\ref{sec:Prospector})
$\dagger$ No host identified.}
\end{deluxetable*} 

% 140713A & 2014-10-23 & LBT/LBC & $g$ & 300 $\times$ 3& 24.124 $\pm$ 0.0914 & 0.164 &This work\\
%  & 2014-10-23 & LBT/LBC & $r$ & 300 $\times$ 4 & 23.854 $\pm$ 0.1898 & 0.113 &This work\\
%  & 2014-10-23 & LBT/LBC & $i$ & 150 $\times$ 8 & 22.645 $\pm$ 0.0858 & 0.084 &This work\\
%  & 2014-10-23 & LBT/LBC & $z$ & 150 $\times$ 6& 22.357 $\pm$ 0.1188 & 0.063 &This work\\
%  & 2018-10-18 & Keck/MOSFIRE & $J$ & 60 $\times$ 27 & 21.932 $\pm$ 0.155 & 0.035 &This work\\
%  & 2019-09-20 & MMT/MMIRS & $H$ & 13 $\times$ 180 & 22.381 $\pm$ 0.239 & 0.022 &This work\\
%  & 2018-10-18 & Keck/MOSFIRE & K$_s$  & 60 $\times$ 29 & 21.689 $\pm$ 0.178 & 0.015 &This work\\
%  & 2016-11-08 & Spitzer/IRAC & 3.6$\mu$m & 36 $\times$ 100 & $21.45 \pm 0.05 $& 0.008 &\cite{Higgins2019}\\
%  & 2016-11-08 & Spitzer/IRAC & 4.5$\mu$m & 36 $\times$ 100 & $21.82 \pm 0.05$ & - &\cite{Higgins2019}

%%%% HOST GALAXY OBS TABLE

We initiated NIR $YJHK$-band observations of the host galaxy in 2016 March, 2020 December, and 2021 January using the Wide-field Camera (WFCAM; \citealt{caa+07}) mounted on the 3.8-m United Kingdom Infrared Telescope (UKIRT) and the MMT and Magellan Infrared Spectrograph (MMIRS) on the MMT telescope. For UKIRT data, we obtained pre-processed images from the WFCAM Science Archive \citep{hcc+08}, which are corrected for bias, flat-field, and dark current by the Cambridge Astronomical Survey Unit\footnote{http://casu.ast.cam.ac.uk/}. For each filter, we co-added the images and performed astrometry relative to 2MASS using a combination of tasks in Starlink\footnote{http://starlink.eao.hawaii.edu/starlink} and IRAF. For MMIRS data, we used a custom pipeline, {\tt POTPyRI}\footnote{\url{https://github.com/CIERA-Transients/POTPyRI}}, to perform bias, flat-fielding, dark corrections, sky subtraction, and co-addition.

The host galaxy is detected in all bands. We performed aperture photometry of the host using IRAF/{\tt phot} with a source radius of $2.5 \times \theta_{\rm FWHM}$ and a background annulus immediately surrounding the host. For photometric calibration, we use sources in the field in common with the SDSS~DR12 and 2MASS catalogs and employ the standard Vega to AB conversions as necessary. The host galaxy is faint, with $r \approx 23.6$~mag, but with red colors ($r-J \approx 1.7$~mag). Our full photometric results are listed in Table~\ref{tab:host}. Using the projected offset from the radio and optical afterglows and the $r$-band magnitude, we calculate a probability of chance coincidence \citep{bkd02}, $P_{cc} \approx 0.003-0.007$, establishing a robust association with GRB\,130131A.

We obtained $2\times 1200$~s of spectroscopy with the Multi-Object Double Spectrograph (MODS) mounted on the LBT on 2015 December 4 UT with the 200l grating, to cover a wavelength range of $\lambda \approx 3200-10000$\AA. The spectral continuum is weakly detected with an average $S/N\approx 1.2$. The spectrum does not exhibit any obvious features to $\lambda \approx 10000$\AA, yielding a tentative lower limit on the redshift of $z \gtrsim 1.3$ (due to the absence of [OII]$\lambda3727$). This is consistent with the redshift upper limit of $z \lesssim 4$ from the detection of the host in the {\it HST} F160W and F606W bands \citep{Chrimes2019}. 

The host galaxy is well detected in \textit{Spitzer}/IRAC $3.6\,\mu$m and $4.5\,\mu$m observations taken on 2016 July 17 (PI: Perley). We downloaded the pipeline processed post-basic calibrated data (pbcd) mosaics and performed photometry using a 3\arcsec\ aperture and 3\arcsec--7\arcsec\ background annulus, masking nearby bright sources from the background region. We applied standard aperture corrections\footnote{\url{https://irsa.ipac.caltech.edu/data/SPITZER/docs/irac/calibrationfiles/ap_corr_warm/}}, and obtain host magnitudes of $\approx 21.2~{\rm mag}$ and $\approx 21.0~{\rm mag}$ at $3.6~\mu{\rm m}$ and $4.5~\mu{\rm m}$, respectively.

\subsection{GRB\,130420B}
\subsubsection{{\it Swift} and Optical Observations}

GRB\,130420B was discovered by BAT on 2013 April 20.54 \citep{2013GCN.14411_130420B_Swiftdet}. The XRT started observations of GRB\,130420B at $\delta t = 54.5~{\rm s}$, finding an uncatalogued X-ray source within the BAT position \citep{2013GCN.14411_130420B_Swiftdet}. Ground-based follow-up at optical wavelengths to search for the optical afterglow of GRB\,130420B reached limiting magnitudes of $R > 22.6~{\rm mag}$ at a mean time of $\delta t \approx 0.02~{\rm days}$ using the 2.4m Gao-Mei-Gu (GMG) telescope \citep{2013GCN.14423_130420B_GMG_limit}.
We interpolate the X-ray light curve to the time of the most constraining optical limit, and we calculate $\beta_{\rm OX} \gtrsim -0.35$, meeting the criterion of a dark burst as determined by \citet{Jakobsson2004}. We find $\beta_{\rm X} = -0.8 \pm 0.2$, leading to $\beta_{\rm OX}-\beta_{\rm X} \gtrsim 0.45$, which does not meet the darkness criteria of \citet{vanderHorst2009}, due to GRB\,130420B's shallow $\beta_{\rm X}$.

\subsubsection{Radio and Millimeter Observations}

We initiated observations of GRB\,130420B on 2013 April 23.14 ($\delta t \approx 2.60~{\rm days}$) with CARMA under program number c0999 (PI: A. Zauderer) at a mean frequency of 85.5~GHz. We do not detect any source within the XRT error circle to a $3\sigma$ limit of $< 436~\mu{\rm Jy}$ (the information presented here supercedes the information from a preliminary reduction reported in \citealt{2013GCN.14444_130420B_radio}).

We initiated observations of GRB\,130420B on 2013 April 23.32 ($\delta t \approx 2.78~{\rm days}$) with the VLA under program 13A-046 (PI: E. Berger) at 6.0~GHz and 21.8~GHz. We do not detect a source within the XRT error circle to a $3\sigma$ limit of $<22.8~\mu{\rm Jy}$ at 6.0~GHz and $<34.8~\mu{\rm Jy}$ at 21.8~GHz (the information presented here supercedes the information from a preliminary reduction reported in \citealt{2013GCN.14444_130420B_radio}). 

\subsubsection{Host Galaxy Search}
We obtained MMT/Binospec $r$-band observations of the field of GRB\,130420B in Nov 2020. While there is a point source directly to the west of the XRT position \citep{Evans2009}, we do not detect any source within the XRT position (90\% confidence) to a $3\sigma$ limit of $r\gtrsim 24.2$~mag. The Binospec image is shown in Figure~\ref{fig:Host_Galaxy_Panel}.

\subsection{GRB\,130609A}
\subsubsection{{\it Swift} and Optical Observations}

GRB\,130609A was discovered by BAT \citep{2013GCN.14828_130609A_SwiftDet}. XRT started observations at $\delta t = 66.8~{\rm s}$, finding an uncatalogued X-ray source within the BAT position \citep{2013GCN.14828_130609A_SwiftDet}. Deep optical and NIR observations to search for the afterglow of GRB\,130609A were taken at a mean time of $\delta t \approx 0.07~{\rm days}$, using the Reionization and Transients Infrared Camera (RATIR) in the $r'i'ZYJH$-bands, reaching limits of $>23.34~{\rm mag}$ to $> 21.06~{\rm mag}$ in $r'-$ and $H-$band, respectively \citep{2013GCN.14831_130609A_Ratir_nondet}. We interpolate the X-ray light curve to the time of the most constraining optical limit ($i' > 23.27~{\rm mag}$ at $\delta t \approx 0.07~{\rm days}$), and we calculate $\beta_{\rm OX} \gtrsim 0.49$, meeting the criterion of a dark burst as determined by \citet{Jakobsson2004}, and confirming the classification suggested by \citet{2013GCN.14830_130609A_P60_non-det}. We find $\beta_{\rm X} = -1.5 \pm 0.2$, leading to $\beta_{\rm OX}-\beta_{\rm X} \gtrsim 1.00$, confirming GRB\,130609A is also dark according to the criterion of \citet{vanderHorst2009}. 

\subsubsection{Radio and Millimeter Observations}

We initiated observations of GRB\,130609A on 2013 June 9.9 UT ($\delta t \approx 0.78~{\rm days}$) with the VLA under program 13A-046 (PI: E. Berger) at 6.0~GHz and 21.8~GHz. We do not detect a source within the XRT error circle to a $3\sigma$ limit of $< 28.5~\mu{\rm Jy}$ at 6.0~GHz and $< 38.4~\mu{\rm Jy}$ at 21.8~GHz \citep{2013GCN.14863_130609A_radio}.

We initiated observations of GRB\,130609A on 2013 June 10.15 ($\delta t \approx 1.02~{\rm days}$) with CARMA under program number c0999 (PI: A. Zauderer) at a mean frequency of 85.5~GHz.
We do not detect any source within the XRT error circle to a $3\sigma$ limit of $<341~\mu{\rm Jy}$ (the information presented here supercedes the information from a preliminary reduction reported in \citealt{2013GCN.14863_130609A_radio}).

\subsubsection{Host Galaxy Search}

We obtained deep MMT/Binospec $r$-band and MMIRS $J$-band observations of the field of GRB\,130609A in Nov and Dec 2020 to search for a host galaxy. The $r$-band imaging reveals a source within the XRT position \citep{Evans2009} with $r=23.29 \pm 0.11$~mag; this source is also weakly detected in $J$-band (labeled ``S1'' in Figure~\ref{fig:Host_Galaxy_Panel}). However, this source has a PSF consistent with being point-like, indicating that it is a foreground star. We identify a second, fainter source to the northwest of the XRT position (labeled ``S2'' in Figure~\ref{fig:Host_Galaxy_Panel}) with $r=24.48 \pm 0.14$~mag that is more clearly extended and is a potential host of GRB\,130609A. Thus, it is difficult to draw any strong conclusions regarding the origin or redshift of GRB\,130609A.

\subsection{GRB\,131229A}
\subsubsection{X-ray and Optical Observations}
\label{sec:131229AXray_and_optical}

GRB\,131229A was discovered by BAT on 2013 December 29.28 \citep{2013GCN.15627_131229A_Swift_det}. The XRT started observations of GRB\,131229A at $\delta t = 93.8~{\rm s}$, finding an uncatalogued X-ray source within the BAT position \citep{2013GCN.15627_131229A_Swift_det}. We initiated deep optical ground-based observations to search for the optical afterglow of GRB\,131229A with the 6.5 m Magellan Clay telescope in $r'i'z'$-bands at a mean time of $\delta t \approx 0.03~{\rm days}$, reaching $z' > 24.3~{\rm mag}$ \citep{2013GCN.15637_131229A_Magellan}. We interpolate the X-ray light curve to the time of the most constraining optical limit, and we calculate $\beta_{\rm OX} \gtrsim 0.24$, meeting the criterion of a dark burst as determined by \citet{Jakobsson2004}, and confirming the classification of \citet{Chrimes2019}. We find $\beta_{\rm X} = -1.2 \pm 0.1$, leading to $\beta_{\rm OX}-\beta_{\rm X} \gtrsim 1.41$, confirming GRB\,131229A is also dark according to the criterion of \citet{vanderHorst2009}.

{\it Chandra} observations of GRB\,131229A (PI:~Levan, ObsID 15195) were initiated at a midtime of 2014 Jan 06.12 ($\delta t = 7.84~{\rm days}$) with a total effective exposure time of 15.05 ks. 
The X-ray afterglow was detected with a count rate of  %(1.06$~\pm~$0.15)$~\times~$10$^{−3}$~s$^{-1}$
$(1.06 \pm 0.15) \times 10^{-3}$ s$^{-1}$ \citep{Chrimes2019}.

\subsubsection{Radio and Millimeter Observations}
\label{sec:131229A_radio}

We initiated observations of GRB\,131229A on 2013 December 30.18 ($\delta t \approx 0.91~{\rm days}$) with the VLA under program 13A-541 (PI: E. Berger) at 6.0~GHz and 21.8~GHz. We do not detect a source within the XRT error circle at 21.8~GHz to a $3 \sigma$ limit of $< 45.6~\mu{\rm Jy}$. We detect a clear source  within the XRT error circle of $3.9\sigma$ significance of $74~\mu{\rm Jy}$ at 6.0~GHz at a position of R.A.=\ra{05}{40}{55.649} ($\pm 0.093 \arcsec$) and Dec=\dec{-04}{23}{47.098} ($\pm 0.115 \arcsec$). We did not re-observe the field, and thus the variability of the source could not be determined, therefore we cannot definitively claim this source as the radio afterglow of GRB\,131229A. To place limits on the presence of a background radio source, we searched the VLA Sky Survey (VLASS, \citealt{VLASS}), and found no source within the XRT localization to a limit of $\lesssim 420~\mu{\rm Jy}$ at 3.0~GHz (with observations taken $\delta t \approx 3.9~{\rm yr}$). 

GRB\,131229A was observed with CARMA at 93.0~GHz at a mean time of $\delta t \approx 1.00~{\rm days}$ after the burst. No millimeter afterglow emission was found within the XRT error circle to a limit of $\lesssim 0.6~{\rm mJy}$ \citep{2013GCN.15638_131229A_CARMA}.

\subsubsection{Host Galaxy Observations}
\label{sec:131229Ahostobs}

The faint host galaxy of GRB\,131229A was first reported in \citet{Chrimes2019} from {\it HST}/F160W imaging, and it is the only detected object within the 90\% XRT position (Figure~\ref{fig:Host_Galaxy_Panel}, \citealt{Evans2009}). We obtained observations of the field with LBT/LBC ($griz$-bands), MMT/MMIRS ($Y$-band), and Magellan/Fourstar ($JK$-bands). We reduced and co-added the data in a similar manner as described before. The host galaxy is not detected in any of our optical imaging to deep limits of $\gtrsim 23.2-24.9$~mag, and is weakly detected in our $Y$, $J$ and $K$-band imaging with $K=22.95 \pm 0.29$~mag. Overall, the host is red, with $r-K \gtrsim 1.8$~mag, is at a $0.41''$ offset from the VLA position and has a $P_{cc} = 1.5 \times 10^{-3}$. Our full photometric results are listed in Table~\ref{tab:host}. From the $K$-band image, we measure a position for the host of R.A.=\ra{5}{40}{55.632} and Dec=\dec{-4}{23}{46.77}.
The deep non-detection of the host in the $i$-band, coupled with a brightness at $Y$-band that is $\gtrsim 1.3$~mag brighter, suggest a 4000\AA\ break in this wavelength regime, with a redshift of $z \gtrsim 1-1.5$.

To place the {\it Chandra} afterglow on the host galaxy image, we perform relative astrometry using three common sources between {\it Chandra} and Magellan $K$-band. We obtained the {\it Chandra} observation from the archive (PI:~Levan; ObsID 15195). We find a tie uncertainty of $\sigma_{{\rm Magellan} \rightarrow {\it Chandra}}=0.13\arcsec$. The corrected position is R.A. = \ra{05}{40}{55.64} and Dec. = \dec{-04}{23}{46.824} with a positional uncertainty of $0.63\arcsec$ (including the uncertainty in the astrometric tie, the positional uncertainty, and the absolute astrometric uncertainty of $0.6\arcsec$). Our {\it Chandra} afterglow position is consistent with the VLA C-band afterglow position and intersects the host galaxy of GRB\,131229A (Figure~\ref{fig:Host_Galaxy_Panel}). 

\subsection{GRB\,140713A}
\subsubsection{{\it Swift} and Optical Observations}
\label{sec:140713AXrayOpticalObs}

GRB\,140713A was discovered by BAT on 2014 July 13.78 \citep{2014GCN.16581SwiftDet}.The XRT began observations starting at $\delta t= 72.8$~s, detecting an uncatalogued X-ray source within the BAT position \citep{2014GCN.16581SwiftDet, 2014GCN.16585XRTRefined}. Deep optical observations to search for the optical afterglow using the ALFOSC instrument on the 2.5-m Nordic Optical Telescope (NOT) were taken at $\delta t \approx 0.14-0.17~{\rm days}$, resulting in $3\sigma$ upper limits on the afterglow flux of $r\gtrsim24.30~{\rm mag}$, $i\gtrsim23.50~{\rm mag}$, and $z\gtrsim22.60~{\rm mag}$ \citep{2014GCN.16587OPTICALNOT,Higgins2019}.  We interpolate the X-ray light curve to the time of the deepest optical limit  ($r > 24.30$~mag at  $\delta t \approx 0.15~{\rm days}$), and we calculate $\beta_{\rm OX} \gtrsim -0.26$, meeting the criterion of a dark burst as determined by \citet{Jakobsson2004}, and confirming the classification of \citet{Higgins2019}. We find  $\beta_{\rm X} = -1.0 \pm 0.2$, leading to $\beta_{\rm OX} - \beta_{\rm X} \gtrsim 0.73$, indicating GRB\,140713A also meets the darkness criterion of \citet{vanderHorst2009}.

\subsubsection{Radio and Millimeter Observations}
\label{sec:140713Aradioobs}
We initiated a series of seven observations of GRB\,140713A starting on 2014 July 15.37 UT ($\delta t \approx 1.59~{\rm days}$) until 2014 August 11.14 UT ($\delta t \approx 28.36~{\rm days}$) with CARMA under program number c0999 (PI: A. Zauderer) at a mean frequency of 85.5~GHz. In the first observation, we detect a clear source within the XRT error circle of $5\sigma$ significance of $\approx 1.5~{\rm mJy}$; the information presented here supercedes the information from a preliminary reduction reported in \citet{2014GCN.16593CARMA}. We report the CARMA afterglow flux densities in Table~\ref{tab:obs}. We obtain a position for the millimeter afterglow of GRB\,140713A of R.A.=\ra{18}{44}{25.403} ($\pm 1.489\arcsec$) and Dec=\dec{+59}{38}{00.97} ($\pm 0.781\arcsec$).

We initiated a series of six observations of GRB\,140713A from 2014 July 17.08 UT ($\delta t \approx 3.36~{\rm days}$) to 2014 September 15.00 UT ($\delta$t $\approx$ 63.21 days), with the VLA (Program number 14A-344, PI: Berger) at 6.0~GHz (side-band frequencies of 4.9 GHz and 7.0 GHz) and 14.7~GHz (side-band frequencies of 13.4 GHz and 15.9 GHz) for all observations. In the first observation, we detected a source within the XRT error circle, and consistent with the CARMA position, with $F_{\nu}\approx0.10~{\rm mJy}$ at 6.0~GHz and $F_{\nu} \approx 0.82~{\rm mJy}$ at 14.7~GHz. Using the observation with the highest signal to noise (the third 14.7~GHz epoch), we derive a position for the radio afterglow of R.A.=\ra{18}{44}{25.481} ($\pm 0.149\arcsec$) and Dec=\dec{+59}{38}{00.69} ($\pm 0.052\arcsec$), an improvement on our CARMA position.

In addition, we initiated  observations of GRB\,1407134A with the Plateau deBure Interferometer (PdBI) at 86.7~GHz as part of a long-term ToO program (Program number S14DD004, PI: A. Castro-Tirado). The PdBI observed the source at six separate epochs across 2014 Jul 18-Sep 14 UT ($\delta t \approx 4.4 - 63.1$ days). We reduced the data with the standard CLIC and MAPPING software distributed by the Grenoble GILDAS group \footnote{https://www.iram.fr/IRAMFR/GILDAS}, and use the carbon star MWC349 as the flux calibrator. We detect a source in all epochs except the final epoch, at a position consistent with the millimeter and radio afterglows. The flux measurements of these observations are listed in Table~\ref{tab:obs}.

To supplement our CARMA, VLA, and PdBI data, we include literature data in our subsequent analysis from the Arcminute Microkelvin Imager (AMI) Large Array (mean frequency of 15.7~GHz) \citep{2014GCN.16603AMIobs, Anderson2018}, 1.4~GHz and 4.8~GHz observations from the Westerbork Synthesis Radio Telescope (WSRT) \citep{Higgins2019}, and 1.4~GHz upper limits from the Giant Metrewave Radio Telescope (GMRT) \citep{2014GCN.16641GMRT}. The AMI and WSRT afterglow observations, along with the NOT optical upper limits, were previously modeled alongside the {\it Swift} X-ray light curve in \citet{Higgins2019}, but the VLA, CARMA, and PdBI observations are presented and modeled for the first time in this work. 

\subsubsection{Host Galaxy Observations}
\label{sec:140713Ahostobs}

Observations by the $10.4$m Gran Telescopio CANARIAS (GTC) telescope at $\delta t \approx 3.1$~days revealed a faint, $r\approx 24$~mag source as the candidate host galaxy \citep{2014GCN.16602HostGal}. Identification and further analysis of this source was also provided in \citet{Higgins2019}. We initiated $griz$-band observations of the host galaxy with the Large Binocular Camera (LBC) on the Large Binocular Telescope (LBT/LBC) on 2014 Oct 23 UT. We reduced and co-added the data in a similar manner as described before. We calibrated the absolute astrometry using sources in common with the Pan-STARRS1 catalog \citep{Chambers2016} using IRAF/{\tt ccmap} and {\tt ccsetwcs}. We performed aperture photometry on these images using IRAF/{\tt phot} using a source radius of $2.5 \times \theta_{\rm FWHM}$ and a background annulus at the host position. For photometric calibration, we use standard stars from PAN-STARRS1 in the field \citep{Chambers2016}, and then convert to the SDSS system using standard transformations \cite{Tonry2012}. We find a host magnitude of $r = 23.8 \pm 0.2$~mag, and the host galaxy photometry in all bands is listed in Table~\ref{tab:host}. 

In addition, we initiated $J$ and K$_s$-band NIR observations of the host galaxy in Oct 2018 using the Multi-Object Spectrometer for Infra-Red Exploration (MOSFIRE) instrument mounted on the Keck I telescope (PI: Fong), We used {\tt POTPyRI}\footnote{https://github.com/CIERA-Transients/POTPyRI} to perform bias, flat-fielding, dark corrections, sky subtraction, and co-addition. We calibrated the absolute astrometry to 2MASS using IRAF/{\tt ccmap} and {\tt ccsetwcs}. The host galaxy is well detected in each of the NIR bands. We calibrated to 2MASS, and converted to the AB system \citep{Blanton2007}; The magnitude values are listed in Table~\ref{tab:host}.

We obtained $3\times1800$~s of host galaxy spectroscopy using the Low Resolution Imaging Spectrometer (LRIS) mounted on the 10m Keck I telescope on 2018 October 6 UT. We used the 400/3400 grism and 400/8500 grating in combination with the D560 dichroic for an effective wavelength range of $\sim$3200-10100\AA.  The raw frames were corrected for bias from the overscan region, flattened, and stitched together using custom methods implemented in {\tt pyraf}\footnote{\url{https://github.com/msiebert1/UCSC_spectral_pipeline}}.  We then extracted one-dimensional spectra of the host galaxy and applied a dispersion correction derived from arc-lamp spectra.  Finally, we applied an atmospheric absorption correction and flux calibration derived from a spectrum of the spectrophotometric standard GD71 obtained on the same night.  We detect a faint continuum (S/N$\sim$10) with several clear emission features. We identify [OII]$\lambda 3727$, H$\beta \lambda 4861$, and [OIII]$\lambda5007$ at a common redshift of $z=0.935 \pm 0.002$. The spectrum is shown in Figure~\ref{fig:140713A_Prospector_SED}.

To supplement our optical and NIR host galaxy observations, we include $3.6$ and $4.5 \mu$m observations from {\it Spitzer}/IRAC photometry, published in \citet{Higgins2019} (Table~\ref{tab:host}). Our $grizJHK_s$ data, as well as the available {\it Spitzer} photometry are subsequently used in our host galaxy modeling.

\subsection{Literature Bursts}

\subsubsection{GRB\,110709B}
\label{sec:110709B_literature}

We gather radio and optical afterglow observations of GRB\,110709B from \citet{Zauderer2013} and X-ray afterglow observations from {\it Swift}. GRB\,110709B has been previously been determined to be a dark burst through the $\beta_{\rm OX} > -0.5$ criterion \citep{Jakobsson2004} in \citet{Zauderer2013}. We gather host galaxy observations of GRB\,110709B from \citet{Zauderer2013}, \citet{Perley2016_I}, and \citet{Selsing2019}.

\subsubsection{GRB\,111215A}
\label{sec:111215A_literature}

We gather radio and optical afterglow observations of GRB\,111215A from \citet{Zauderer2013} and \citet{VanderHorst2015}, and X-ray afterglow observations from {\it Swift}. GRB\,111215A has previously been determined to be a dark burst through the $\beta_{\rm OX} > -0.5$ criterion \citep{Jakobsson2004} in \citet{Zauderer2013}. We gather host galaxy observations of GRB\,111215A from \citet{VanderHorst2015}, and assume a redshift of $z = 2.012$ \citep{VanderHorst2015, Chrimes2019}.

\subsubsection{GRB\,160509A}
\label{sec:160509A_literature}

The radio, optical, and X-ray afterglow of GRB\,160509A has previously been modeled with our modeling framework in \citet{Laskar2016_160509A_lab16}, and we utilize the afterglow parameters within for our discussion (Section~\ref{sec:Discussion}). The optical afterglow of GRB\,160509A was heavily obscured, with a line-of-sight extinction of $A_{V, \rm GRB} \approx 3.4~{\rm mag}$, indicating it is likely a dark burst. Using the optical afterglow detections in $r'$-band at $\delta t \approx 0.25~{\rm days}$ \citep{Laskar2016_160509A_lab16}, we calculate $\beta_{\rm OX} \approx 0.03$, meeting the darkness criterion of \citet{Jakobsson2004}, and classifying GRB\,160509A as a dark burst. Additionally, we find $\beta_{\rm X} = -1.0\pm 0.1$, leading to $\beta_{\rm OX} - \beta_{\rm X} \approx 0.98$. Thus, GRB\,160509A is also classified as a dark burst according to the \citet{vanderHorst2009} criterion.

The host galaxy of GRB\,160509A was previously observed with {\it HST}/WFC3 in the F110W and F160W filters on 2017 July 5 (PI: Kangas) \citep{Kangas2020_160509A}. We retrieved, aligned, and drizzled the individual exposures for each band using the {\it HST} reduction pipeline {\tt hst123} \citep{Kilpatrick21_gw170817,hst123}.  We also included a sky subtraction step as part of {\tt astrodrizzle} \citep{drizzlepac} to remove large-scale background emission near the host galaxy.  Using an elliptical aperture and the tabulated {\it HST} zeropoints, we calculated the F110W and F160W brightness of the host galaxy (Table~\ref{tab:host} and Figure~\ref{fig:Host_Galaxy_Panel}).

The host galaxy is also clearly detected in \textit{Spitzer}/IRAC $3.6~\mu{\rm m}$ observations taken on 2017 Nov 05 (PI: Perley). We downloaded and reduced the individual basic calibrated data frames (i.e., {\tt cbcd}) using {\tt photpipe} \citep{Rest05,Kilpatrick18_16cfr}, including alignment, flux calibration, and optimal stacking to a pixel scale of 0.3\arcsec/pix.  We performed final PSF photometry in the stacked frames using a custom version of {\tt DoPhot} \citep{Schechter93}.  The host galaxy appears point-like in the stacked IRAC frames and is clearly separated from a nearby galaxy of similar brightness seen in the {\it HST} frames (Figure~\ref{fig:Host_Galaxy_Panel}).  Therefore, we use the {\tt DoPhot} photometry calibrated using the appropriate warm {\it Spitzer} sensitivity function for IRAC and obtain a magnitude of $\approx 19.6~{\rm mag}$ at $3.6~\mu{\rm m}$ (Table~\ref{tab:host}).

We supplement these host galaxy observations with the Keck/LRIS $g'$- and $r'$-band measurements taken at $\delta t \approx 28.2~{\rm days}$, where afterglow modeling found the host galaxy to dominate the flux (\citealt{Laskar2016_160509A_lab16}; Table~\ref{tab:host}). 

%%%%%%%%%%%%%%% TABLE
\tabletypesize{\normalsize}
\begin{deluxetable*}{c|ccccc}
\tablecolumns{5}
\tablewidth{0pc}
\tablecaption{Forward Shock Parameters
\label{tab:GAMMA_bestfit_stat_alt}}
\tablehead {
\colhead {GRB}                &
\colhead{110709B}   &
\colhead {130131A }          &
\colhead {140713A}         \\
\colhead{Env.} &
\colhead{Wind} &
\colhead{ISM} &
\colhead{ISM} 
}
\startdata
$p$ &  $2.01$  &  $2.67$  &  $2.02$  \\
 &  $2.02^{+0.01}_{-0.01}$  &  $2.45^{+0.15}_{-0.11}$  &  $2.02^{+0.01}_{-0.01}$  \\
\hline
$E_{\rm K}$ &  $2.91 \times 10^{-2}$  &  $1.02 \times 10^{-1}$  &  $8.75 \times 10^{-2}$  \\
$(10^{52}~{\rm erg})$ &  $4.33^{+3.17}_{-1.64} \times 10^{-2}$  &  $5.74^{+3.32}_{-3.28} \times 10^{-2}$  &  $9.19^{+1.37}_{-1.05} \times 10^{-2}$  \\
\hline
$A_*/n_{0}$ &  $1.45 \times 10^{-2}$  &  $7.83 \times 10^{-1}$  &  $1.91 \times 10^{0}$  \\
$(-/{\rm cm}^{-3})$ &  $2.65^{+2.46}_{-1.25} \times 10^{-2}$  &  $1.29^{+0.82}_{-0.59} \times 10^{0}$  &  $2.24^{+0.79}_{-0.49} \times 10^{0}$  \\
\hline
$\epsilon_e$ &  $4.77 \times 10^{-1}$  &  $4.64 \times 10^{-1}$  &  $1.04 \times 10^{-1}$  \\
 &  $3.75^{+1.78}_{-1.21} \times 10^{-1}$  &  $5.49^{+2.12}_{-1.7} \times 10^{-1}$  &  $8.81^{+4.03}_{-2.62} \times 10^{-2}$  \\
\hline
$\epsilon_B$ &  $1.66 \times 10^{-1}$  &  $1.33 \times 10^{-5}$  &  $5.57 \times 10^{-1}$  \\
 &  $4.0^{+14.26}_{-3.27} \times 10^{-2}$  &  $5.55^{+52.57}_{-3.97} \times 10^{-5}$  &  $4.4^{+2.28}_{-1.69} \times 10^{-1}$  \\
\hline
$t_{\rm jet}$ &  $2.38$  &  $0.79$  &  $9.51$  \\
$({\rm day})$ &  $2.44^{+0.21}_{-0.19}$  &  $0.86^{+0.24}_{-0.18}$  &  $9.4^{+0.71}_{-0.71}$  \\
\hline
$\theta_{\rm jet}$ &  $1.02$  &  $3.65$  &  $17.25$  \\
$({\rm deg})$ &  $1.14^{+0.13}_{-0.12}$  &  $4.67^{+0.67}_{-0.62}$  &  $17.49^{+0.52}_{-0.48}$  \\
\hline
$A_{V, \rm GRB}$ &  $\gtrsim 5.0$ &  $ \approx 2.4$ &  $\gtrsim 3.1$ \\
$({\rm mag})$ &  $\gtrsim 5.1^{a}$ &  $2.3^{+0.1}_{-0.1}$ &  $\gtrsim 3.1^{a}$ \\
\enddata
\tablecomments{
The top row for each parameter corresponds to the best fit forward shock value from our MCMC modeling. The bottom row for each parameter corresponds to the summary statistics from the marginalized posterior density functions (medians and 68\% credible intervals) \\
$^{a}$ $A_{V, \rm GRB}$ value from afterglow model using median values \\
}
\end{deluxetable*}
%%%%%%%%%%%%%%% TABLE

\section{Alternative MCMC Afterglow Models}
Here we present the alternative MCMC afterglow models for the bursts whose afterglow observations did not distinguish between a wind or ISM environment through preliminary analytical arguments (GRB\,110709B, 130131A, and 140713A).

\subsection{GRB\,110709B}
\label{appendix:110709B}
\begin{figure}
\centering
\includegraphics[width=0.45\textwidth]{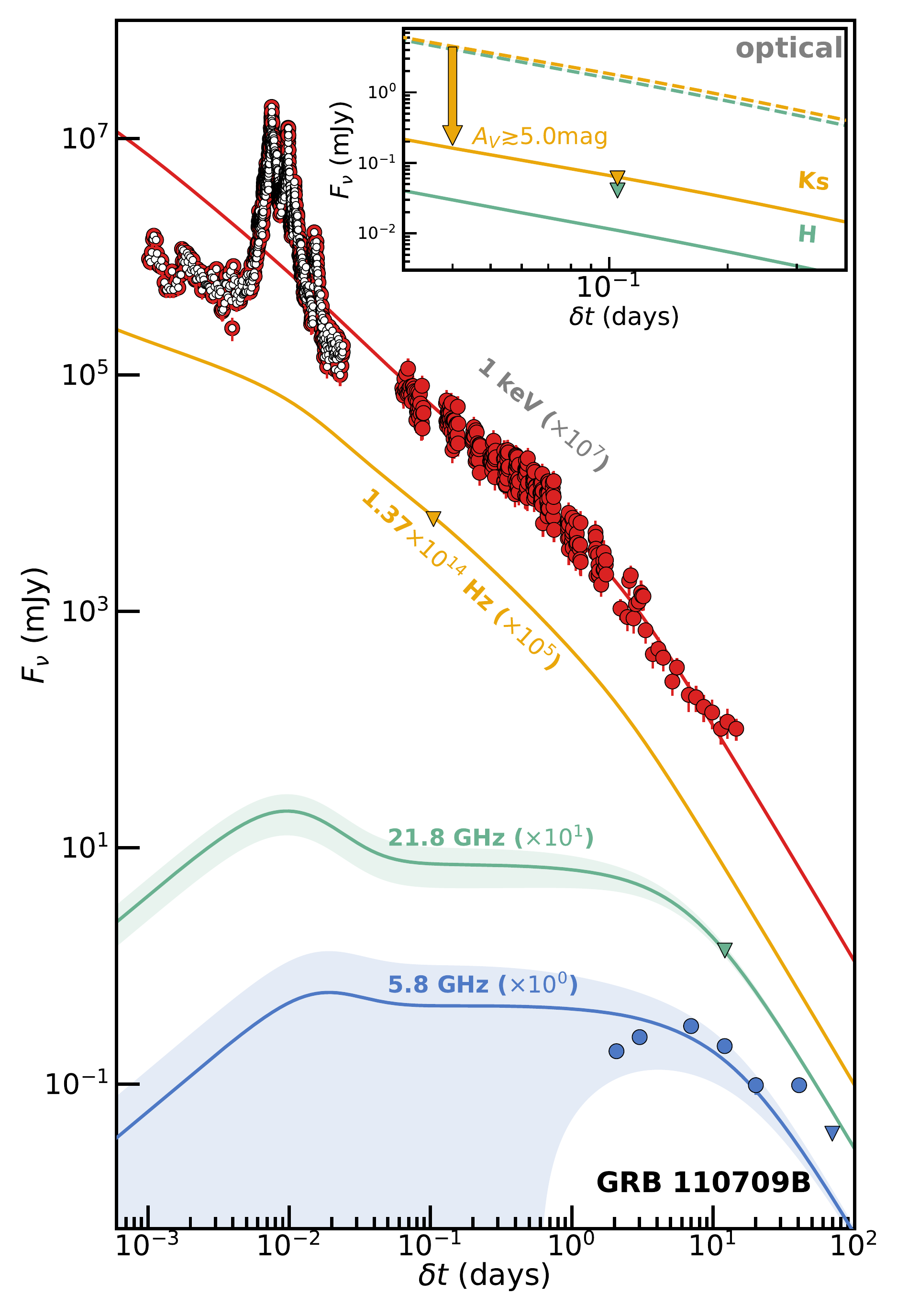}
\caption{X-ray, optical, radio afterglow light curves of GRB\,110709B, together with the best-fit forward shock model in an wind environment (lines). Circles represent literature data \citep{Zauderer2013}, and triangles represent $3\sigma$ upper limits. Open symbols indicate data that are not included in the fit, and shaded regions represent variability due to scintillation. The inset shows the model $K_s$-band and $H$-band light curves (solid lines) as well as the non-extinguished models (dashed lines), indicating $A_{V, \rm GRB}\gtrsim 5.0$~mag to explain the upper limits.
} 
\label{fig:lc110709B_wind}
\end{figure}

We present the best fit (highest likelihood) wind environment model for GRB\,110709B in Figure~\ref{fig:lc110709B_wind} and list the model parameters as well as the summary statistics from the marginalized posterior density functions (medians and 68\% credible intervals) in Table~\ref{tab:GAMMA_bestfit_stat} in Table~\ref{tab:GAMMA_bestfit_stat_alt}.

Our best fit wind model for GRB\,110709B finds a later $t_{\rm jet} \approx 2.4~{\rm days}$ compared to our ISM model (Section~{\ref{sec:110709B_MCMCModeling}}). While the wind model better fits the X-ray light curve at $\delta t \gtrsim 5~{\rm days}$, the marginal statistical preference for the ISM model is a result of the ISM model better matching the X-ray light curve at $\delta t \lesssim 5~{\rm days}$, where the majority of the X-ray data exist. The best fit parameters for our wind model are the same as \citet{KangasFruchter2021} within $2\sigma$, though our $E_{\rm K}$ and $A_*$ are orders of magnitude different than those found by \citet{Zauderer2013}, likely due to the inclusion of IC effects in our model. Our best fit wind model requires $A_{V, \rm GRB} \gtrsim 5.01~{\rm mag}$ to match the GROND optical/NIR afterglow upper limits.

%%%%% 110709B MODELING FIGURE
%%%%% 111215A MODELING FIGURE

\subsection{GRB\,130131A}
\label{appendix:130131A}

\begin{figure*}
\centering
\includegraphics[width=\textwidth]{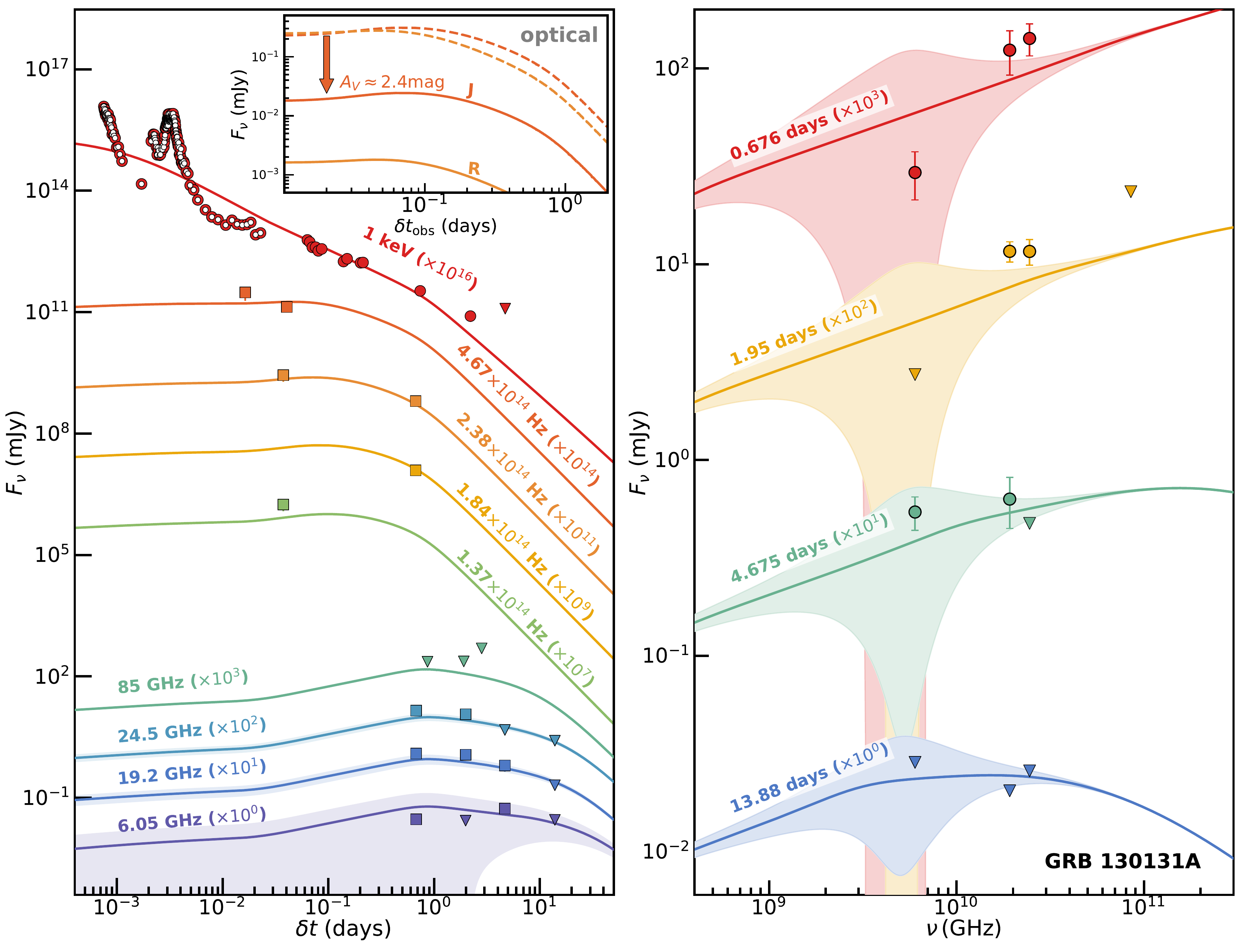}
\caption{{\it Left:} X-ray, optical, millimeter and radio afterglow light curves of GRB\,130131A, together with the best-fit forward shock model in an ISM environment (lines). Squares represent data newly reported here, and triangles represent $3\sigma$ upper limits. Open symbols indicate data that are not included in the fit, and shaded regions represent variability due to scintillation. The inset shows the model $J$- and $R$-band light curves (solid lines) as well as the non-extinguished models (dashed lines), indicating $A_{V, \rm GRB} \approx 2.4$~mag to explain the detections.
{\it Right:} Radio and millimeter spectral energy distributions (SEDs) of the afterglow of GRB\,130131A from $\delta t \approx 0.7$-$13.9$ days, together with the best-fit forward shock model (lines).
} 
\label{fig:lcspec130131A_ISM}
\end{figure*}

We present the best fit wind environment for GRB\,130131A model in Figure~\ref{fig:lcspec130131A_ISM} and list the model parameters as well as the summary statistics from the marginalized posterior density functions (medians and 68\% credible intervals) in Table~\ref{tab:GAMMA_bestfit_stat_alt}.

Our ISM model of GRB\,130131A finds an earlier jet break time of $t_{\rm jet} \approx 0.8~{\rm days}$ compared to our wind model. As a consequence, while the radio afterglow of GRB\,130131A is better fit by the ISM model, the X-ray light curve is under-predicted by the ISM model at $\delta t \gtrsim 1~{\rm day}$, resulting in the statistical preference for the wind model of GRB\,130131A. Our ISM model finds that the extinction required to match the optical and NIR observations is $A_{V, \rm GRB}\approx 2.41~{\rm mag}$. 

%%%%% 130131A MODELING FIGURE
%%%%% 130131A MODELING FIGURE

\subsection{GRB\,140713A}
\label{appendix:140713A}

\begin{figure*}
\centering
\includegraphics[width=\textwidth]{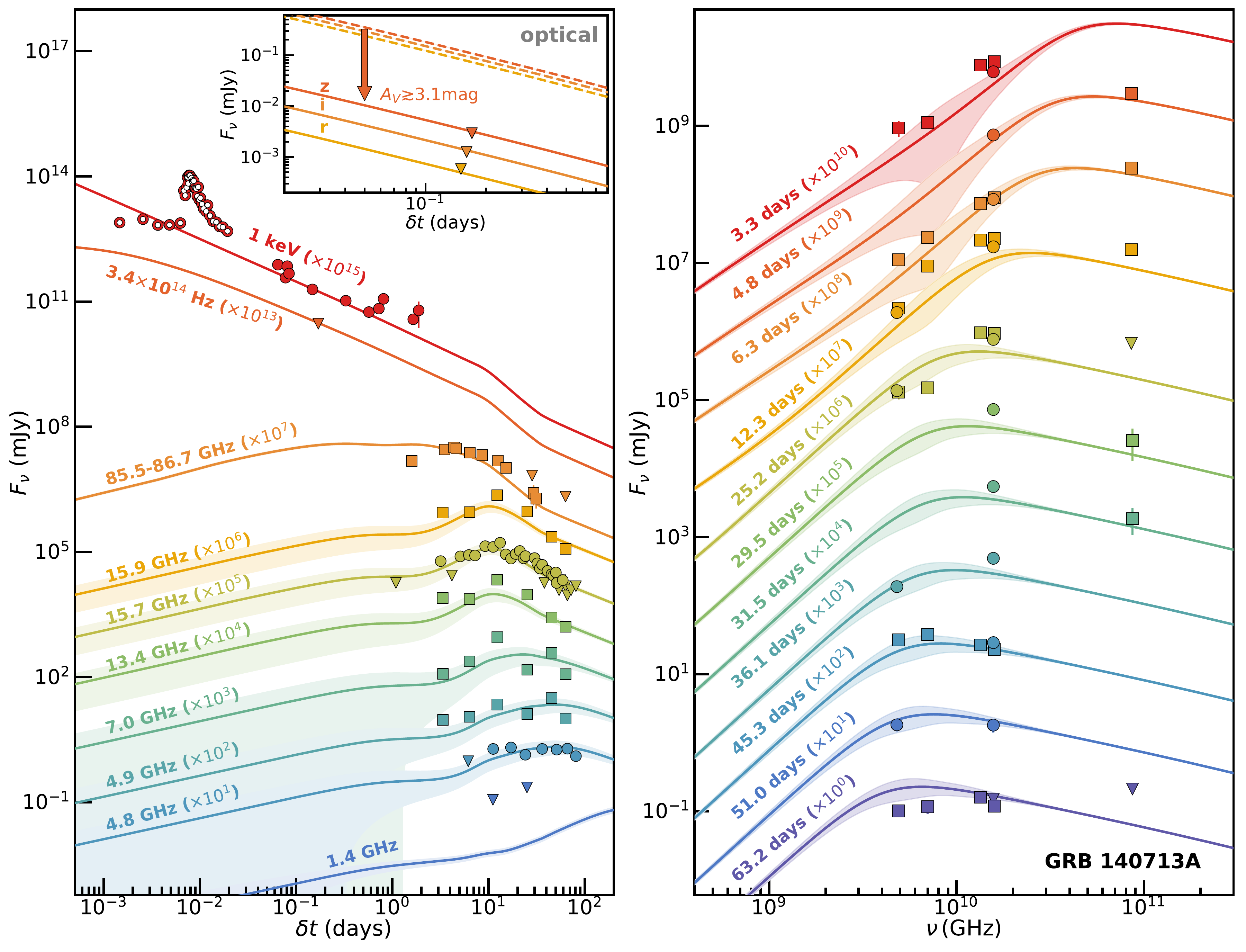}
\caption{{\it Left:} X-ray, optical, millimeter and radio afterglow light curves of GRB\,140713A, together with the best-fit forward shock model in an ISM environment (lines). Squares represent data newly reported here, circles represent literature data (\cite{Higgins2019}), and triangles represent $3\sigma$ upper limits. Open symbols indicate data that are not included in the fit, and shaded regions represent variability due to scintillation. The inset shows the model $r$-, $i$-, and $z$-band light curves (solid lines) as well as the non-extinguished models (dashed lines), indicating $A_{V, \rm GRB}\gtrsim3.1$~mag to explain the upper limits. 
{\it Right:} Radio and millimeter spectral energy distributions (SEDs) of the afterglow of GRB\,140713A from $\delta t \approx 3.0$-$31.5$ days, together with the best-fit forward shock model (lines).} 

\label{fig:lcspec140713A_ISM}
\end{figure*}

We present the best fit ISM environment model for GRB\,140713A in Figure~\ref{fig:lcspec140713A_ISM} and list the model parameters as well as the summary statistics from the marginalized posterior density functions (medians and 68\% credible intervals) in Table~\ref{tab:GAMMA_bestfit_stat} in Table~\ref{tab:GAMMA_bestfit_stat_alt}.

Similar to the wind environment model (Section~\ref{sec:140713A_MCMCModeling}), the X-ray light curve for the ISM model of GRB\,140713A is under-predicted at $\delta t \gtrsim 0.7~{\rm days}$. While the K$_u$-band observations are better fit by the ISM model, the significant statistical preference for the wind model is attributed to the better fit C-band and 3mm observations light curves. Our ISM model for GRB\,140713A requires $A_{V, \rm GRB} \gtrsim 3.12~{\rm mag}$ necessary to match the optical limits. This limit is the same as that derived by \citet{Higgins2019} for an SMC-like galactic extinction model.

%%%%% 140713A MODELING FIGURE
%%%%% 140713A MODELING FIGURE

% \end{appendices}

\newpage 
\bibliographystyle{aasjournal}
\bibliography{library}

% \clearpage

\end{document}